\PassOptionsToPackage{dvipsnames}{xcolor}
\documentclass[acmsmall]{acmart}

\AtBeginDocument{%
  \providecommand\BibTeX{{%
    \normalfont B\kern-0.5em{\scshape i\kern-0.25em b}\kern-0.8em\TeX}}}

\setcopyright{acmcopyright}
\copyrightyear{2024}
\acmYear{2024}
\acmDOI{XXXXXXX.XXXXXXX}

\acmJournal{TOSEM}
\acmVolume{1}
\acmNumber{1}
\acmArticle{1}
\acmMonth{10}

\usepackage{tikz}

\usepackage{multirow}
\usepackage{blindtext}
\usepackage{hyperref}
\usepackage{hyperxmp}
\usepackage{listings}
\usepackage{tcolorbox}

\usepackage{svg}

\begin{document}

\title{Do Current Language Models Support Code Intelligence for R Programming Language?}


\author{Zixiao Zhao}
\email{zixiaosh@student.ubc.ca}
\orcid{0009-0008-6947-2723}
\affiliation{%
  \institution{University of British Columbia}
  \city{Kelowna}
  \state{BC}
  \country{Canada}
}

\author{Fatemeh H. Fard}
\email{fatemeh.fard@ubc.ca}
\orcid{1234-5678-9012}
\affiliation{%
  \institution{University of British Columbia}
  \city{Kelowna}
  \country{Canada}
}

\renewcommand{\shortauthors}{Zhao and Fard}

\begin{abstract}

Recent advancements in developing Pre-trained Language Models for Code (Code-PLMs) have urged many areas of Software Engineering (SE) and brought breakthrough results for many SE tasks. Though these models have achieved the state-of-the-art performance for SE tasks for many popular programming languages, such as Java and Python, the Scientific Software and its related languages like R programming language have rarely benefited or even been evaluated with the Code-PLMs. Research has shown that R has many differences with other programming languages and requires specific techniques.  
In this study, we provide the first insights for code intelligence for R. For this purpose, we collect and open source an R dataset, and evaluate Code-PLMs for the two tasks of code summarization and method name prediction using several settings and strategies, including the differences in two R styles, \textit{Tidy-verse} and \textit{Base} R. 
Our results demonstrate that the studied models have experienced varying degrees of performance degradation when processing R programming language code, which is supported by human evaluation. Additionally, not all models show performance improvement in R-specific tasks even after multi-language fine-tuning. The dual syntax paradigms in R significantly impact the models' performance, particularly in code summarization tasks. Furthermore, the project-specific context inherent in R codebases significantly impacts the performance when attempting cross-project training. 
Interestingly, even when Large Language Models like CodeLlama and StarCoder2 are used for code generation, the Pass@K ($K=1,5,10$) results lags signigicantly behind Python scores. 
Our research shows that R as a low resource language requires different techniques to collect a high quality data. Specifically separating the two R styles has a great impact on the results and the separate dataset could increase the performance of the models.
Our research sheds light on the capabilities of Code-PLMs and opens new research directions for researchers and practitioners for developing code intelligence tools and techniques for R. 
With R's widespread use and popularity, the results of our study can potentially benefit a large community of R developers, both in research and industry.

\end{abstract}



\received{01 October 2024}
\received[revised]{2024}
\received[accepted]{10 April 2024}

\maketitle

\section{Introduction}\label{sec:intro}



The R programming language was first developed by statisticians Ross Ihaka and Robert Jetman at the University of Auckland, New Zealand, in the early 1990s as a personal project~\cite{Rlanguage}. 
R was originally designed as a statistical programming language and a statistical teaching tool by combining the syntax of S, which is widely used among statisticians, with the semantics of Scheme ~\cite{Rlanguage}. 
The use and impact of the R language are growing every year as the number of users and application scenarios increase ~\cite{RImpotent}. 
Version 1.0.0 of R appeared on the CRAN (The Comprehensive R Archive Network)\footnote{https://cran.r-project.org/} in 2000, and by the time this paper is written (2024), the current version of R is 4.4.1 and there are $\sim21,423$ available widely used packages on CRAN. 

With the growing size of the community and applications of R, it is ranked 6th in the PYPL Popularity of Programming Language~\footnote{https://pypl.github.io/PYPL.html} and 17th in the popularity of programming languages~\footnote{https://spectrum.ieee.org/top-programming-languages-2024}. R's extensive library of packages, ranging from data manipulation to advanced statistical modeling, makes it one of the best choices in teaching~\cite{advancedR, RImpotent}, IT Sector~\cite{Rit}, scientific fields~\cite{MedeirosMirra2023, ma15175910, ningthoujam2023r, ABDOLLAHI2022101044}, and Finance~\cite{Rrisk}. 
Though R is studied from various Software Engineering perspectives \cite{zanella2020social, huang2022empirical, PLAKIDAS2017119, Vidoni2022b}, developing automation tools to support R developers, such as comment generation, is still in its infancy, compared to other languages like Java or Python. 


The development of pre-trained language models has revolutionized the field of software engineering and development~\cite{comparison2023}. These models, trained on massive datasets using deep learning techniques, possess the ability to generate human-like text. The significance of pre-trained language models in software engineering and development cannot be overstated~\cite{Zhu2019AutomaticCS}. They facilitate code autocompletion and generation, enhancing developers' productivity and accuracy. 


However, even though the R language has unparalleled importance in related fields, very limited research on code intelligence in R has been conducted. A conventional GPT-2~\cite{GPT2} model is used for the code completion task in R~\cite{CodeCompletionModel}. Though a large number of pre-trained models have been successfully developed in recent years and have proven their effectiveness in various software engineering (SE) tasks \cite{roberta,bert, CodeBERT, GraphCodeBERT,codet5}, to the best of our knowledge, no model has attempted to extend their effectiveness to the R language. 
While progress has been made in code intelligence for R~\cite{CodeCompletionModel}, there exists a gap in the literature, limiting the realization of its full potential. 

More recently, Large Language Models (LLMs) have demonstrated remarkable capabilities across various software engineering tasks~\cite{GPT2, GPT3, starcoder, codellama}. However, recent studies indicate that these models exhibit inconsistent performance across different programming languages~\cite{10.1145/3697010}. The performance disparity in different language is particularly evident in the R programming language, where the unique characteristics of the language pose specific challenges for LLMs. The coexistence of multiple syntax paradigms in R, i.e., \textit{Tidy-verse} and \textit{Base}, along with its heavy reliance on project-specific contexts and specialized libraries, could create additional complexities that current LLMs struggle to handle effectively. 

In this study, we intend to explore to which the current techniques are applicable to R. 
However, as there is no current dataset, we first collect a dataset of high-quality R packages from GitHub repositories (inclusion/exclusion criteria described in Section~\ref{sec:datacollection}). 
Then, we consider four pre-trained models, CodeBERT~\cite{CodeBERT}, GraphCodeBERT~\cite{GraphCodeBERT}, UniXCoder~\cite{unixcoder}, and CodeT5~\cite{codet5} to empirically study their performance for two tasks of code summarization and method name prediction, with various programming styles and settings.
The results of the study are compared across models and with other programming languages. We then investigate an R-specific approach based on different programming styles to evaluate its effect on the models' performance. 
In this study we also investigate the performance of LLMs for the R language on code generation.
\textbf{This paper is the second Stage of our accepted TOSEM Registered Paper\footnote{https://dl.acm.org/journal/tosem/registered-papers}, in which we are reporting the results along with the discussions.}

Our results reveal several key findings regarding the models and their capabilities for the R programming language. 
First, the models studied consistently experience varying degrees of performance degradation when processing R code compared to other programming languages. 
Second, our experiments demonstrate that multi-language fine-tuning does not uniformly improve performance across all PLMs for R-specific tasks, contrary to expectations. 
Third, we find that different syntax of the R styles in \textit{Tidy-verse} and \textit{Base} code significantly impact the performance, which affects the results more for the code summarization task.
The challenges extend beyond syntax to project-specific contexts inherent in R codebases, which pose a substantial challenge when attempting cross-project training.
This is further complicated by the observation that combined syntax approaches generally underperform compared to either individual syntax paradigms.
Our experiments with LLMs also show the low Pass@K scores for code generation, falling much behind the scores obtained for Python. 
These insights emphasize the need for more nuanced approaches to improving model performance for R programming, beyond simply increasing training data or applying multi-language fine-tuning strategies.

The contributions are as follows:
\begin{itemize}
    \item We open source an R dataset to support other research related to code intelligence in R\footnote{https://zenodo.org/records/13871742}. 
    \item We empirically evaluate the performance of different pre-trained models for R, for the specified tasks. 
    \item We provide an R-specific approach and study its effect on the models' performance. 
\end{itemize}

It is worth noting that though the foundation models (i.e., Large Language Models) in the past few months have found their way into SE, there is rarely a study or application available to use them for R. Additionally, their usage is costly if one wants to call their APIs. Therefore, their usage is not available for all, and fine-tuning them is not possible due to being large, besides their hidden ability for this programming language. LLMs have some shortcomings \cite{miceli2023larger}, and still studying the smaller Code-PLMs can benefit the community.  
Therefore, we have studied the Code-PLMs and also added results for code generation using LLMs. 

\textbf{Significance: }
This study provides valuable insights into the effectiveness of using pre-trained code models for R and highlights the need for further research in this field. The results of our work can be used as a reference for future research and provide a foundation for the development of intelligent tools for the R community. 
This is of specific importance as research shows that many scientists develop/use scientific software and ``better software leads to better research'' \cite{betterSoftBetterResearch}.
Our work thus can provide insights to support R developers, superficially in the context of program comprehension and updating comments, which could lead to better maintainability. This is one of the two important aspects for scientific software developers, and R is considered as a language used for scientific software development~\cite{ScientificSoftSMS}. 
We provide discussions and implications at the end, which help shape the future research for code intelligence in R, including the need to curate R datasets and develop separate techniques to increase the performance of the models, including LLMs, for the two R styles. 

The rest of this paper is as follows. In Section \ref{sec:literature}, we provide an overview of the related research. We cover the methodology in Section \ref{sec: methodology}. The result of our study is shown in Section \ref{sec:results}. We discuss the results in Section \ref{sec:discussions} and the implications and threats to validity in Sections \ref{sec:implications} and \ref{sec:threats}, respectively. Finally, we conclude this study in Section \ref{sec:conclusion}.


\section{Related Works} \label{sec:literature}

In this section, we provide some background information and related work to our study.

\subsection{Pre-trained Language Model for Code}

In recent years, Pre-trained Language Models (PLMs) that are pre-trained on code-- we refer to these models as Code-PLMs-- have achieved great performance in different software engineering tasks~\cite{codet5, CodeBERT, GraphCodeBERT, unixcoder}. PLMs are self-supervised deep learning models that have been previously trained on large amounts of text data to perform various natural language processing tasks. These models capture knowledge during the pre-training phase, which is then transferred to different downstream tasks, mainly by fine-tuning the model on specific downstream tasks.

Pre-trained models, specifically transformer-based models~\cite{Attention}, have found significant applications in various software engineering tasks, including code summarization, bug detection, and code translation~\cite{CodeBERT,unixcoder}. 
Examples of these models are CodeBERT~\cite{CodeBERT}, GraphCodeBERT ~\cite{GraphCodeBERT}, CodeT5~\cite{codet5}, and UniXCoder~\cite{unixcoder}. 
PLMs outperform many of the state-of-the-art models on several datasets and tasks and have been used extensively in multiple software engineering tasks~\cite{CodeBERT,codesearchnet, comparison2023}. 

The use of pre-trained language models in software engineering offers several advantages, such as reducing the amount of labeled data required for training and increasing the efficiency of training~\cite{PTM}. They also provide a strong baseline for evaluating the performance of models and can be used to develop novel techniques in this field.


Due to the advantages of the PLMs for source code, there are works that investigate the usage and capabilities of Code-PLMs. 
\citet{whatdotheycapture} studied the features captured by CodeBERT and GraphBERT through attention analysis, probing, and syntax tree induction, on Python, Java, and PHP. 
\citet{troshin2022probing} introduces probing tasks to discover the syntactic and semantic information of various Code-PLMs. 
\citet{romain} uses probing tasks to study the Code-PLMs, but does not include the R language. 
\citet{ahmed2023towards} studies the Code-PLMs specifically by evaluating the models' ability to learn semantics through objective and straightforward evaluation.


With all those studies on different areas and tasks in software engineering, to the best of our knowledge, no study has attempted to apply PLMs to the R programming language. 

\subsection{Scientific Programming Languages}

Scientific programming languages play a pivotal role in the realm of data analysis, statistical modeling, and visualization, catering to the needs of researchers, statisticians, and data scientists~\cite{evolutionR}. Among the diverse array of programming languages utilized in scientific domains, R stands out as a powerful and popular choice due to its robust statistical capabilities and extensive ecosystem~\cite{lifeR}.
One of the key contributors to R's widespread adoption is its vibrant and dedicated community, which has actively expanded the language's functionality. 
Over the past decade, there has been a lot of research devoted to expanding the applicability and functionality of the R language. 
Notable contributions include the development of ggplot2~\cite{villanueva2019ggplot2}, a sophisticated and flexible data visualization package that enables the creation of high-quality graphics with ease. The \textit{Tidy-verse}~\cite{wickham2019welcome}, a collection of R packages designed to enhance data manipulation and analysis, has further streamlined workflows, promoting code readability and efficiency. RStudio, an integrated development environment (IDE) specifically tailored for R, has played a pivotal role in facilitating a user-friendly programming experience. Offering features like syntax highlighting, code completion, and interactive data visualization.
Benefiting from the fact that these contributions have made the R language more applicable, many researchers from other fields have incorporated the R language into their analyses~\cite{de2022acquisition, zhao2021propensity}.
\citet{evolutionR} studied how R evolved, and showed that the emergence of the \textit{Tidy-verse} is one of the primary causes behind the evolution of R. 
At the time of writing this manuscript, there is no research on the applicability and commonality of pre-trained models and large language models over the R language.

\subsection{Scientific Software, Jupyter Notebooks, and R Studies}





Recent research has shown substantial progress in code generation for popular scientific programming languages like Python and Julia. Due to Python's extensive open-source ecosystem, large language models like Codex~\cite{humanevaluating} have leveraged a large number of repositories to train and achieve high performance on widely adopted benchmarks (e.g., HumanEval~\cite{humanevaluating} and MultiPL-E~\cite{MultiPL-E}). 
Similarly, Julia benefits from growing popularity in high-performance computing and data science, making it a recurring target in the state-of-the-art multilingual code-generation evaluations~\cite{Athiwaratkun2023}.

The challenges of R in this landscape are derived from its design and the constraints of the ecosystem. Although functional meta-programming via packages like \textit{rlang} \footnote{https://rlang.r-lib.org/index.html} offers flexibility, the R language presents unique challenges. 
R is a mainstay for statistical computing and data analysis, but R's public codebase is comparatively smaller, and many open-source R repositories cater to highly specialized use cases. As a result, LLMs trained on general-purpose code datasets may struggle to capture R-specific paradigms (e.g., vectorized operations, specialized data-wrangling functions), leading to lower performance relative to Python and Julia~\cite{humanevaluating}. 
This discrepancy highlights the problem of ``low-resource languages'' in code generation, where the scarcity of high-quality training data hinders a model's performance~\cite{cassano2024knowledge}.
In software engineering, several attempts and perspectives have been aimed at advancing code intelligence in R. However, to the best of our knowledge, no research has been conducted explicitly exploring the potential of pre-trained language models in this context.
 
\citet{CodeCompletionModel} develop a deep learning model for code completion in R. 
\citet{Sharma2022} studied Self-Admitted Technical Debt (SATD) in R software packages, and \citet{Chandramouli2022} presented analyzeR, a SonarQube plugin for analyzing object-oriented R packages.
\citet{zanella2020social} study the R packages on CRAN through a social network perspective. They report that the performance of R packages can be explained as a flow of information in the obtained network. 
\citet{huang2022empirical} conducts an empirical study on the R ecosystem and found that the R development has been done by software engineers and from other disciplines such as biology, environmental, and medical science, other than statistics. 

In another work, \citet{PLAKIDAS2017119} investigates the evolution of R, and \citet{DependencyVersioningR} discusses the dependency versioning of R and demonstrates the issue and the solution through three use cases. 
\citet{RDesign} applies static and dynamic analysis to assess the design of R.
\citet{impactofR} studies how the impact of R packages could be modeled through their dependencies and their contributors' network. 

Package dependency in R and CRAN packages~\cite{packageDependency}, distribution of R packages~\cite{EMSE}, characterizing the bugs in R and Python~\cite{ahmed2023characterizing}, and Studying the API breaking changes in a specific field of R packages~\cite{chowdhury2023empirical} are among other research on R. \citet{{babu2022qa4r}} develops a question answering system for R and \cite{wrenn2023dependently} conducts the dependently typing for R vectors, arrays, and matrices.

Another closely researched area is scientific software and scientific computing. \citet{reproducibility} discussed the reproducibility in scientific computing and \citet{betterSoftBetterResearch} reports that 91\% of scientists surveyed online use scientific software, and at least 38\% spend one-fifth of their time developing scientific software. The author discusses that better software leads to better research. 


Other works also study the data science practices in Jupyter notebooks. 
\citet{liu2021haconvgnn} develops a model to generate code documentation for Jupyter Notebooks. 
\citet{zhang2022coral} collects a dataset and builds a model to classify the cells of Jupyter notebooks in data science projects. 
\citet{subtleBugs} develop a Jupyter notebook extension based on program synthesis and and test selection to generate documentations that help users find bugs for data wrangling. 
\citet{splitting} study the data cleaning practices in 30 Jupyter notebooks. 
\citet{errorIdentification} examine Jupyter notebooks to define error identification strategies of Python notebooks. Their work is a replication of a similar work with a user study to find these strategies for R Markdown notebooks~\citet{AbdelGhani2022LooksOT}. 
Maintenance~\cite{JupyterMaintenance}, bug analysis~\cite{de2022bug}, structure improvement~\cite{structureImprovement}, and understanding documentation practices~\citet{wang2021makes} are among other research explored in Jupyter notebooks.

\textbf{Differences among our work and the current literature.} 
Among the related works, the closest ones are the approaches that use deep learning to detect SATD~\cite{Sharma2022} or complete code in R~\cite{CodeCompletionModel}. The work multiple\_E~\cite{cassano2022multiple} proposed a translator that is able to translate the HumanEval~\cite{humanevaluating} benchmark dataset to other programming languages, including R. Still, none of these or other research studies the ability of Code-PLMs for different tasks in R or considers the two R styles. 
Though the Jupyter notebook supports the R programming language, the current studies mainly focus on Python-based libraries, excluding R. 
The other research on the R programming language is related to empirical studies to gain a better insight into its different aspects. 
Our work differs from all the current literature by first providing a dataset for code summarization and method name prediction, and then exploring the capabilities of Code-PLMs for the R programming language, in various settings including intra- and cross-projects and considering its different styles, \textit{Tidy-verse} and \textit{Base}.

\section{Methodology} \label{sec: methodology}

In this section, we will review the research questions, data collection process, studied tasks, and Code-PLMs, along with our approach to answer each of the research questions. 


\subsection{Research Questions}
The research questions we investigate and the answers in this research are as follows.


\textbf{RQ1: What is the performance of the existing Code-PLMs on R language?}

Many studies show that Code-PLMs obtain state-of-the-art performance for programming languages on the CodeSearchNet dataset. However, to what extent these models can accomplish the tasks for the R language is not defined. 
Therefore, in this RQ, we evaluate the performance of the Code-PLMs for R. 
The answer to this RQ reveals the capability of the code-PLMs for the R language, which is not explored in the existing research.

\textbf{RQ2: Does the multilingual training in the fine-tuning phase increase the performance of the models compared to monolingual training for R? }

Previous research~\cite{Ahmed} has shown that multilingual training in the fine-tuning phase increases the performance of the Code-PLMs for code summarization and method name prediction. In this RQ, we investigate whether such an effect is observed in the case of R. 
Answering RQ2 helps identify the effect of multilingual training for other languages beyond the six languages covered in the CodeSearchNet dataset (R in our study). This finding adds value to the current literature to assess the effectiveness of different approaches for low-resource languages, in this case, R. 

\textbf{RQ3: What is the effect of intra- or cross-project training?}

Previous research~\cite{mcmillan2019, ahmed-fewshot} studied the effect of different dataset designs on the performance of the code summarization models. In this RQ, we investigate the outcome of models based on choosing the training and test functions from the same or different R repositories, also known as intra- or cross-project data selection. 
Although intra- or cross-project effects have been explored for other languages, it has not been explored for R. The results can shed light on designing experiments and the dataset to increase the performance of the models.

\textbf{RQ4: Can we improve the performance of the models based on different programming styles in R?}

Two of the popular programming styles in R are \textit{Tidy-verse} and \textit{base R}. 
The \textit{Tidy-verse} is a collection of R packages designed for data science \footnote{\url{https://www.tidyverse.org/}}. 
Though both are used frequently, the syntax of each one is different from the other~\cite{advancedR}. 
Here, we investigate the effect of different programming styles on the performance of Code-PLMs. 
The results can benefit developing models for code intelligence tasks for the R developers, for both \textit{base} and \textit{Tidy-verse}. 


\textbf{RQ5:} \textbf{What is the performance of Large Language Models on the selected tasks for R?}

Recent advances in foundation models or Large Language Models have shown promising results for some of the software engineering tasks~\cite{GPT3, codellama, codet5}. Accordingly, in this RQ, we will explore the performance of LLMs for the studied tasks for R.
Currently, there is limited research on assessing the capabilities of LLMs for low-resource languages, including R. With the interest in using LLMs, the results of RQ5 will shed light on the potential of these models for code intelligence in R, compared to the code-PLMs.

\subsection{Data Collection}

\subsubsection{Overview}

\begin{figure}[htb]
  \includegraphics[width=0.95\textwidth]{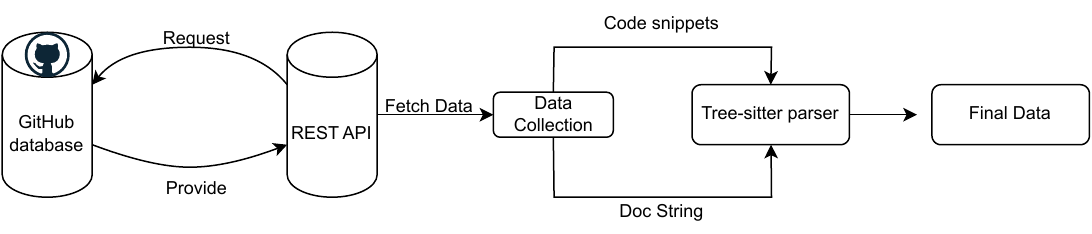}
  \caption{Data collection workflow}
  \label{fig:flow}
\end{figure}

Figure \ref{fig:flow} presents an overview of our methodology for preparing the data. First, we collect GitHub repositories that include R files, considering the inclusion/exclusion criteria that we discuss next. Then, the eligible code snippets are parsed into code tokens using the tree-sitter parser. In the end, we followed the Roxygen2 documentation structure to match the code snippets and their corresponding natural language description to generate a dataset where each line of the data contains a pair of Programming Language (PL) and its corresponding Natural Language (NL). 

\subsubsection{Scraping GitHub Repositories} \label{sec:datacollection}
The API we used to collect the data is the GitHub REST API\footnote{\url{https://docs.github.com/en/rest?apiVersion=2022-11-28}}, which allows users to search for repositories based on specific criteria. Based on previous work~\cite{i/e, codesearchnet}, we devised a similar selection strategy for the R language to ensure that the included code repositories are active, original, and relevant. To ensure the objectivity of the selection, we list the following Inclusion/Exclusion criteria adopted from previous works~\cite{i/e, codesearchnet}.

Inclusion criteria:

\begin{itemize}
    \item The project should be a public project with an open-source license on GitHub.
    \item The project needs to be an R package and have a proper R package structure. For example, it needs to contain a DESCRIPTION file.
    \item The project needs to be an active and up-to-date project. We flirted the repositories where the last commit was made within a year. 
\end{itemize}

Exclusion criteria:
\begin{itemize}
    \item The project should be a non-personal project. So, all personal projects are excluded. 
     \item The project should not be archived. When a repository is archived, its issues, pull requests, code, labels, and other components of the project become read-only. The owner and collaborators also can not update the project. Therefore, all archived repositories are excluded. 
\end{itemize}

\subsubsection{Data}

To create the dataset for software engineering tasks, we formatted our data following the CodeSearchNet dataset~\cite{codesearchnet} structure, where each line of the JSON file contains one data example, including the function name, the code tokens, the natural language description present in their respective documentation, and the doc Strings. 
Each data point also contains metadata on the function, such as the repository it was extracted from.
In order to do so, we performed a series of preprocessing and filtering on the raw data. The following discusses preprocessing and filtering in detail. 

A total of $1,664$ GitHub repositories were collected using the GitHub API following the inclusion and exclusion criteria discussed above. 
We then pre-processed the data to obtain the required input for the software engineering tasks. Roxygen2\footnote{\url{https://roxygen2.r-lib.org/}}, a popular documentation system for R that is similar to JavaDoc, was used to parse the functions into Abstract Syntax Trees (ASTs) and comments. 
We first make sure that these repositories are written in R and have an Roxygen framework. The Roxygen framework is needed to ensure that the code comments can be extracted. 
For this purpose, we first inspect the file types of the source codes in our collected data. If the package includes \texttt{.R} files, typically found in an R folder in the root directory, it indicates the use of the R language.
Then, we look for Roxygen2 comments within these \texttt{.R} files. Roxygen2 is widely used and it manifests in the source code as specific comments, starting with the \texttt{`\#'} symbol, followed by the \texttt{`@'} symbol, like \texttt{`\# @param'} or \texttt{`\# @return'}. 
If we observe such comments, it suggests that the package employs Roxygen2 for documentation.
Lastly, we examine the \texttt{DESCRIPTION} file in the package's root directory. 
This file contains various metadata about the package, such as version number, maintainer, description, and package dependencies. 
If Roxygen2 is listed in the \texttt{Imports} or \texttt{Suggests} field, it provides a further indication that the R package employs Roxygen2. 
This process allowed us to extract the necessary information from the R code to form the input for the pre-trained models.



\begin{table}[h]
\centering
\begin{tabular}{llll}
\toprule
Programming   language & Training & Dev    & Test   \\\midrule
Ruby                   & 24,927   & 1,400  & 1,261  \\
JavaScript             & 58,025   & 3,885  & 3,291  \\
Java                   & 164,923  & 5,183  & 10,955 \\
Go                     & 167,288  & 7,325  & 8,122  \\
PHP                    & 241,241  & 12,982 & 14,014 \\
Python                 & 251,820  & 13,914 & 14,918 \\\midrule
R\textsubscript{Base}     & 8,291 & 1,031 & 1,031 \\
R\textsubscript{Tidy}     & 24,380 & 3,047 & 3,048 \\
R\textsubscript{Combine}  & 32,671 & 4,078 & 4,084 \\
\bottomrule
\end{tabular}
\caption{ CodeXGLUE and R dataset statistics}
\label{table:datastat}
\end{table}

\subsubsection{Parser}

Roxygen is a documentation generator for R code that uses a parser to analyze R code and extract information that can be used to create documentation~\cite{Vidoni2022b}. Roxygen can use the information in the function to generate documentation for the function. For example, it can extract the names and default values of the function's arguments, as well as information about the return value of the function. Roxygen can also use ASTs to identify any errors or potential problems in the code, such as syntax errors or unused variables. 

In this study, we utilize Roxygen to achieve two primary objectives. First, we employ Roxygen to match code snippets from R language packages with their corresponding natural language descriptions. 
Second, we leverage Roxygen and the dependency information present in R language packages to classify them into either the \textit{base} R or the \textit{Tidy-verse} version.
The process of using Roxygen to match code snippets with their corresponding natural language descriptions can be outlined as follows. Within R language packages that adhere to the Roxygen framework, there exist two directories: \texttt{R} and \texttt{Rd} (R document). For each \texttt{R} file, we specifically identify the method names within the code and search for matching method names in the \texttt{Rd} directory. Upon finding a match between a function name in the code and its counterpart in an \texttt{Rd} file, we establish a linkage. This enables us to obtain a comprehensive mapping between the code tokens within a function and the corresponding natural language description associated with that function.




Furthermore, we utilize Roxygen and Tree-sitter\footnote{\url{https://tree-sitter.github.io/tree-sitter/}}-- GitHub's universal parser-- to classify an R language package as belonging to either base R or the tidy-verse. As previously described, the tidy-verse is a collection of R language packages that provide additional functionality and convenience on top of the R language itself. In R packages that potentially utilize the tidy-verse, their dependency environment includes the importation of R packages contained within the tidy-verse collection.
To determine the classification, we employ Tree-sitter to parse and analyze the structure of the description file within an R language package. Specifically, we focus on the import section, which contains the necessary information. If this section includes the importation of the tidy-verse collection, we classify the R language package as belonging to the tidy-verse. Conversely, if the import section does not include the tidy-verse, the package is classified as base R.

After this processing, the dataset was reduced to 1,583 repositories, pertaining to 40,833 functions (i.e., code-comment pairs).
The statistics of our collected dataset are in Table~\ref{table:datastat}.
The full R corpus is divided into two style-specific subsets—\(\mathrm{R}\textsubscript{Base}\) and \(\mathrm{R}\textsubscript{Tidy}\) (Tidyverse); their union forms the aggregated dataset \(\mathrm{R}\textsubscript{Combine}\), i.e., \(\mathrm{R}\textsubscript{Combine}= \mathrm{R}\textsubscript{Base} \cup \mathrm{R}\textsubscript{Tidy}\). The other rows of the table demonstrate the CodeSearchNet dataset. Following the previous works~\cite{codesearchnet}, we split the R\textsubscript{Base}, R\textsubscript{Tidy}, and R\textsubscript{Combine} with 8:1:1, as shown in Table~\ref{table:datastat}. 
Note that we have ensured that there are no duplicate functions in the dataset.

\subsection{Tasks}
In the following, we explain the selected tasks for our experiments. 

\textbf{Code Summarization} is a technique in software engineering that aims to automatically generate a concise and informative summary of the code ~\cite{CodeBERT, GraphCodeBERT}. Code summarization has the advantage of being fast, consistent, and reproducible, and can help developers easily understand the semantics of the source code~\cite{Zhu2019AutomaticCS}. 
This can be particularly useful in large and complex software projects where it can be difficult to keep track of the different components and understand how they interact with each other.
In recent years, code summarization has gained significant attention in the software engineering community, with several research studies exploring the effectiveness of various summarization techniques and models~\cite{bert,codesearchnet}. However, previous research in code summarization has mainly focused on programming languages that are widely used in firms, such as Java and Python ~\cite{CodeBERT, GraphCodeBERT,roberta}, leaving a gap in our understanding of code summarization for other programming languages, such as R.
Therefore, we study code summarization as one of the tasks. The dataset we collected includes both the methods and their natural language comments, written by developers. Therefore, we can use the collected dataset, and it is ready to train the models for code summarization. 

\textbf{Method Name Prediction} or extreme code summarization refers to the task of predicting or generating appropriate names for methods (functions or procedures) in programming code~\cite{Allamanis2016ACA}. 
Given a code snippet, the model aims to generate a suitable name that accurately represents the intended behavior or functionality of the method. This task is important in software development as well as code comprehension because well-named methods can improve code readability, maintainability, and understandability~\cite{Allamanis2016ACA}.
Method Name Prediction often involves training models on large-scale code repositories to learn patterns and common naming conventions used in programming. The models can then use this knowledge to generate plausible method names for given code contexts.
This task has applications in various code-related tasks, such as code completion, code summarization, and code search, where generating informative method names can greatly assist developers in writing or understanding code more efficiently~\cite{Methodnameimport}. It has also been studied in multiple previous research~\cite{Ahmed, code2seq, code2vec}. 
Therefore, we chose this task in our study. 
The collected dataset is valid for the method name prediction task, as given the code without the method name, and the natural language comments, we can train and evaluate the models to generate the method name.

\textbf{Code Generation} aims to generate a code snippet from a given natural language description. \citet{ridnik2024code} proposed AlphaCodium, which performed well on the code generation benchmark.  
The model is given as input a description written in natural language, and the goal is to generate the code that corresponds to the description. Developers often take code from blogs, posts, articles, and other websites and adapt it to their context. The code generation task is important as it provides a better replacement solution by generating code for users based on natural language descriptions.
We select code generation to assess the capabilities of the models for a different category of code intelligence tasks (natural language to Code task). As there is no code generation for R, we will follow the HumanEval~\cite{humanevaluating} dataset, which is in Python. 
We use MultiPL-E~\cite{MultiPL-E} to evaluate the LLM's performance on R. MultiPL-E~\cite{MultiPL-E} is a multi-language benchmark that translates the HumanEval dataset from Python into 18 other programming languages. We sample 20 completions per prompt at temperature 0.2 with top-p 0.9 following~\cite{MultiPL-E} for zero-shot samples. Note that typically 50 is needed to compute Pass@1, but~\citet{MultiPL-E} reports that 20 completions would give a stable estimate for the Pass@1 score. 

We also run the expressions in the few-shot setting to compare different prompting techniques as well. We conducted one, three, five, and ten-shot experiments following the greedy approach (0 temperature) under three different prompting approaches: simple R, BM25~\cite{BM25}, and embedding-based using voyage-code-3 model. The details are discussed in Section~\ref{sec:rq5-results}.

\subsection{Studied Code-PLMs}

In this section, we will briefly discuss the chosen benchmarking pre-trained models. We choose these models based on their architecture, pre-training tasks, and performance, and we follow the recommendation of a recent empirical study for selecting them~\cite{comparison2023}. 
\citet{comparison2023} conducted empirical studies on various code intelligence tasks and code-PLMs and provided recommendations for the model choices for each task. Accordingly, we choose CodeBERT, GraphCodeBERT, UniXCoder, and CodeT5 in our study. Note that at the time of this study, to the best of our knowledge, there is no research on evaluating the models for the R language. Therefore, we follow the recommendations of~\cite{comparison2023} and choose the following models, in addition to conducting experiments with large language models, GPT-3.5-Turbo and CodeLlama.

\textbf{CodeBERT}~\cite{CodeBERT} is a language model specifically designed for programming languages. CodeBERT's architecture is based on the Transformer model, similar to the original GPT model, but with modifications to accommodate programming code.
CodeBERT is pre-trained using two primary tasks: masked language modeling (MLM) and replaced token detection (RTD). For MLM, which has been shown to be effective in literature~\cite{bert}, random tokens in the input code are masked, and the model is trained to predict the masked tokens. For RTD~\cite{electra}, the objective is to use both bimodal and unimodal data for training.
CodeBERT has shown promising results in various programming language-related tasks, such as natural language code retrieval and code summarization. It has achieved state-of-the-art performance at the time of publication in code-related benchmarks and competitions, demonstrating its effectiveness in understanding and generating code.

\textbf{GraphCodeBERT}~\cite{GraphCodeBERT} extends CodeBERT~\cite{CodeBERT} by incorporating the graph-based representation of code. It represents code snippets as abstract syntax trees or control flow graphs (CFGs) to capture the structural information of the code. GraphCodeBERT follows BERT\cite{bert} and uses multi-layer bidirectional Transformers as the backbone. In addition to source code, GraphCodeBERT incorporates paired comments during pre-training to enhance the model's ability to handle code-related tasks that involve natural language, such as natural language code search \cite{CodeBERT}. Moreover, it includes data flow, represented as a graph, as an input component for the model.
GraphCodeBERT employs similar pre-training tasks as CodeBERT. It also incorporates Edge Prediction and Node Alignment to enhance the model's understanding of relationships between entities mentioned in the text. By pre-training the model on a task that involves predicting edges between entities or concepts, the model can potentially learn to capture the semantic connections and dependencies between them.
GraphCodeBERT has demonstrated improved performance over CodeBERT~\cite{CodeBERT} in tasks that require a deeper understanding of the code structure. It has been successful in tasks such as code translation, code clone detection, and code refinement.

\textbf{UniXCoder}~\cite{unixcoder} is a model designed for multilingual code-mixing scenarios, where programming code contains multiple languages. It aims to handle code written in multiple programming languages within the same codebase.
UniXCoder employs a multimodal architecture that combines the strengths of both Transformer models and recurrent neural networks (RNNs). It consists of a hybrid Transformer-RNN model, where the Transformer processes the textual code tokens, and the RNN handles the language-related features.
UniXCoder utilizes pre-training tasks similar to CodeBERT to learn code-related representations. Additionally, it incorporates tasks specific to multilingual code-mixing scenarios, such as language identification and code-switching prediction.
UniXCoder has demonstrated effectiveness in handling multilingual code-mixing scenarios and achieving high performance in code-related tasks involving multiple programming languages.

\textbf{CodeT5}~\cite{codet5} is another variant of the T5 (Text-To-Text Transfer Transformer) model specifically designed for code-related tasks. It combines the power of T5's text-to-text transfer learning with code-specific pre-training. CodeT5 utilizes the T5 Transformer architecture~\cite{codet5}, which consists of a transformer encoder-decoder framework. The model takes code as input and generates code-based outputs for various downstream tasks.
CodeT5~\cite{codet5} employs a text-to-text transfer learning paradigm, where a variety of code-related tasks are cast into a unified text generation format. These tasks include code translation, code summarization, code completion, and more. The model is trained to generate the target code given the corresponding input and task description.
CodeT5~\cite{codet5} has achieved strong performance in multiple code-related tasks, surpassing previous state-of-the-art models. CodeT5~\cite{codet5} has shown effectiveness in code generation, code summarization, and code translation tasks across various programming languages.



\textbf{CodeLlama-7B}. CodeLlama-7B is a model from the Code Llama~\cite{codellama} model series, which is a large programming domain-oriented language model based on Llama 2~\cite{llama2}, which allows for the direct generation or understanding of code using textual hints (Prompts). CodeLlama has the ability to perform code completion by generating up to 100k tokens~\cite{codellama}. Additionally, CodeLlama has the ability to generate zero-sample instructions for programming tasks. In addition, CodeLlama has zero-sample instruction adherence for programming tasks, including code generation. This small-sample learning capability is worth noting when testing code to accomplish tasks in small resource languages like R.~\citet{codellama} claim that CodeLlama can be used for a wide range of programming tasks and is one of the best performing LLM at public programming tasks, making developers' workflows faster and more efficient, and lowering the learning threshold for programming. Therefore, we select CodeLlama to explore the RQ5 answer.

\textbf{StarCoder2-base}.
In recent years, the field of code generation has advanced significantly with the development of Large Language Models (LLMs) tailored for code-related tasks. One of the most prominent models is StarCoder~\cite{starcoder}, along with its base variant, StarCoderBase, both part of the BigCode project\footnote{\url{https://huggingface.co/bigcode}}.StarCoder and StarCoder-base are models trained on permissively licensed data from GitHub, covering over 80 programming languages, Git commits, GitHub issues, and Jupyter notebooks. StarCoder-Base, which has approximately 15 billion parameters, was trained on 1 trillion tokens. StarCoder-Base was further fine-tuned on 35 billion Python tokens to create StarCoder, specifically optimized for Python-related tasks.

StarCoderBase surpasses many existing open-source Code LLMs on popular programming benchmarks and competes with closed models like OpenAI's code-cushman-001 (the Codex model behind early versions of GitHub Copilot). With a context length exceeding 8,000 tokens, StarCoder can handle large inputs, making it ideal for tasks like acting as a technical assistant, autocompleting code, modifying code based on instructions, or providing natural language explanations of code snippets. Given StarCoder's performance and versatility, StarCoder models offer a strong foundation for community adoption and customization across various applications, including code generation. Therefore, we use StarCoder-base in our experiments in RQ5.

\textbf{Codegen-2B-multi.}
We should note that initially, we intended to use ChatGPT as one of the LLMs for code generation. Since GPT-3.5 (chatgpt) is not open-source, we used CodeGen-2B-multi~\cite{CodeGen} as an alternative for the code generation task. 
CodeGen-2B-multi is a large-scale transformer model trained on both natural language and multiple programming languages, allowing it to generate code from natural language descriptions. 
CodeGen-2B-multi is open-source, making it a valuable resource for research and development. The model is pre-trained on natural language datasets and fine-tuned on multilingual code datasets, enabling it to perform well across various programming languages for tasks like code completion and synthesis. Therefore, we used it for the experiments in RQ5.


\subsection{Approach}
\label{sec: Approach}
In this section, we will provide a detailed experimental setup of each of the studied RQs.

\textbf{RQ1:} 
This RQ is motivated by the lack of research and understanding of the performance of Code-PLMs for the R programming language. To address this challenge, we compared the four Code-PLMs for each of the languages on the CodeXGLUE benchmark (which is the same data as CodeSearchNet for these tasks) and the R dataset. To achieve the comparison, we trained each of the models on each of the programming languages shown in the first six rows of Table~\ref{table:datastat} and also on the R\textsubscript{Combine} datasets separately, and then tested on the test split of each one. Therefore, we can evaluate them when trained and tested on each of the languages in a mono-lingual setting.

\textbf{RQ2:}
For this RQ, we intend to evaluate the effect of multi-lingual fine-tuning for R. To achieve this, we combined the training splits of all the seven languages (i.e., Ruby, JavaScript, Java, Go, PHP, Python, and R\textsubscript{Combine}) and then train each of the Code-PLMs on this multi-lingual dataset. We tested the performance of each model on the test set of each of the programming languages separately. The training and testing were done for code summarization and method name prediction, and the results are reported accordingly.


\textbf{RQ3:}
For this RQ, we use the R\textsubscript{Combine} dataset, and we split the functions of the training and test sets based on the repository they belong to. 
For the intra-project setting, we split the training and test sets such that we have some functions from \textit{every} repository in the training set and some of its functions in the test set. Note that we take care not to have duplicate functions in the training and test sets. 
For the cross-project setting, the repositories are split such that their functions only belong to the training or test set. Therefore, the model does not have functions from the same repository in the training and test sets.

\textbf{RQ4:}
As mentioned earlier in section~\ref{sec: methodology}, there are two different opinionated programming styles in R, the \textit{Base} version, and \textit{Tidy-verse}. 
The \textit{Base} R is the fundamental R programming language, and it has been the only dialect for a long time. \textit{Base} R can be more stable when compared with \textit{Tidy-verse}~\cite{advancedR}. Tidy-verse, on the other hand, is a collection of packages that work as an add-on to base R that provides extra functionality and makes the workflow and syntax more user-friendly. Despite the benefits and user groups of each, the syntax can be very different. Listing~\ref{lst:tidy} provides an example of \textit{Tidy-verse}, and a \textit{Base} R example is shown in Listing~\ref{lst:base}. 
Based on these differences, we hypothesize that separating the datasets might affect the obtained scores in a positive way. 

Therefore, for this RQ, we repeat the experiments of RQ1--RQ3, but with R\textsubscript{Base} and R\textsubscript{Tidy} datasets separately. 
To have a fair comparison, we decrease the size of the R\textsubscript{Tidy} to include the same number of records as the R\textsubscript{Base}. 
This will be in addition to having the experiments on the R\textsubscript{Tidy} dataset with its original number of records. 

\textbf{RQ5:} To answer this research question, we run experiments using UniXCoder~\cite{unixcoder}, as well as Codegen-2B~\cite{CodeGen}, {CodeLlama}~\cite{codellama} and StarCoder2~\cite{starcoder}. We exclude other models as they are not developed for code generation. 
For code generation, as stated above, the models will be prompted to generate code. We prompted the LLMs in a zero-shot and few-shot setting and compare their results with the full fine-tuned models. 
For code summarization and method name prediction tasks, we selected 100 samples from the test dataset (same as the ones used for human evaluation, described in Section~\ref{sec:humaneval}).


\begin{minipage}{.45\textwidth}
\begin{lstlisting}[caption=Tidy-verse, label={lst:tidy}, escapeinside={(*}{*)}]
mtcars %>% (*\label{lst:mtcars}*)
  group_by(cyl, gear) %>% 
  summarise(mpg.mean = mean(mpg),
            mpg.sd = sd(mpg),
            wt.mean = mean(wt),
            wt.sd = sd(wt)) %>% 
  ungroup()
\end{lstlisting}
\end{minipage}\hfill
\begin{minipage}{.45\textwidth}
\begin{lstlisting}[caption=Base R, label={lst:base}, escapeinside={(*}{*)}]
mtcars_by <- by(mtcars, 
   INDICES = list(mtcars$cyl, mtcars$gear),
   FUN = function(x){
     data.frame(cyl = unique(x$cyl),
                gear = unique(x$gear),
                mpg.mean = mean(x$mpg),
                mpg.sd = sd(x$mpg),
                wt.mean = mean(x$wt),
                wt.sd = sd(x$wt))
   })

do.call(rbind, mtcars_by)
\end{lstlisting}
\end{minipage}


\subsection{Evaluation Metrics}

For code summarization, following previous works~\cite{CodeBERT, codet5}, we evaluate the results using the smoothed BLEU-4 score~\cite{lin-och-2004-orange}. The smoothed BLEU-4 score is a technique used in the field of machine translation to assess the quality of translations. It is an extension of the BLEU (Bilingual Evaluation Understudy) metric, which measures the similarity between a machine-generated translation and one or more reference translations. The BLEU score can be calculated using the following formula~\ref{bleu}:

\begin{equation}
BLEU = BP * exp(\sum_{n=1}^{N}W_{n}logP_{n})
\label{bleu} 
\end{equation}

where BP is the brevity penalty and $P_{n}$ is the modified precision score, let c be the length of the candidate and r be the effective reference corpus length, BP can be calculated using:

\begin{equation}
BP= \left\{
\begin{array}{lr}
1 & \textsf{if}\ c>r\\
e^{1-r/c} & \textsf{if}\ c \leq r \\
\end{array} \right.
\label{BP} 
\end{equation}

When reporting the smoothed BLEU score, the model with the higher score is considered a better model for this task. A higher score is considered better. 

For method name prediction, we follow the previous work~\cite{Ahmed}, and report the F1-score to evaluate the classification models, as defined below. 

\begin{equation}
F1 = 2 * \frac{precision * recall}{precision + recall}
\label{F1} 
\end{equation}

Where $precision = \frac{True positive}{True positive + False positive}$ and $
recall = \frac{True positive}{True positive + False negative}$. 
Here, True positive indicates that the token that is part of the name is been predicted as part of the function name. When reporting the F1 score, the model with the higher F1 score is considered a better model for this task.


For code generation, following the metrics used in code generation studies using HumanEval~\cite{codellama}, we report the pass@K. Pass@K is an evaluation metric used to measure the probability that at least one correct answer is included among the top K responses generated by a model, calculated as follows~\cite{humanevaluating}:

\begin{equation}
\operatorname{pass} @ k:=\mathbb{E}_{\text {Problems }}\left[1-\frac{\left(\begin{array}{c}
n-c \\
k
\end{array}\right)}{\left(\begin{array}{c}
n \\
k
\end{array}\right)}\right]
\end{equation}

For instance, if $k$ is set to $5$, pass@5 signifies the probability that at least one correct answer is present within the first five responses generated by the model. This metric helps assess whether the model's output contains the correct result among multiple possibilities, providing a measure of the reliability of the model's output. We will apply the metric for $k \in \{1, 5, 10\}$. 

\textbf{Statistical tests:} Following the previous works~\cite{Ahmed}, we apply the Wilcoxon signed-rank test to compare the results of the models for all the tasks.

\subsection{Human Evaluation} \label{sec:humaneval}

Other than automatic evaluation metrics, we also conduct a human evaluation, following the protocol used in previous studies~\cite{apicontext2com, lamner}. We randomly select $100$ samples from the test dataset for each task. Each sample will include code, ground truth values, and the generated output (code comments or method names) for all models. 
Two Computer Science graduate students, who are both familiar with R, conducted the evaluation.
Each record was ranked by two reviewers, and the average value was reported. 
Each sample was evaluated based on the following criteria, on a Likert scale of $1$ to $5$, where $1$ shows the lowest and $5$ shows the highest.

For code summarization, we considered the following aspects: 
\begin{itemize}
    \item Informativeness (I), which considers how well the generated comment covers the key points of the code.
    \item Relevance (R), which evaluates the consistency of the provided details in the comments and compares them with the details in the code.
    \item Fluency (F), which assesses English fluency in terms of the comments being well-written and grammatically correct.
\end{itemize}

For method name prediction, we considered the semantic similarity of the ground truth and the generated names.
To avoid bias, we hide whether the text (comment or method name) is generated by a model or is written by a human. Moreover, the evaluators were educated before scoring. We provided examples and explanations for each criterion before they started the evaluation.

\subsection{Experimental Setup}

\noindent
All experiments were run on the UBC Advanced Research Computing Sockeye and Digital Research Alliance
of Canada. Each job required a compute node with 1 CPU core, 48\, GB of system memory, and two
NVIDIA~Tesla~V100~32\,GB GPUs. We followed the hyper-parameter settings described in the original
papers, training the models with a learning rate of 0.00005. The \texttt{source\_length}
was set to 256, and the \texttt{target\_length} was 128 for code summarization. For method name
prediction, the \texttt{target\_length} was adjusted to 10. We employed a \texttt{RandomSampler},
ensuring that training samples---including those from different programming languages---were shuffled
at each epoch to avoid biases in data ordering.

\noindent
The batch size was fixed at 32 for CodeBERT~\cite{CodeBERT} and GraphCodeBERT~\cite{GraphCodeBERT},
and 48 for UniXCoder~\cite{unixcoder} and CodeT5~\cite{codet5}, mirroring the original paper
settings for a fair comparison across different programming languages. Finally, all experiments
were executed on Dell~EMC~R440 GPU compute nodes to ensure consistent and reproducible results.

\section{Results} \label{sec:results}

The obtained results are presented in this section.


\subsection{RQ1: Performance of Code-PLMs in Mono-Lingual Fine-tuning Setting on R}

\subsubsection{Code Summarization}

\begin{table}[h]
    \centering
    \begin{tabular}{|c|c|c|c|c|}
        \hline
          \textbf{Source} &\textbf{CodeBERT} &\textbf{GraphCodeBERT}& \textbf{UniXCoder} & \textbf{CodeT5}\\
        \hline
        JS & 14.67 & 14.84  & 15.44  & \textbf{16.17}\\
        Go & 18.12 & 18.48 & \textbf{19.26} & 19.25\\
        Java & 18.65 & 19.24 & 20.14 & \textbf{20.53} \\
        PHP & 25.36  & 25.75  & \textbf{26.52}  & 26.13\\
        Ruby & 11.54 & 11.76 & 15.01 & \textbf{15.42} \\
        Python & 18.29  & 18.31& 19.17 & \textbf{19.32} \\
        Combined\textsubscript{six} & 19.15 & 19.20& \textbf{19.89} &19.73 \\
        \hline
        
        R\textsubscript{Combine} & 11.80 & \underline{11.02} &  \textbf{18.43}  & 15.42\\
        \hline
        
        R\textsubscript{Tidy} & 17.92 & 10.38 & \textbf{18.93} & 16.64 \\
        
        R\textsubscript{Base} & \textbf{\underline{29.55}} &  9.56 &  \underline{20.93} & \underline{17.61}\\
        \hline
        
    \end{tabular}
    \caption{Performance for all the baselines for the six programming languages on code summarization, R styles, and the combined R dataset under monolingual training. The `Combined' represents the combined dataset of the six programming languages from CodeXGLUE. `Rcombine' shows the results when the data from both styles of R are combined together. Best results in each row are \textbf{bold}. For the three R datasets, we also \underline{underlined} the best results in each column (only among the R datasets).}
    \label{tab:sum-mon}
\end{table}

Table~\ref{tab:sum-mon} presents the BLEU-4 scores for code summarization, for all languages. 
The Combined\textsubscript{six} represents the results when we combined all of six languages from CodeSearchNet.
Row R\textsubscript{combine} shows the dataset that we combine both \textit{Tidy-verse} and \textit{Base} R datasets. 

We first replicated the summarization task by fine-tuning the pre-trained model for six languages, following the reported set of instructions. This step is done to ensure the reliability of our results and avoid errors that might have been introduced by the hardware or the system environment. For all six languages and models, our replicated results are close to the models' reported values. 
For all models, the same training setup is used for R. 

As shown in Table~\ref{tab:sum-mon}, 
for most of the six languages, JavaScript (JS), Go, Java, PHP, Ruby, and Python, CodeT5 has the best performance. For some of the languages, UniXCoder has close scores, all being higher than CodeBERT and GraphCodeBERT. 
When the models are tested on R\textsubscript{Combine}, we observe that UniXCoder has the highest performance, being $18.43$, followed by CodeT5 BLEU-4 score of $15.42$. CodeBERT and GraphCodeBERT both have lower scores in the range of $11$ to $12$ BLEU-4 scores.

It is worth noting that GraphCodeBERT~\cite{GraphCodeBERT} is primarily designed to model the structural relationships in code (e.g., how different components of code, like functions, variables, and classes, are related). This is beneficial for tasks that rely heavily on understanding the syntactic and semantic structure of code, such as code completion or defect detection. For summarization, where the goal is to generate natural language descriptions from code, a model needs to excel in both code understanding and natural language generation. 
Though we tokenized the R dataset using the same tool, Tree-sitter, GraphCodeBERT continues to struggle with R. One possible reason could be that R, as a programming language, is fundamentally different from other languages. This difference means that R does not share the same expressions or patterns, making it challenging for the model to learn the data effectively during training, which might be a culprit for the low performance of GraphCodeBERT on R.


\subsubsection{Method Name Prediction}
\begin{table}[h]
    \centering
    \begin{tabular}{|c|c|c|c|c|}
        \hline
          \textbf{Source} & \textbf{CodeBERT} & \textbf{GraphCodeBERT} & \textbf{UniXCoder} & \textbf{CodeT5}\\
        \hline
        JS & 0.37 &  \textbf{0.39} & 0.34 & 0.34\\
        Go & \underline{\textbf{0.52}} &  \textbf{0.52} & \underline{0.49} & 0.40\\
        Java &  0.51 &   \textbf{0.52} & 0.45 & 0.42\\
        PHP & 0.50 &  \underline{\textbf{0.54}} & 0.42 & \underline{0.47}\\
        Ruby & 0.42 &  \textbf{0.45} & 0.41 & 0.33\\
        Python & 0.44 &  \textbf{0.46} & 0.43 & 0.40\\
        Combined\textsubscript{six}  & 0.43 &  \textbf{0.46} &  0.35 & 0.38\\
        \hline
        R\textsubscript{Combine} & 0.25 &  \textbf{0.32} &  0.26 & 0.29\\
        \hline
        R\textsubscript{Tidy} & 0.26 &  \textbf{0.31} &  0.25 & 0.29\\
        R\textsubscript{Base} & 0.26 &  \textbf{0.32} &  0.23 & 0.30\\
        
        \hline
    \end{tabular}
    \caption{F1-score of the models for the six programming languages, R styles, and combined R dataset on name prediction under monolingual fine-tuning. Best results in each row are \textbf{bold} and best results in each column are \underline{underlined}.}
    \label{tab:name-mon}
\end{table}

We study the performance of fine-tuning all models using R\textsubscript{Tidy}, R\textsubscript{Base}, and R\textsubscript{combine}. As we did for code summarization, we re-trained the baseline models to get performance numbers for each of the six programming languages in the codeSearchNet~\cite{codesearchnet} dataset and report all the results. 
Table~\ref{tab:name-mon} shows the F1 scores for all programming languages for method name prediction.
GraphCodeBERT outperforms CodeBERT, UniXCoder, and CodeT5 in all languages except Go, for which GraphCodeBERT and CodeBERT achieve the same score. 
The performance of GraphCodeBERT demonstrates that GraphCodeBERT’s graph-based approach is more
effective than the other three baseline models in handling method name prediction. 
On the other hand, UniXCoder does not continue to maintain its performance in code summarization, instead, UniXCoder shows its struggle for method name prediction, being the lowest in JS, PHP, 
R\textsubscript{Tidy}, and R\textsubscript{Base}. 
UniXCoder, which is designed to handle multiple code-intelligence tasks in a unified manner, shows varied results across languages.
Its lower scores compared with those of the other models may stem from its broader generalization approach, which may not specialize as effectively in code understanding.
Likewise, CodeT5 failed to maintain strong performance on this task and was the worst performer in the four remaining languages—Go, Java, Python, and Ruby.

The R language datasets present significant challenges for all models. 
The scores are around $0.25$ for most models and around $0.32$ for the best model (i.e., GraphCodeBERT). 
Interestingly, for this task, there is no difference between the R\textsubscript{Tidy}, R\textsubscript{Base}, and R\textsubscript{combine}; and all R datasets have exact or very close scores. 
This indicated the need to handle R’s domain-specific syntax and statistical focus by further adaptation or specialized pre-training models that work for R.

\subsection{RQ2: Performance of Code-PLMs in Multi-lingual Fine-Tuning Setting on R}
In this section, we discuss the multilingual setup for RQ2. 
\subsubsection{Code Summarization}


\begin{table}[h]
    \centering
    \begin{tabular}{|c|c|c|c|c|c|c|c|c|}
        \hline
           & \multicolumn{2}{|c|}{\textbf{CodeBERT}} & \multicolumn{2}{|c|}{\textbf{GraphCodeBERT}} & \multicolumn{2}{|c|}{\textbf{UniXCoder}} & \multicolumn{2}{|c|}{\textbf{CodeT5}} \\
        \hline
          \textbf{Language} & \textbf{BLEU} & p-value & \textbf{BLEU} & p-value & \textbf{BLEU} & p-value & \textbf{BLEU} & p-value\\
        \hline
        JS & 15.1 & $0.008^*$ & 15.77 & $<0.001^*$ & 15.63 & 0.452 & \underline{\textbf{16.03}} & $0.036^*$ \\
        
        Go & 18.38 & 0.672 & \textbf{19.15} & $<0.001^*$ & 18.31 & $<0.001^*$ & 6.16 & $<0.001^*$ \\
        
        Java & 18.73 & $0.003^*$ & \textbf{19.84} & $<0.001^*$ & 19.58 & $<0.001^*$ & 7.39 & $<0.001^*$ \\
        
        PHP & \underline{25.34} & 0.192 & \underline{\textbf{26.35}} & $<0.001^*$ & \underline{25.97} & $<0.001^*$ & 7.39 & $<0.001^*$ \\
        
        Ruby & 14.78 & $<0.001^*$ & \textbf{14.98} & $<0.001^*$ & 14.31 & $0.004^*$ & 5.63 & $<0.001^*$ \\
        
        Python & 18.59 & $<0.001^*$ & \textbf{19.01} & $<0.001^*$ & 18.88 & $<0.001^*$ & 7.76 & $<0.001^*$ \\
        \hline
        
        R\textsubscript{Combine} & 7.24 & $<0.001^*$ & 9.65 & $<0.001^*$ & \textbf{14.34} & $<0.001^*$ & 2.33 & $<0.001^*$\\
        \hline
        R\textsubscript{Tidy} & 8.4 & $<0.001^*$ & 11.14 & 0.766 & \textbf{18.79} & 0.077 & 2.23 & $<0.001^*$\\
        R\textsubscript{Base} & 9.02 & $<0.001^*$ & 12.02 & $0.003^*$ & \textbf{20.02} & 0.759 & 2.11 & $<0.001^*$\\
        \hline
        
    \end{tabular}
    \caption{Performance of all the baselines for the six programming languages on code summarization, R styles, and the combined R dataset under multi-lingual fine-tuning. Best results in each row are \textbf{bold} and best results in each column are \underline{underlined}.}
    \label{tab:sum_multi}
\end{table}

Table~\ref{tab:sum_multi} represents the results for code summarization under multi-lingual fine-tuning. In this setting, the performance of the baseline models contradicted with previous findings~\cite{Ahmed}, where multi-lingual fine-tuning typically improved model performance.
For UniXCoder, and CodeT5, multilingual fine-tuning actually leads to performance degradation. UniXCoder experiences a drop in performance for most programming languages, and CodeT5 shows slight performance degradation only in Ruby, with a significant drop in all other languages, indicating that the effectiveness of multilingual fine-tuning varies depending on the model. This drop in performance is also reported in previous works~\cite{wang}, and dropped even further with the addition of R language to the dataset. 
Specifically, for CodeBERT, the BLEU-4 scores increased for JavaScript, Go, and Ruby by $7\%$, $1.4\%$, and $28\%$, respectively. The improvement is significant for JavaScript and Ruby, showing a modest improvement across these languages. 
However, other languages exhibited mixed results. For example, PHP and Java showed almost no change, and for the R language, the combined R dataset (R\textsubscript{Combine}) demonstrated a $39\%$ performance drop compared to the monolingual setting.

When examining GraphCodeBERT, there was a consistent trend of performance improvement across all languages except for combined R dataset (R\textsubscript{Combine}), which is decreased by $1.4$ BLEU-4 score. All six languages of CodeSearchNet data have about one BLEU score improvement, except Ruby, which jumps to $14.98$ from $11.76$, seeing the highest increase.
The performance on R datasets was more variable, with R\textsubscript{Base} seeing a notable increase of $25\%$ (p-value equal to 0.003), while R\textsubscript{Tidy} only increases $7\%$ (with p-value equal to 0.766, which is not significant). Here we can see that although GraphCodeBERT performed well under multi-lingual fine-tuning, the combination of R\textsubscript{Tidy} and R\textsubscript{Base} would hurt the effects of fine-tuning while testing them, as shown in the results of R\textsubscript{Combine}.

In multi-lingual setting, UniXCoder~\cite{unixcoder} on the other hand, shows reduced BLEU scores for all languages, albeit slightly.
Particularly for R\textsubscript{Combine}, while it had previously demonstrated strong performance in monolingual settings, its performance dropped by 22\% (with p-value less than 0.001). 
For R\textsubscript{Base} and R\textsubscript{Tidy}, the performance dropped slightly. 
This observation again highlights the negative effects of combining R\textsubscript{Tidy} and R\textsubscript{Base} datasets during testing.



Lastly, CodeT5 demonstrated very low performance on all languages, except JavaScript. 
This performance drop is more noticeable for R datasets, where the BLEU scores are around $2$. 
The low scores could be attributed to CodeT5's text-to-code architecture, which may not generalize as well in multi-lingual fine-tuning. While CodeT5 performed well in the monolingual setting, its performance dipped significantly during fine-tuning on the R datasets. This points to a potential limitation in its ability to handle diverse programming paradigms across different languages. It is worth pointing out that CodeT5 also took the most time to fine-tune, and during training, we observed that the model had a hard time converging to the point where the model showed a drop in performance on the dev set. 
Combined with the performance of other baseline models, this is a good indication that pre-trained models such as CodeT5 have some shortcomings in learning a language such as R, which has its own unique syntax and structure, especially for R\textsubscript{Combine}.

To evaluate the significance of the differences in model performance, we employed the Wilcoxon signed-rank test, a non-parametric statistical test commonly used for comparing paired samples. Wilcoxon signed-rank test was chosen due to its suitability for comparing two related samples, making it ideal for performance comparisons across different models on the same dataset. The p-value is computed by applying the Wilcoxon signed test on mono-lingual fine-tuning result and multi-lingual fine-tuning for code summarization and method name prediction. Under this setting the p-value represents whether the observed differences in performance are statistically significant or not.
To control for the risk of false positives due to multiple comparisons, we applied the Benjamini-Hochberg (B-H) correction following a previous study~\cite{Ahmed}. The B-H correction is a widely used method for controlling the false discovery rate (FDR), which refers to the expected proportion of incorrectly rejected null hypotheses (false positives) among all rejected hypotheses. By adjusting p-values obtained from the Wilcoxon signed-rank test, the B-H correction allows us to maintain a balance between discovering true effects and limiting the inclusion of false positives when performing multiple comparisons.

The p-values in Table~\ref{tab:sum_multi} that are $<0.05$ and show the numbers are statistically significant from mono-lingual setting, are shown with an asterisk (*). 
In the table, some results are not significant based on their p-value. These include JavaScript for UniXCoder, Go for CodeBERT, PHP for CodeBERT, R\textsubscript{Tidy} for GraphCodeBERT and UniXCoder, and R\textsubscript{Base} for UniXCoder, these higher p-values indicate that the differences in BLEU scores for these models are not statistically significant.
Most results, though, show a significant difference for R.

In summary, multi-lingual fine-tuning presented varied impacts across models and languages. While certain models like GraphCodeBERT excelled in handling multi-lingual, all models struggled with R\textsubscript{Combine}. This suggests the disadvantages of the baseline model, particularly when dealing with diverse languages and paradigms like those in the R language.

\subsubsection{Method Name Prediction}
\begin{table}[h]
    \centering
    \begin{tabular}{|c|c|c|c|c|c|c|c|c|}
        \hline
           & \multicolumn{2}{|c|}{\textbf{CodeBERT}} & \multicolumn{2}{|c|}{\textbf{GraphCodeBERT}} & \multicolumn{2}{|c|}{\textbf{UniXCoder}} & \multicolumn{2}{|c|}{\textbf{CodeT5}} \\
        \hline
         \textbf{Model} & \textbf{F1-score}& p-value & \textbf{F1-score}& p-value & \textbf{F1-score}& p-value & \textbf{F1-score}& p-value\\
        \hline
        JS & 0.13& $<0.001^*$ &  0.15 & $<0.001^*$ & 0.16 & $<0.001^*$ & \underline{\textbf{0.32}} & $<0.001^*$\\
        
        Go & 0.13& $<0.001^*$ &  0.13& $<0.001^*$ & \textbf{0.14}& $<0.001^*$ & 0.12 & $<0.001^*$\\
        
        Java &  0.15& $<0.001^*$& \textbf{0.16}& $<0.001^*$ & \textbf{0.16}& $<0.001^*$ & 0.115 & $<0.001^*$\\
        
        PHP & \underline{\textbf{0.21}}& $<0.001^*$ &  0.20& $<0.001^*$ & 0.19& $<0.001^*$ & 0.107 & $<0.001^*$\\
        
        Ruby & 0.14& $<0.001^*$ &  0.15& $<0.001^*$ & \textbf{0.18}& $<0.001^*$ & 0.112 & $<0.001^*$\\
        
        Python & 0.16& $<0.001^*$ &  0.16& $<0.001^*$ & \textbf{0.19}& $<0.001^*$ & 0.114 & $<0.001^*$\\
        \hline
        
        R\textsubscript{Combine} & 0.20& $<0.001^*$ &  \textbf{0.28}& $<0.001^*$ &  0.23 & $<0.001^*$ & 0.10 & $<0.001^*$\\
        \hline
        R\textsubscript{Tidy} & \underline{0.21}& $<0.001^*$&  \textbf{0.38}& $<0.001^*$ &  0.23 & 0.227 & 0.097 & $<0.001^*$\\
        R\textsubscript{Base} & \underline{0.21}& $<0.001^*$ &  \underline{\textbf{0.41}}& $0.002^*$ &  \underline{0.24}& $0.008^*$ & 0.095 & $<0.001^*$\\
        
        \hline
    \end{tabular}
    \caption{Result of method name prediction for RQ2. Best results in each row are bold, and best results in each column are underlined. The p-values compared to mono-lingual setting are shown for each model. P-values smaller than $0.05$ show a statistically significant difference and are marked with an asterisk (*).}
    \label{tab:name_multi}
\end{table}

Table~\ref{tab:name_multi} shows the result of all the baselines for the six programming languages, R styles, and combined R dataset on method name prediction under multilingual training. 
We observe a significant drop in performance for all four models and the six languages of the CodeSearchNet dataset (the p-value is less than 0.001 for nearly all of the models and languages).
Multilingual training generally provides models with a broader understanding of diverse languages by exposing them to a combined dataset, which helps in cross-linguistic generalization but may dilute performance on individual languages due to the need to balance between multiple linguistic patterns.
For all six languages of the CodeSearchNet dataset, and for all models, the performance has dropped to $0.10s$ with only a few close to $0.2$. 


In examining the R styles, especially R\textsubscript{Tidy} and R\textsubscript{Base}, the results under multilingual training are notably different. GraphCodeBERT performs surprisingly well in these cases, boosting the performance; particularly under multilingual training, where it scores $0.38$ for R\textsubscript{Tidy} and $0.41$ for R\textsubscript{Base}, compared to $0.31$ and $0.32$ in a mono-lingual setting, which is statistically significant. 
This means the performance is improved by $22.58\%$ and $31.25\%$ for each of the R\textsubscript{Tidy} and R\textsubscript{Base}, respectively.
However, the performance of CodeBERT and UniXCoder drops slightly under multi-lingual fine-tuning, but the p-value still shows that the change is significant since it is less than 0.05.
For R\textsubscript{Combine}, the performance is dropped slightly in a multi-lingual setting for all models, suggesting that combining the two R styles for method name prediction is not helpful, though separate styles would benefit in a multi-lingual setting.

For R\textsubscript{Combine}, R\textsubscript{Tidy}, and R\textsubscript{Base}, all p-values are less than $0.05$ (marked with an asterisk), indicating statistically significant differences compared to the monolingual setting, with one exception. The exception is UniXCoder's performance on R\textsubscript{Tidy}, where the p-value of $0.227$ suggests no statistically significant difference.
Notably, GraphCodeBERT consistently shows the best performance across all R datasets, with F1-scores of $0.28$, $0.38$, and $0.41$ for R\textsubscript{Combine}, R\textsubscript{Tidy}, and R\textsubscript{Base}, respectively, outperforming all other languages. These results are all statistically significant ($p < 0.05$).

Overall, the comparison between multilingual training and monolingual fine-tuning reveals that while multilingual training can offer some cross-linguistic generalization, especially for models like GraphCodeBERT, monolingual fine-tuning consistently yields higher performance. The notable differences in performance across different languages and models also highlight the varying degrees to which each model can generalize across languages, with CodeT5, for example, showing the largest gap between multilingual and monolingual settings, struggling more with cross-linguistic generalization compared to others.



CodeT5, built on the T5 architecture, frames code-related tasks as text-to-text problems. This approach allows for a unified model capable of handling various software engineering tasks, including both code summarization and method name prediction.
Our results show that each model's architecture influences its performance differently across tasks. For both tasks, we found that CodeT5's specific architecture is experiencing a significant drop in performance with a drop of over $50$\% and a p-value less than $0.001$. 
This suggests that a unified framework that converts language problems into a text-to-text format will extraordinarily affect the performance of the model under multi-lingual training. 
It is worth pointing out that CodeT5 also has an early stop mechanism, which will automatically stop fine-tuning if the training results do not improve in relation to a specific round to reduce the fine-tuning time. When fine-tuning under a multi-lingual setup, the model takes much longer than mono-lingual fine-tuning, and more epochs to stop. This also suggests that converting the problem to a text-text problem under multi-lingual fine-tuning poses an extra challenge for the model to converge.

Although multilingual fine-tuning can incorporate knowledge from a wide array of languages, thereby exposing the model to diverse syntactic structures and potentially improving generalization, R has domain-specific characteristics—particularly statistical functions and specialized libraries—that may not align with the broader patterns learned from other programming languages. This mismatch can dilute the relevance of learned representations, preventing the model from fully capturing R’s unique semantics. In other words, while multilingual data might grant a broader contextual understanding, it risks overshadowing the specialized needs of R code, which can ultimately hinder performance when domain-specific nuances are critical, which is supported by our result in this section.

\subsection{RQ3: Effect of Intra- and Cross-Project Training}

\subsubsection{Code Summarization}

\begin{table}[h]
    \centering
    \begin{tabular}{|c|c|c|c|c|c|}
        \hline
         & \textbf{Dataset/Model}& \textbf{CodeBERT} & \textbf{GraphCodeBERT} & \textbf{UniXCoder} & \textbf{CodeT5} \\
        \hline
        \multirow{3}{3em}{\textbf{Cross-project}} & 
        R\textsubscript{Combine} & 4.12   & 4.28  & \textbf{5.95} & 5.9 \\
        &R\textsubscript{Tidy} & 3.48   & 3.51   & 4.35 & \textbf{5.19} \\
        &R\textsubscript{Base} & 4.13   & 3.68   & \textbf{6.58} & 5.66 \\
        
        \hline
        \multirow{3}{3em}{\textbf{Intra-project}} & 
        R\textsubscript{Combine} & 16.56   & \underline{11.91}  & \textbf{19.82} & \underline{18.36} \\
        &R\textsubscript{Tidy} & 18.78   & 10.76   & \textbf{18.94} & 17.64 \\
        &R\textsubscript{Base} & \textbf{\underline{31.18}}   & 9.46   & \underline{21.45} & 16.87 \\
        \hline
    \end{tabular}
    \caption{Result of code summarization for RQ2 Best results in each row are \textbf{bold} and best results in each column are \underline{underlined}.}
    \label{tab:withinS}
\end{table}

Table~\ref{tab:withinS} reports the code-summarization results under cross-project training for CodeBERT, GraphCodeBERT, UniXCoder, and CodeT5.
In cross-project training, all code snippets from a given project are assigned exclusively to the training, validation, or test set, ensuring that the model is tested on data from projects it has never seen.

For the \textit{Base} R style, UniXCoder outperforms the other models with a BLEU-4 score of $6.58$, followed by CodeT5 with $5.66$, CodeBERT with $4.13$, and GraphCodeBERT with $3.68$. 
In the \textit{Tidy-verse} style, CodeT5 again leads with a score of $5.19$, while UniXCoder is close behind at $4.35$. CodeBERT and GraphCodeBERT perform similarly in this style, scoring $3.48$ and $3.51$, respectively. 
When considering the combined R dataset, which includes both \textit{Tidy-verse} and \textit{Base} R, UniXCoder maintains the highest performance with a BLEU-4 score of $5.95$, with CodeT5 closely trailing at $5.9$. GraphCodeBERT slightly outperforms CodeBERT in this combined scenario, scoring $4.28$ compared to $4.12$.

These results demonstrate that UniXCoder mostly delivers superior performance across different R programming styles in a cross-project training scenario, particularly excelling in \textit{Base} R tasks. 
CodeT5 also shows competitive results, particularly in the \textit{Tidy-verse} style. Though UniXCoder and CodeT5 have the highest scores in the cross-project setting, their performance has dropped significantly, more than $11$ BLEU-4 scores compared to Table~\ref{tab:sum-mon}.
Similar to RQ1, \textit{Base} R and the combined version have higher scores compared to \textit{Tidy-verse}. 
This finding aligns with previous work on intra- vs.\ cross-project settings~\cite{10.1145/2884781.2884839}, where intra-project outperforms cross-project settings in most cases.

In contrast, the intra-project results, where the models are tested on the same project they were trained on, demonstrate higher performance. CodeBERT, for example, achieves a significant improvement on the R\textsubscript{Base} dataset, scoring $31.18$, an improvement from $4.13$ BLEU-4. 
We observe the highest improvement for CodeBERT, followed by CodeT5, compared to other models. 
Among the two R styles and R\textsubscript{Combined}, R\textsubscript{Base} scores increase more than the others. 

Overall, the intra-project results underscore the importance of project-specific training, as models tend to perform significantly better when tested on data with a similar distribution. This is particularly evident from the large boosts in the models' performances from cross-project to intra-project scenarios. 
GraphCodeBERT, however, although it still improves, does not gain as much, suggesting that it may struggle more with the nuances of project-specific contexts.
This aligns with our previous findings from RQ1, where GraphCodeBERT had the lowest score among all models.
CodeT5 and UniXCoder, while performing well across both settings, show a slightly smaller margin of improvement, indicating a more consistent but less specialized approach to the task.


\subsubsection{Method Name Prediction}
 \begin{table}[h]
    \centering
    \begin{tabular}{|c|c|c|c|c|c|}
        \hline
         &\textbf{Dataset}& \textbf{CodeBERT} & \textbf{GraphCodeBERT} & \textbf{UniXCoder} & \textbf{CodeT5} \\
        \hline
        \multirow{3}{3em}{\textbf{Corss-project}} 
        & R\textsubscript{Combine} & 0.17 & 0.18  & 0.19 & \textbf{0.20} \\
        & R\textsubscript{Tidy} & 0.14   & 0.17  & 0.15 & \textbf{0.17} \\
        & R\textsubscript{Base} & 0.16   & 0.16   & 0.11 & \textbf{0.19} \\
        
        \hline
        \multirow{3}{3em}{\textbf{Intra-project}} 
        &R\textsubscript{Combine} & \underline{0.27}   & \underline{0.27}  & 0.19 & \underline{\textbf{0.30}} \\
        &R\textsubscript{Tidy} & 0.26   & \underline{0.27}   & 0.18 & \underline{\textbf{0.30}} \\
        & R\textsubscript{Base} & 0.26   & \underline{0.27}   & \underline{0.20} & \underline{\textbf{0.30}} \\

        \hline
    \end{tabular}
    \caption{The result for method name prediction using cross-project training and Intra-project training. Best results in each row are \textbf{bold} and best results in each column are \underline{underlined}.}
    \label{tab:RQ3_name}
\end{table}

Table~\ref{tab:RQ3_name} represents the results for method name prediction in intra- and cross-project settings. The results highlight the consistent trend observed in prior studies involving intra and cross-project settings~\cite{cross}. Models generally perform better in intra-project settings compared to cross-project settings. This pattern is evident across all four models when applied to the R programming datasets.


In the cross-project training scenario for R\textsubscript{Combine}, CodeT5 achieved the highest prediction accuracy with a score of $0.20$, followed closely by UniXCoder at $0.19$. CodeBERT and GraphCodeBERT lagged slightly behind, with scores of $0.17$ and $0.18$, respectively.

In the intra-project training context, the performance improved overall, with the best result again coming from CodeT5, which reached $0.30$. Both CodeBERT and GraphCodeBERT scored equally at $0.27$, while UniXCoder recorded a lower score of $0.19$. 
There is a notable improvement in performance when moving from cross-project to intra-project training for R\textsubscript{Combine}, with scores increasing by 0.10 for CodeBERT and GraphCodeBERT, and CodeT5. CodeT5 shows the best performance with R\textsubscript{Combine} in both settings.

In the intra-project setting, where the model is trained and tested on data from the same project, the F1 scores are significantly higher across all datasets and models. For instance, CodeT5 achieves an F1 score of 0.30 on the R\textsubscript{Base} dataset, compared to 0.19 in the cross-project setting. Similarly, UniXCoder's performance on the R\textsubscript{Base} dataset jumps from 0.11 in the cross-project setting to 0.20 in the intra-project setting. This improved performance indicates that models can better capture the idiosyncrasies and specific coding practices within a single project, leading to more accurate predictions.

\subsection{RQ4: Code-PLM Performance When Separating R Style Datasets}

\begin{figure}[h]
    \centering
    \includegraphics[width=1\textwidth]{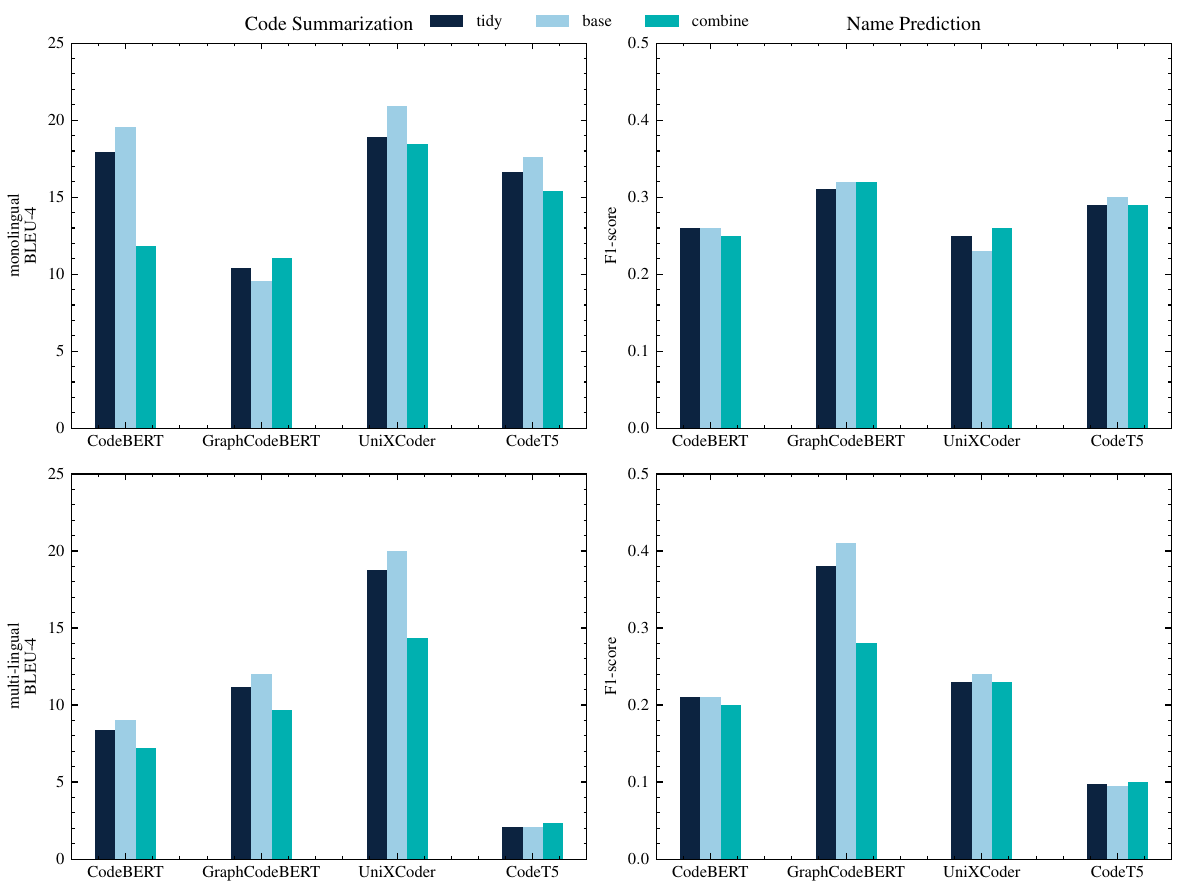}
    \caption{Performance of baseline models on code summarization and method name prediction tasks across different programming styles in R under monolingual and multilingual setups.}
    \label{fig:RQ4_1}
\end{figure}

In this section, we examine the performance of four baseline models fine-tuned on three distinct R programming datasets for two tasks of {code summarization} and {name prediction}, as shown in Figure~\ref{fig:RQ4_1}. The goal is to evaluate how different styles of R and varying training setups impact the models' ability to accurately summarize code and predict method names.
We also examined the statistic significance of the difference between R\textsubscript{Combine} and the two styles, R\textsubscript{Tidy} and R\textsubscript{Base}. Since the population size is not the same across R\textsubscript{Combine}, R\textsubscript{Tidy}, and R\textsubscript{Base}, we take random samples from R\textsubscript{Combine} to match the size of the other two. 

\begin{itemize}
    
    \item For code summarization under monolingual training, the differences in performance are statistic significant for CodeBERT and CodeT5, the differences are only significant between R\textsubscript{Combine} and R\textsubscript{Base} for GraphCodeBERT and UniXCoder. When looking at name prediction under monolingual fine-tuning, the difference in performance is only statistically significant between R\textsubscript{Combine} and R\textsubscript{Base} for UniXCoder.
    
    \item For code summarization under multi-lingual fine-tuning, the differences in performance are statistically significant for all models except CodeT5, where the performance dropped 5\% and 9.4\% for R\textsubscript{Tidy} and R\textsubscript{Base}, respectively. For name prediction under multi-lingual fine-tuning, similar trends were found, where the difference in performance is only statistically significant for one model, which is GraphCodeBERT.
    
    \item For code summarization under cross-project training, the difference in performance is statistically significant between R\textsubscript{Combine} and R\textsubscript{Tidy} for CodeBERT, GraphCodeBERT, and UniXCoder. The differences are more significant under intra-project training, where all of them are statistically significant.
    
    \item For method name prediction using cross-project training, the difference in performance is significant for CodeBERT and UniXCoder. It is only significant for GraphCodeBERT and CodeT5 between R\textsubscript{Combine} and R\textsubscript{Base} and R\textsubscript{Combine} and R\textsubscript{Tidy}, respectively. Unlike code summarization, the differences are less significant under intra-project training, where all of them are not statistically significantly different.
    
\end{itemize}

In the \textbf{code summarization} task for the monolingual setting, the results indicate that models trained on the \textit{Base} R dataset outperform those trained on the \textit{Tidy-verse} and the combined datasets. The \textit{Base} R models produced the highest scores for all baseline models except GraphCodeBERT, indicating that these models handle \textit{Base} R syntax more effectively when summarizing code. 
The improvement in performance for \textit{Base} R could be a benefit from a syntax structure that is simpler and closer to other programming languages. \textit{Tidy-verse}-based models, while still performing relatively well, did not match the performance of R\textsubscript{Base} models, suggesting that the abstract and functional style of \textit{Tidy-verse} poses additional challenges for the models during the fine-tuning phase. 
The combined dataset, which merges both \textit{Base} R and \textit{Tidy-verse}, resulted in the lowest performance. 
This outcome implies that combining two distinct paradigms introduce complexities that interfere with the model’s ability to generate concise and accurate summaries. The performance shows that the two styles introduce conflicting patterns to the models and make the models struggle to reconcile. 

In contrast, in the \textbf{name prediction} task, the performance differences between the datasets are minimal. Whether the models were trained on \textit{Base} R, \textit{Tidy-verse}, or the combined dataset, the performance remained relatively similar. 
Previous work~\cite{Ahmed} shows that the identifier plays the most important role in code tasks. This consistency indicates that method name prediction is less sensitive to the specific R style used, but relies more on the identifiers to capture the key information. 
Developers would use method names across \textit{Base} R and \textit{Tidy-verse} that share enough similarities to enable the models to generalize effectively across these paradigms. 
The combined dataset, although slightly weaker, did not exhibit the performance drop seen in code summarization. 
The consistency across different R programming styles reinforces the observation that name prediction is less dependent on variations in R programming styles, but mainly dependent on the identifiers. 

When shifting to the \textbf{multilingual setup}, the performance for both tasks showed various degrees of decline. 
In the \textbf{code summarization} task, the same pattern observed in the monolingual setup persisted: models trained on \textit{Base} R achieved the highest scores for most baseline models, while the combined dataset performed the worst. Contradicting previous research~\cite{Ahmed}, the multilingual environment did not improve the performance of the models; it introduces additional complexity, as the models must handle syntax and structures from multiple programming languages that has different syntax and structure in many ways. 
This added burden likely impedes their ability to focus on the nuanced differences between \textit{Base} R and \textit{Tidy-verse}, making it more difficult to generate accurate code summaries, especially when trained on a mix of both paradigms.

Similarly, in the \textbf{multilingual name prediction} task, the models’ performance remained consistent across datasets, mirroring the monolingual setting. The scores displayed little variation, with \textit{Base} R, \textit{Tidy-verse}, and combined datasets all yielding nearly identical results, except for GraphCodeBERT. 
This further confirms that name prediction is a relatively straightforward task for these models, even when trained in a multilingual setting. It suggests that method names in R are relatively easy for the models to generalize, regardless of the specific programming style or the presence of multiple languages in the training set.
Similar to code summarization, the scores are dropped in multilingual settings.

To conclude, this section's analysis reveals several key trends. First, for code summarization, models trained exclusively on \textit{Base} R tend to perform better than those trained on \textit{Tidy-verse} or a combination of both, particularly in a monolingual setup. The \textit{Tidy-verse} introduces additional complexity, which the models do not handle as effectively, and combining both styles into one dataset exacerbates this issue.
In contrast, for method name prediction, the models are largely unaffected by the specific R paradigm, with shows similar performance across all datasets and setups, indicating that this task is less sensitive to the underlying coding style. Finally, while the multilingual setup introduces some performance degradation, particularly in code summarization, the models maintain consistent performance in name prediction, demonstrating their ability to generalize method names across different programming languages and paradigms. These insights highlight the importance of choosing the right dataset and fine-tuning setup for specific tasks, especially in complex environments like R programming.

\subsubsection{Model Performance by Code Length}
\begin{figure}[h]
    \centering
    \includegraphics[width=1\textwidth]{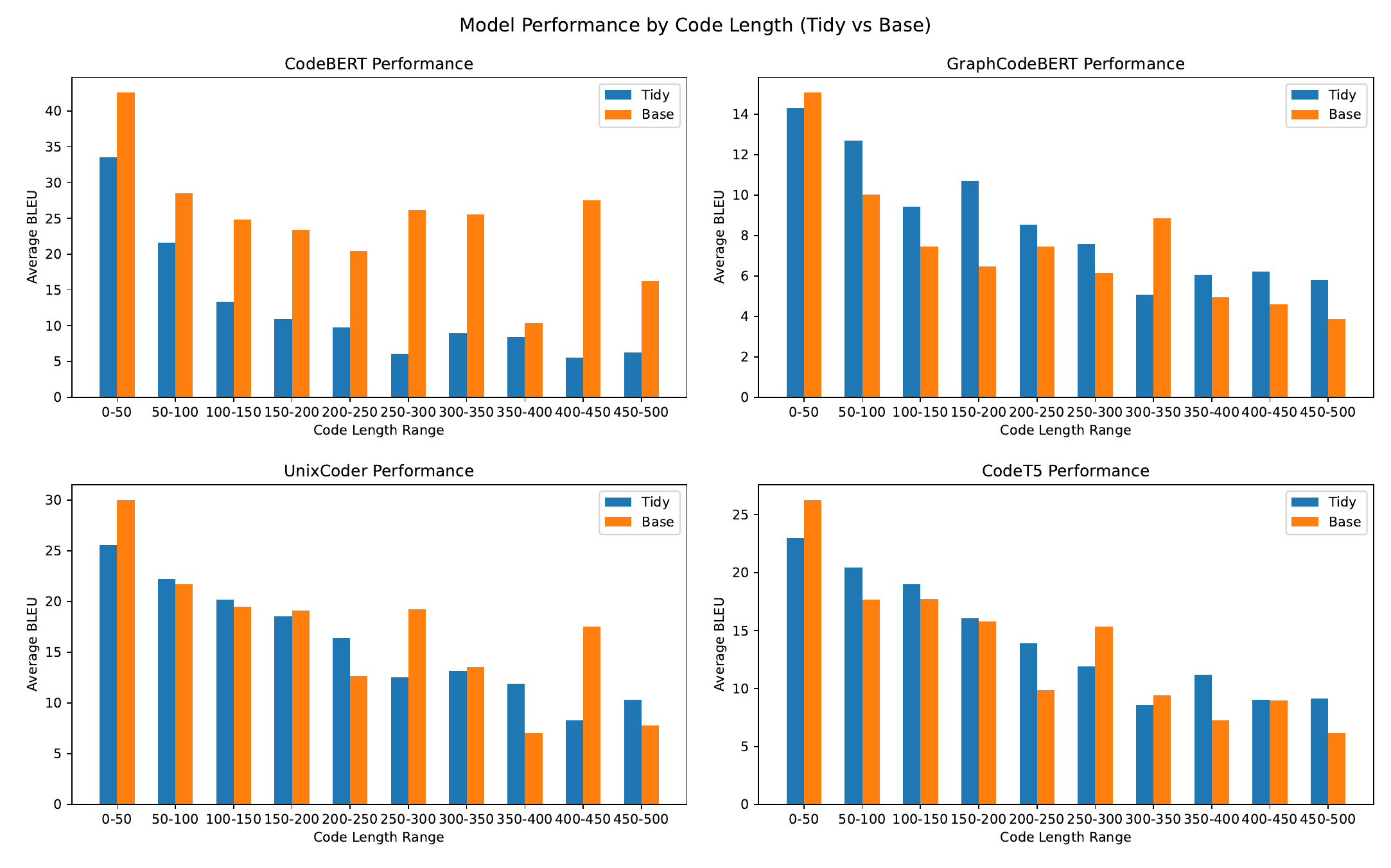}
    \caption{Performance of baseline models on Tidy and Base across different code lengths in R under monolingual setups.}
    \label{fig:code_length}
\end{figure}

Figure \ref{fig:code_length} compares the performance of four baseline models—CodeBERT, GraphCodeBERT, UnixCoder, and CodeT5—across code segments of varying lengths in R, while contrasting the \textit{Tidy-verse} and \textit{Base} versions of the dataset for code summarization. A clear trend emerges in which shorter code snippets (for instance, those under 50 tokens) achieve higher BLEU scores. This outcome is not surprising, as shorter snippets frequently lack the level of complexity found in longer blocks of code, making it easier for the models to learn mappings between source code and summary. Fewer tokens also reduce the chance of encountering convoluted control flow or deeply nested function calls, leading to more straightforward patterns that the models can accurately capture.

As the code length increases beyond around 150 or 200 tokens, performance begins to decrease, reflecting the additional challenges posed by complex logic, multiple layers of function calls, and generally higher token diversity. When code is longer, the model must handle a broader context and maintain a more extensive memory of previous tokens to accurately capture meaning. The chart further shows that, within the same length range, the \textit{Tidy-verse} data often confers a noticeable advantage over \textit{Base} data (except for CodeBERT). \textit{Tidy-verse} data removes inconsistencies such as irregular spacing or extraneous comments, thereby reducing noise and helping the model focus on core functional elements of the code. This benefit becomes even more significant for longer snippets, where seemingly minor inconsistencies can accumulate and hinder the model’s ability to process and summarize effectively. 
Additionally, \textit{Tidy-verse} is focused on data science, providing a consistent and intuitive syntax (such as the pipe operator and a series of well-defined functions) that makes data cleaning, transformation, and visualization more efficient and readable\footnote{\href{https://tidyverse.tidyverse.org/articles/paper.html}{https://tidyverse.tidyverse.org/articles/paper.html}}. This design approach is similar to the usage of data science libraries like pandas in Python, as both emphasize code readability and coherent operations. Therefore, the syntax and usage could be more specific compared to \textit{Base R}, and this difference might affect the results. 

\subsection{RQ5: LLM Results} \label{sec:rq5-results}

\begin{table}[h]
    \centering
    \begin{tabular}{c|cc|cc|cc}
        \hline
           & \multicolumn{2}{c}{\textbf{K=1}} & \multicolumn{2}{c}{\textbf{K=5}} & \multicolumn{2}{c}{\textbf{K=10}} \\
        \textbf{Models} & R & Python & R & Python & R & Python \\ \hline
        UniXCoder & 0 & 0 & 0 & 0 & 0 & 0 \\
        codegen-2B-multi & 4.32 & 10.18 & 6.89 & 19.51 & 7.76 & 26.83\\
        CodeLlama-7B & 20.50 & 25.98 & 27.74 & 48.17 & 33.5 & 57.93\\
        StarCoder2-7B & 19.65 & 28.9 & 28.35 & 51.21 & 35.4 & 66.46\\
        \hline
    \end{tabular}
    \caption{Pass@k on HumanEval for selected LLMs. The temperature is set to 0.8.}
    \label{tab:code_gen}
\end{table}

The results for code generation on MultiPL-E~\cite{MultiPL-E} and Humaneval~\cite{humanevaluating} are shown in Table~\ref{tab:code_gen}.
UniXCoder achieves 0 for both R and Python for all metrics, showing it is not able to produce executable code, in contrary to the LLMs.
The performance disparity between R and Python is a key finding of this study. Across all evaluated models, R consistently yields lower pass@k scores compared to Python. For instance, the top-performing StarCoder2-7B model achieves a pass@10 score of $35.4\%$ for R, which, while impressive, falls significantly short of its $66.46\%$ score for Python. This gap is consistently observed across different model sizes and architectures, suggesting a systemic challenge in R code generation. While UniXCoder fails to generate valid solutions for both languages, achieving 0\% pass rates across all K values, larger models show more promise but with notable language-dependent differences. CodeLlama-7B and StarCoder2-7B demonstrate superior performance, particularly for Python, with StarCoder2-7B achieving the highest pass rates of 28.9\% at K=1 and 66.46\% at K=10 for Python code generation.

The improvement in pass@k scores as k increases is particularly noteworthy for R. For StarCoder2-7B, the score nearly doubles from k=1 (19.65\%) to k=10 (35.4\%). This suggests that allowing multiple generation attempts could significantly enhance the practical utility of these models in R programming environments. It also implies that the models have a broader understanding of potential R solutions but may require multiple attempts to articulate the correct one. The performance trends across model sizes are consistent for both R and Python, with larger models generally performing better. For instance, the jump from codegen-2B-multi to CodeLlama-7B shows a more substantial relative improvement for both R and Python. This performance improvement is expected, as the model has more attempts to create a correct code, which is not specific to R. 

\begin{table}[ht!]
\centering
\begin{tabular}{lcccccc}
\toprule
\textbf{few-shot-example} & \textbf{Model} & \textbf{Zero Shot} & \textbf{1-Shot} & \textbf{3-Shot} & \textbf{5-Shot} & \textbf{10-Shot} \\
\midrule
\multirow{3}{4em}{\textbf{Simple R function}}
&codegen-2B-multi & 4.3 & 4.3 & 5.6 & 5.6 & 0.6 \\
&CodeLlama-7B     & 21.1 & 17.4 & 18.6 & 11.8 & 0.0 \\
&StarCoder2-7B    & 20.5 & 19.3 & 22.4 & 22.4 & 9.9 \\
\hline
\multirow{3}{3em}{\textbf{BM25}}
&codegen-2B-multi & 4.3 & 1.24 & 0.0 & 0.0 & 0.0 \\
&CodeLlama-7B     & 21.1 & 7.45 & 0.0 & 0.0 & 0.0 \\
&StarCoder2-7B    & 20.5 & 7.45 & 0.0 & 0.0 & 0.0 \\
\midrule
\multirow{3}{3em}{\textbf{Embedding}}
&codegen-2B-multi & 4.3 & 1.86 & 1.86 & 0.62 & 0.0 \\
&CodeLlama-7B     & 21.1 & 10.56 & 12.42 & 3.73 & 0.0 \\
&StarCoder2-7B    & 20.5 & 15.53 & 11.80 & 8.70 & 0.0 \\
\bottomrule
\end{tabular}
\caption{Model Performance at various shot counts applied to R. Results represent pass@1. The temperature is set to 0, and we follow a greedy approach.}
\label{tab:shot-comparison}
\end{table}


Previous research shows that adding examples as context to the models improves their performance for code generation~\cite{nashid2023retrieval}. 
Therefore, we conducted additional experiments, in which we provided the models with additional context, being 1, 3, 5, and 10 samples. 
We applied three settings to add additional context. In the first setting, the examples are chosen from R tutorials and/or written by an R expert. These are simple examples showing, for example, how to import a csv file or read a column from a tabular dataset.
Then these examples are used as the additional context to the model. 
In the second setting, we used the BM25 algorithm to choose the top $K$ similar samples from our collected dataset, where $k \in \{1,3,5,10\}$ and chose them as the context in the few-shot setting. In the third setting, we calculated the cosine similarity between the embeddings of our collected dataset and the question prompt, and chose the top $K$ similar samples from the dataset as the context in the few-shot setting.

Table \ref{tab:shot-comparison} presents a performance comparison between three few-shot strategies—reusing the same, simple R function as the examples for all queries, BM25, and by cosine similarity between the embeddings. 
In the first setting, each of the three models produces higher scores compared to using BM25, reaching as high as $22.4$ (3-shot using StarCoder2-7B). This indicates that even a single, generic example can still provide enough information for the models to solve the tasks in the HumanEval setting. 
By contrast, the BM25-based approach shows consistently low performance, with most scores dropping to zero for all but the zero-shot column. Two main reasons explain this performance gap. First, the code snippets retrieved via BM25, though superficially related to the query, often do not match the HumanEval tasks in either style or goal, meaning the models do not benefit much from these examples. Second, the retrieved examples can be lengthy, so the models fail to utilize the content effectively. 
We see the same trend for the embedding-based approach, where the model shows consistently decreased performance. However, the embedding-based approach outperformed the BM25-based approach, where CodeLlama-7B and StarCoder2-7B achieved measurable improvements at different shot settings, for example, achieving $11.80$ for CodeLlama-7B and $12.42$ for StarCoder2-7B in 3-shot compared to $0$ of BM25. Embedding methods appear more adept at selecting relevant context than keyword-based BM25, showing that vector similarities capture semantic relationships beyond surface-level token matches. 
Compared to the zero-shot approach, providing 3 examples in the prompt helps improve the scores slightly. 
However, the Simple R function has similar scores with the embedding setting. 
Simple R function provides a simpler and more focused demonstration within the context window of the model, which helps generate the code for the test dataset.
Another interesting observation is that adding more examples does not improve the performance, as shown by the performance drop for 5-shot for CodeLlama-7B and all models in the 10-shot setting. 
We attribute this behavior to the lengthy input provided to the model. 

\subsection{Human evaluation}

\begin{table}[ht]
\centering
\begin{tabular}{c c c c}
\hline
\textbf{Model} & \textbf{Informativeness} & \textbf{Relevance} & \textbf{Fluency} \\ 
\hline
CodeBERT & 2.83 & 3.07 & 3.76\\
GraphCodeBERT & 3.09 & 3.28 & 3.80\\
UniXCoder & 3.09 & 3.20 & 3.78\\
CodeT5 & 3.53 & 3.59 & 3.94\\
Ground truth & 3.99 & 3.97 & 4.26\\ 
\hline
\end{tabular}
\caption{Human evaluation result for code summarization}
\label{tab:human_sum}
\end{table}

Table~\ref{tab:human_sum} shows the human evaluation scores received by each model for the code summarization task. Each model is evaluated from three aspects: informativeness, relevance, and fluency.

\textbf{Informativeness:} CodeT5 performs the best among the models with a score of $3.53$, which is notably closer to the ground truth ($3.99$) than the other models. This suggests that CodeT5 is more effective at extracting and conveying key information from the code. GraphCodeBERT and UniXCoder tie at $3.09$, indicating they perform similarly in this aspect, while CodeBERT lags slightly behind at $2.83$. The gap between the best-performing model (CodeT5) and the ground truth (0.46) indicates that there is still room for improvement in capturing the full informational content of the code.

\textbf{Relevance:}
This aspect assesses how well the generated summaries align with the actual functionality and purpose of the code. Again, CodeT5 leads the pack with a score of $3.59$, closest to the ground truth of $3.97$. GraphCodeBERT follows at $3.28$, with UniXCoder close behind at $3.20$, and CodeBERT at $3.07$. The scores suggest that while all models produce relevant summaries to some degree, CodeT5 is noticeably better at generating summaries that accurately reflect the code's purpose. The smaller gap between models in this category compared to Informativeness suggests that achieving relevance could be easier than capturing all informative details.

\textbf{Fluency:}
This metric evaluates the linguistic quality and readability of the generated summaries. All models perform relatively well in this aspect, with scores clustered more closely together and closer to the ground truth. CodeT5 again leads with $3.94$, close to the ground truth of $4.26$. The other models follow closely: GraphCodeBERT ($3.80$), UniXCoder ($3.78$), and CodeBERT ($3.76$). The high and closely grouped scores suggest that all these models are quite proficient at generating fluent and readable summaries, with less differentiation between them in this aspect compared to Informativeness and Relevance.

These results indicate that CodeT5 consistently outperforms the other models across all three aspects, coming closest to the ground truth in each category. This suggests that CodeT5 has made significant strides in understanding and summarizing R code. The scores also reveal that while these models have achieved impressive fluency in their summaries, there is still a noticeable gap in their ability to fully capture the informative content and relevance of the code compared to human-generated summaries (ground truth).

\begin{table}[ht]
\centering
\begin{tabular}{c c}
\hline
\textbf{Model} &   \textbf{Semantic Similarity} \\ \hline
CodeBERT & 2.53 \\
GraphCodeBERT & 2.66 \\
UniXCoder & 2.27 \\
CodeT5 & 2.83\\
Ground truth & 2.66 \\ \hline
\end{tabular}
\caption{Human evaluation result for name prediction}
\label{tab:human_name}
\end{table}


Table~\ref{tab:human_name} shows the human evaluation scores for the baseline models. Semantic similarity measures how well the predicted names capture the meaning and purpose of the code elements they are naming. It is a crucial aspect of name prediction, as semantically appropriate names significantly enhance code readability and maintainability.
CodeT5 stands out as the top performer with a score of $2.83$, which is notably higher than the ground truth score of $2.66$. This is an intriguing result, suggesting that CodeT5 is capable of generating names that human evaluators find even more semantically appropriate than the original names in some cases. This could indicate that CodeT5 has developed a strong understanding of code semantics and naming conventions, possibly even surpassing human performance in certain contexts. 
On the other hand, the low score received by the ground truth also indicates that some ground truth answers would make more sense when given the background information of the project.

GraphCodeBERT follows with a score of $2.66$, which interestingly matches the ground truth score exactly. This suggests that GraphCodeBERT's name predictions are, on average, as semantically similar to the code's purpose as the original names. 
CodeBERT comes in third with a score of $2.53$, which is relatively close to the ground truth. While not matching the performance of CodeT5 or GraphCodeBERT, it still demonstrates a decent ability to generate semantically appropriate names.
UniXCoder shows the lowest performance among the models with a score of $2.27$, which is noticeably below the ground truth and other models. This suggests that UniXCoder may have more difficulty capturing the semantic nuances required for effective name prediction in this context.
The relative performance of these models in name prediction aligns somewhat with their performance in code summarization (from Table~\ref{tab:human_name}), with CodeT5 leading in both tasks. This consistency suggests that the underlying capabilities that make a model good at summarizing code also contribute to its ability to predict semantically appropriate names.



\section{Discussions} \label{sec:discussions}

In this section, we examine several factors influencing the observed performance in our study, particularly in the context of R programming. Key points include the choice of models employed, the impact of the dataset, the models' generalization capabilities, and the effectiveness of different training strategies.

\subsection{R as a Low Resource Language}

\begin{figure}[h]
    \centering
    \includegraphics[width=0.9\textwidth]{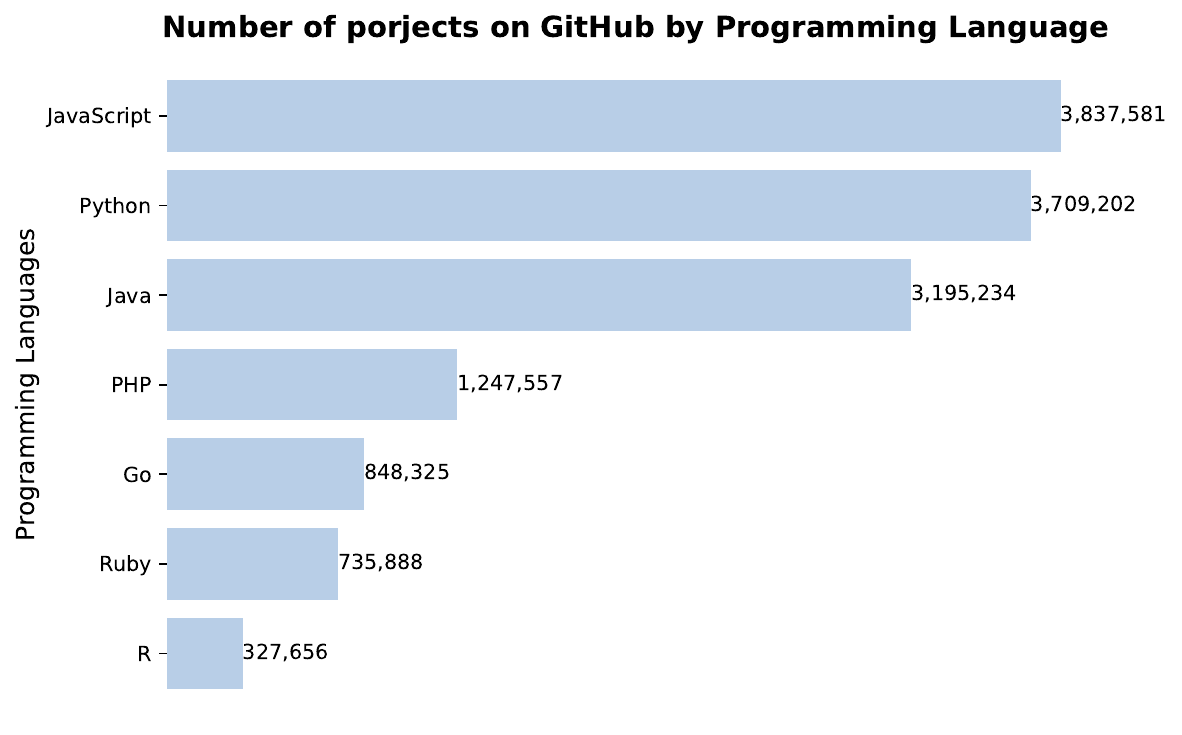}
    \caption{number of available projects on GitHub. The result is obtained by requesting GitHub through the REST API with the same query statements, except programming language.}
    \label{fig:d1}
\end{figure}

As one of the driving forces behind this early-stage research, we are keenly aware that R, as a low-resource language, faces significant challenges in both training and fine-tuning. For example, in the process of dataset collection, R may not yield as many high-quality training sets as mainstream languages like Python. Also, during the course of this research, we recognize that most current pre-trained models and large language models require vast amounts of text and code snippets for training. Moreover, the specificity and uniqueness of the R language make it difficult for R to appear on a large scale in corpora used for training. For example, on GitHub, a well-known hosting platform, we used the same search criteria to harvest datasets. The results, comparing R with other mainstream languages, are presented in Figure~\ref{fig:d1}. 
From Figure~\ref{fig:d1}, JavaScript, Python, and Java have approximately 10 times more GitHub repositories than R.
Moreover, we can see that the available R projects are half of the size compare to Ruby, which is considered a low-resource language in the CodeSearchNet~\cite{codesearchnet} dataset. 
The relatively low number of R packages on GitHub compared to other popular programming languages presents significant challenges for training effective language models and LLMs focused on R.

The smaller dataset also poses difficulties for transfer learning and fine-tuning approaches. We can see from Table~\ref{tab:sum_multi} and Table~\ref{tab:name_multi}, multi-lingual fine-tuning would not increase the performance, even though this approach has been showed to be effective in previous studies~\cite{Ahmed}. 

When adapting general-purpose language models to R-specific tasks, there's a risk of over-fitting due to the limited examples available. This could hinder the model's ability to generalize effectively and may lead to sub-optimal performance when dealing with novel R code or projects that integrate multiple programming languages. The reduced diversity in coding styles and patterns within the smaller pool of R packages might also impact the model's understanding of the broader R ecosystem, including package dependencies and usage patterns.

\begin{minipage}{.45\textwidth}
\begin{lstlisting}[caption=POV-base, label={lst:POV-base}, escapeinside={(*}{*)}]
function(x, filter.dups = FALSE) {
  checkmate::assertMatrix(x, mode = "numeric", min.cols = 1L, min.rows = 1L)
  
  df = as.data.frame(t(x))
  
  colnames(df) = paste0("y", seq_len(ncol(df)))
  
  if (filter.dups){
    df = filterDuplicated(df)
  }
  
  return(df)
}
\end{lstlisting}
\end{minipage}\hfill
\begin{minipage}{.45\textwidth}
\begin{lstlisting}[caption=POV-tidy, label={lst:POV-tidy}, escapeinside={(*}{*)}]
function(x, filter.dups = FALSE) {
  checkmate::assertMatrix(x, mode = "numeric", min.cols = 1L, min.rows = 1L)

  df <- as_tibble(t(x))

  df <- df %>% rename_with(~ paste0("y", seq_len(ncol(df))))

  if (filter.dups) {
    df <- df %>% distinct()
  }

  return(df)
}
\end{lstlisting}
\end{minipage}

\begin{table}[h!]
\centering
\begin{tabular}{|c|p{10cm}|}
\hline
\textbf{Model} & \textbf{Content} \\ \hline
CodeBERT & Base: Convert a matrix to a data frame \newline Tidy: matrix to data.frame \\ \hline
GraphCodeBERT & Base: Convert a matrix to a data frame \newline Tidy: matrix to data.frame \\ \hline
UniXCoder & Base: lifecycle - experimental options : alt = Experimental \newline Tidy: rename matrix to df \\ \hline
CodeT5 & Base: Converts a numeric matrix to a dataframe \newline Tidy: rename matrix to df \\ \hline
CodeLlama2 & Base: This function takes a numeric matrix as input and returns a data frame \newline Tidy: This function is used to convert a matrix to a tibble \\ \hline
StarCoder2 & Base: transform the data into a data frame with one row per sample and one column per feature \newline Tidy: The matrix is transposed, so that the rows are the observations and the columns are the variables. The variables are renamed to `y1`, `y2` \\ \hline
\end{tabular}
\caption{Summarizations generated by the models}
\label{tab:model_summarizations}
\end{table}

\subsection{Impact of Tidy vs. Base}

The dichotomy between \textit{Tidy-verse} and \textit{Base} R in the R programming presents an interesting challenge for language models and LLMs trained on R code. This division could significantly impact model performance and introduce unique considerations for models working with R.
\textit{Tidy-verse}, a collection of popular R packages designed for data science, introduces a distinct syntax and philosophy that differs from \textit{Base} R. This split in coding styles and approaches within the R community can affect language models in several ways. Models trained on a dataset that includes code from both R styles, may struggle to maintain consistency in their outputs, potentially mixing idioms. This is more challenging, specifically interfering with code generation by mixing both paradigms. This could lead to confusion for users expecting pure \textit{Tidy-verse} or \textit{Base} R solutions.
For example, the summaries generated by CodeLlama2 for both \textit{Tidy} and \textit{Base} show a good understanding of the code's functionality, but the \textit{Tidy} version shows less information about the function compared to the \textit{Base} version.

Listings~\ref{lst:POV-base} and ~\ref{lst:POV-tidy} are the \textit{Base} and \textit{Tidy} versions of the same function. The function processes a numeric matrix by first validating the input to ensure it meets the minimum criteria. It transposes the matrix, converts it into a data frame, and renames the columns systematically as $y_1, y_2, \dots, y_n$. Optionally, it filters duplicated rows based on user specification, ensuring that the data is structured and ready for further analysis or modeling. 
The \textit{Base} R implementation, as shown in the Listing~\ref{lst:POV-base} example, follows a more traditional, imperative style, relying on \textit{Base} R functions like \texttt{as.data.frame()}, \texttt{t()}, and \texttt{filterDuplicated()}. In contrast, the \textit{Tidy-verse} version, listing~\ref{lst:POV-tidy}, embraces a more expressive, declarative approach, utilizing \texttt{as\_tibble()}, \texttt{rename\_with()}, and \texttt{distinct()} from the \textit{Tidy-verse} packages.

We provided a list of summarizations from each model in Table~\ref{tab:model_summarizations}, where they are asked to summarize a \textit{Base} and \textit{Tidy} version of the same function shown in Listings~\ref{lst:POV-base} and ~\ref{lst:POV-tidy}.
The pre-trained language models are trained for code summarization, and the LLMs are doing this task under a zero-shot setup.
The performance of the analyzed models reflects their varying levels of exposure to and familiarity with these R programming idioms. CodeBERT, GraphCodeBERT and CodeT5 demonstrate a stronger grasp of the underlying concepts when compared to UniXCoder, as evidenced by their precise summaries for both the \textit{Base} R and \textit{Tidy-verse} versions. However, the summaries provided by those models are very short in general, missing key information about the function's underlying task. For example, only CodeT5 in its summary for the \textit{Base} version mentioned that the matrix need to be numeric. Moreover, the word ``rename'' is not accurate about the function's functionality.
In contrast, when we look at the summaries provided by the LLMs, one of the main difference is that LLM's summaries are longer, and contain more information. For instance, CodeLlama provided information about the input and output of the function. Moreover, StarCoder provides summaries about what happened during the execution. On the other hand, StarCoder did not mention that the matrix is transposed into a dataframe when summarizing \textit{Tidy} version, and missed the numeric part for \textit{Base} version.

Furthermore, all models' ability to capture project-level context and constraints, such as domain-specific requirements, implicit dependencies, and established coding conventions, appears to be an area for potential improvement. All of the models missed the part about \texttt{checkmate::assertMatrix}, where it is a function from a larger context. The summaries generated by the models, while accurate in conveying the general purpose of the functions, lack the nuanced understanding of how these code snippets would integrate within the larger context of a real-world R project.
This is an existing challenge for other programming languages, as shown in previous studies in the existance of unfamiliar libraries or private repositories \cite{zhoudocprompting,jimenez2023swe}. Previous work~\cite{YUN2024112149} also shows that including context information would increase the performance for the code summarization task.

Moreover, from the examples provided in listing~\ref{lst:POV-base} and listing~\ref{lst:POV-tidy}, we can see that the prevalence of \textit{Tidy-verse} in modern R programming might skew model performance. If the training data heavily favors \textit{Tidy-verse} code, the model may underperform when dealing with \textit{Base} R tasks, and vice versa. This imbalance could result in biased code suggestions or completions, potentially steering users towards one paradigm over the other unintentionally, especially when \textit{Tidy-verse} is heavily used in some domains while barely used in others. The model's ability to understand and generate idiomatic R code could vary significantly depending on whether it is working with \textit{Tidy-verse} or \textit{Base} R constructs.

Finally, building upon our empirical investigation, we uncover an intriguing relationship between data volume and model performance across R syntax paradigms. Despite \textit{Tidy} R syntax having approximately three times more training data than \textit{Base} R, our results indicate that this substantial data advantage does not translate into consistently superior model performance. 
Analysis across multiple models reveals that \textit{Base} R syntax often achieves comparable or marginally better results: UniXCoder and CodeBERT demonstrate slightly higher performance with \textit{Base} R (20.93\% and 19.55\% respectively) compared to \textit{Tidy} R (18.93\% and 17.92\%). 
Only GraphCodeBERT shows a minor advantage when processing \textit{Tidy} syntax (10.38\% vs. 9.56\% for \textit{Base}). 
Notably, the combined syntax approach generally underperforms compared to either individual syntax, suggesting increased complexity when models must handle both paradigms simultaneously. 
These findings challenge the assumption that larger training datasets inherently lead to better performance, indicating that the relationship between syntax paradigms and model effectiveness is more nuanced. Factors such as syntax complexity, consistency, or the model's architectural bias may play more significant roles than data quantity alone in determining performance outcomes, emphasizing that the quality of the data matters, we discussed earlier.

\subsection{Intra- and Cross-Project}

Analyzing the effects of intra- and cross-project contexts on a model's performance for the R language is complex and nuanced. This aspect of model training presents unique challenges and opportunities that can significantly impact the effectiveness of pre-trained models and LLMs working with R.

Cross-project context presents a greater challenge. R is used across various domains, from bioinformatics to finance, each with its own specialized libraries, data structures, and best practices. A model trained on a broad spectrum of R projects might produce generalized code that fails to account for domain-specific requirements or optimizations. For example, code generated for a bioinformatics project might not inherently understand the need for efficient large-scale data processing that is crucial in genomics research.
Another significant challenge lies in understanding and maintaining project-specific logic and business rules. Many R projects in professional settings implement complex analytical pipelines or decision-making processes that are unique to their use case. These often rely on domain knowledge that is not explicitly coded but is crucial for correct implementation.

\subsection{Discussion on Model Behavior}
All of our baseline models (CodeBERT, GraphCodeBERT, UniXCoder, and CodeT5) differ in how they learn and leverage code representations. First, each model's \emph{pretraining objective} shapes what it emphasizes: BERT-like objectives (as in CodeBERT and GraphCodeBERT) encourage strong token-level encoding, while T5-style objectives (as in CodeT5) promote sequence-to-sequence generation, naturally aligning with summarization tasks. Second, \emph{language-specific syntax} affects performance: more verbose or structurally consistent languages (e.g., Java) often suit models that rely on explicit tokens and clear code patterns, whereas dynamic or less standard languages (e.g., Ruby, or different R styles) can pose unique challenges. Third, \emph{internal attention patterns} vary by architecture: GraphCodeBERT’s additional graph representation may highlight data flows, whereas CodeT5’s text-to-text design focuses on coherent generation. Finally, \emph{fine-tuning data coverage} can cause certain models to overfit to common constructs or underperform on outlier patterns. Together, these factors may explain performance fluctuation across different programming languages. However, a comprehensive study separate from this work is required to delve into the exact causes and their effects.

\section{Implications and Future Directions} \label{sec:implications}

The findings of this study have several implications for R and scientific software developers and researchers.

First, the insights of our research contribute to the methodological improvement of utilizing Code-PLMs in R. 
We have demonstrated the performance of Code-PLMs when integrating these models into the R ecosystem. 
As discussed, one of the issues is the lack of training datasets for R. 
Researchers and developers working on language models for R could employ several strategies to address this issue. Augmenting the GitHub dataset with high-quality code from other reputable sources, such as CRAN or Bioconductor, could significantly expand the training data. 
Additionally, applying data augmentation techniques could artificially increase the dataset's size and diversity. 
Leveraging transfer learning from models trained on larger programming language datasets, combined with careful domain-specific fine-tuning, could help bridge the gap caused by the limited R-specific data. These approaches, while not eliminating the underlying data scarcity issue, can help improve model performance and ensure more robust and accurate language models for R programming.

Furthermore, researchers and practitioners can devise strategies to enhance model performance, specifically with the lack of huge datasets. 
This could also include optimizing the selection process and adaptation of Code-PLMs or LLMs for different tasks in R. 
As observed in RQ5, the LLMs' performance on code generation for R is falling behind Python. 
Specific techniques could be developed to generate acceptable code in R. The code generation and creating a benchmark in \textit{Tidy-verse} style is also another area of research.


Second, our study provides valuable guidance for R language users who seek to leverage the current Code-PLMs. The comparison and evaluation of different pre-trained models in the R environment shed light on their relative strengths and weaknesses. This information empowers practitioners to make informed decisions when selecting the most suitable model for their specific use cases, considering factors such as model accuracy and compatibility with R programming styles. 
The practitioner and researcher should be aware of the limitations and strengths of the models when intend to use them for R. 

As discussed earlier, the \textit{Base} and \textit{Tidy} styles have different syntax, and combining them in the training dataset could lower the performance. Careful curation of training data to ensure balanced representation of both \textit{Tidy-verse} and \textit{Base} R code is crucial, even separating them for specific tasks. Additionally, models might benefit from explicit tagging or context-awareness to distinguish between \textit{Tidy-verse} and \textit{Base} R code segments. This could enable more accurate and context-appropriate code generation and completion. Ultimately, the goal would be to develop models that can seamlessly work with both paradigms, understanding their strengths and appropriately suggesting or generating code that aligns with the user's preferred style or the specific requirements of the task at hand.

Another implication for researchers is exploring ways to develop datasets and techniques to improve the performance of models for project-specific or R specific domains. 
One potential solution is to develop models that can intake and process broader project contexts, e.g., by analyzing entire codebases or project documentation before making suggestions or generating code. This could help the model align its outputs more closely with project-specific conventions and requirements. Another approach might involve creating domain-specific models or fine-tuning general R models for particular fields of application. This could help in generating more relevant and optimized code for specific types of R projects, whether they are in finance, healthcare, or scientific research.

The integration of pre-trained models in R opens up opportunities for their application in other low-resource (i.e., the amount of available data is limited) programming languages and software engineering tasks. Researchers and practitioners working on these topics can draw upon our findings to explore and exploit the potential of pre-trained models within the R ecosystem, leading to improved solutions and advancements in their respective areas, for other code-intelligent tasks and other under-studied programming languages, including scientific software.


\section{Threats to Validity} \label{sec:threats}

\textbf{External Validity:} We only conduct experiments in R and for code summarization and method name prediction, and for code generation using LLMs only. Limiting the number of tasks is mainly due to reducing the bias of building high-quality data for other tasks. Though the approach can be utilized for other code-related tasks in R, we cannot expand the conclusions to other programming languages or tasks. However, one can benefit greatly from the strategy we introduced for separating the two R styles, namely separating the \textit{Base} R and \textit{Tidy-verse}.

An additional threat to validity arises from our exclusive reliance on GitHub repositories. While GitHub was selected to facilitate direct identification of Roxygen2 parser usage and to streamline quality-focused filtering, depending solely on GitHub may overlook specialized R usage scenarios, particularly in fields like bioinformatics or finance, where code is often shared on CRAN or Bioconductor. As a result, our dataset might not fully capture these domain-specific practices.

\textbf{Internal validity:} The internal validity in our work could be related to the dataset collection process and pre-processing steps. 
In our study, the quality of data is a significant concern. The code and associated documentation within R packages can vary widely in quality. 
Some packages may be meticulously maintained with high-quality code and comprehensive documentation, providing valuable insights for model training. However, others may be less well-maintained, with misleading, inaccurate, or sparse documentation. 
This inconsistency can lead to a model that is trained on a mixture of high and low-quality data, potentially reducing its effectiveness and accuracy. To mitigate this issue, it is crucial to implement robust data cleaning and preprocessing steps, possibly including manual review and curation of the dataset.

To achieve this, we consulted constantly with an R expert, who has more than ten years of research and development experience in R, ensuring the quality of the collected dataset and the soundness of the approach, specifically the decision on \textit{Base} or \textit{Tidy-verse} R repositories. 
We also manually evaluated the collected data at different steps to ensure our developed process leads to the desired data, e.g., the code and comments are tokenized correctly.

For the tokenization part, we used several tools, and each failed in one part. Finally, we found a reliable grammar rule for R that could be used with a Tree-sitter. 
The obtained dataset was investigated by the first author and another student, ensuring that the code and comments were separated correctly. 
These steps were taken in addition to following the advice of the R expert and previously published works to select the inclusion and exclusion criteria to scrape the related GitHub repositories. 

There is an uneven spread between \textit{Base} R and \textit{Tidy-verse} packages in the dataset. This could lead to a model that is biased towards the coding style and conventions of one over the other, limiting its effectiveness across the diverse R ecosystem. 
Additionally, as mentioned earlier, the syntax of the two styles is different, which can affect the performance of the models. 
Therefore, we added RQ4 to evaluate the effect of separating the data related to each programming style.
To avoid any issues with the fair comparison of R styles, we conducted manual evaluations and used the same number of records for R\textsubscript{Base} and R\textsubscript{Tidy}. 

Another internal threat is related to selecting and training the models. We followed the recommendation of a recent paper \cite{comparison2023} to select the Code-PLMs for code summarization and method name prediction. 
The author who trained the models has extensive experiments with running deep learning models. Therefore, we anticipate minimal threat related to this point. 

\textbf{Construct validity:} We use the commonly used evaluation metrics for code summarization and method name prediction. 
For code generation we also used a benchmark dataset and the Pass@K metrics, which are well studied in the related literature. 
These metrics are well known in NLP and SE research and therefore, we do not anticipate such validity threats. 

Another threat could be related to the multilingual training of the models, investigated in RQ2. Although the difference in the distribution of the training data could affect the results, previous research has shown the benefits of multi-lingual training for code summarization and method name prediction~\cite{Ahmed}. We also compare the results of multilingual and monolingual fine-tuning, therefore, anticipating low threats related to this setting.


\section{Conclusion and Future Studies} \label{sec:conclusion}

In this study, we explored the capabilities of the Code-PLMs for the two tasks of code summarization and method name prediction for R. This is the first attempt towards code intelligence in R for the mentioned tasks. We also examined LLMs for the code generation task for the R programming language.
Both LLMs and Code-PLMs consistently experience varying degrees of performance degradation when processing R code compared to other programming languages. 
The challenge stems from the different syntax of the R styles as well as its project-specific and package-based nature. Interestingly, increasing the dataset size or adding other programming languages for multi-lingual fine-tuning does not necessarily help in improving the performance.

The significance of this study extends far beyond its immediate findings about R language processing. As software development increasingly relies on AI-assisted programming tools, our research addresses a critical gap in understanding how current state-of-the-art language models handle diverse programming languages and paradigms. The limitations we uncovered in processing R code have broader implications for the development of code understanding and generation models that can handle different programming languages. As organizations increasingly adopt AI tools for software development, our findings underscore the importance of carefully evaluating these tools' capabilities across different programming languages and contexts. The limitations we have identified in R processing could potentially extend to other specialized or less common programming languages, making our study an important reference point for future research in this area.

\section*{Acknowledgment}

This research is supported by a grant from Natural Sciences and Engineering Research Council of Canada RGPIN-2019-05175.\\
We also sincerely appreciate the help of Dr. Melina Vidoni, who has more than ten years of research and development in R, for her guidance and feedback to curate the R dataset.

\bibliographystyle{ACM-Reference-Format}
{\footnotesize\bibliography{references}}


\begin{thebibliography}{93}


\ifx \showCODEN    \undefined \def \showCODEN     #1{\unskip}     \fi
\ifx \showDOI      \undefined \def \showDOI       #1{#1}\fi
\ifx \showISBNx    \undefined \def \showISBNx     #1{\unskip}     \fi
\ifx \showISBNxiii \undefined \def \showISBNxiii  #1{\unskip}     \fi
\ifx \showISSN     \undefined \def \showISSN      #1{\unskip}     \fi
\ifx \showLCCN     \undefined \def \showLCCN      #1{\unskip}     \fi
\ifx \shownote     \undefined \def \shownote      #1{#1}          \fi
\ifx \showarticletitle \undefined \def \showarticletitle #1{#1}   \fi
\ifx \showURL      \undefined \def \showURL       {\relax}        \fi
\providecommand\bibfield[2]{#2}
\providecommand\bibinfo[2]{#2}
\providecommand\natexlab[1]{#1}
\providecommand\showeprint[2][]{arXiv:#2}

\bibitem[Abdel-Ghani et~al\mbox{.}(2022)]%
        {AbdelGhani2022LooksOT}
\bibfield{author}{\bibinfo{person}{Amal Abdel-Ghani}, \bibinfo{person}{Kelly Bodwin}, \bibinfo{person}{Amelia~A. McNamara}, \bibinfo{person}{Allison~S. Theobold}, {and} \bibinfo{person}{Ian G.~Flores Siaca}.} \bibinfo{year}{2022}\natexlab{}.
\newblock \showarticletitle{“Looks Okay to Me”: A Study of Best Practice in Data Analysis Code Review}.
\newblock \bibinfo{journal}{\emph{Bridging the Gap: Empowering and Educating Today’s Learners in Statistics. Proceedings of the Eleventh International Conference on Teaching Statistics}} (\bibinfo{year}{2022}).
\newblock


\bibitem[Abdollahi et~al\mbox{.}(2022)]%
        {ABDOLLAHI2022101044}
\bibfield{author}{\bibinfo{person}{Masoud Abdollahi}, \bibinfo{person}{Babak Farjad}, \bibinfo{person}{Anil Gupta}, {and} \bibinfo{person}{Quazi~K. Hassan}.} \bibinfo{year}{2022}\natexlab{}.
\newblock \showarticletitle{CMIP6-D\&A: An R-based software with GUI for processing climate data available in network common data format}.
\newblock \bibinfo{journal}{\emph{SoftwareX}}  \bibinfo{volume}{18} (\bibinfo{year}{2022}), \bibinfo{pages}{101044}.
\newblock
\showISSN{2352-7110}
\urldef\tempurl%
\url{https://doi.org/10.1016/j.softx.2022.101044}
\showDOI{\tempurl}


\bibitem[Ahmed et~al\mbox{.}(2023a)]%
        {ahmed2023characterizing}
\bibfield{author}{\bibinfo{person}{Shibbir Ahmed}, \bibinfo{person}{Mohammad Wardat}, \bibinfo{person}{Hamid Bagheri}, \bibinfo{person}{Breno~Dantas Cruz}, {and} \bibinfo{person}{Hridesh Rajan}.} \bibinfo{year}{2023}\natexlab{a}.
\newblock \showarticletitle{Characterizing Bugs in Python and R Data Analytics Programs}.
\newblock \bibinfo{journal}{\emph{arXiv preprint arXiv:2306.08632}} (\bibinfo{year}{2023}).
\newblock


\bibitem[Ahmed and Devanbu(2022)]%
        {Ahmed}
\bibfield{author}{\bibinfo{person}{Toufique Ahmed} {and} \bibinfo{person}{Premkumar Devanbu}.} \bibinfo{year}{2022}\natexlab{}.
\newblock \showarticletitle{Multilingual Training for Software Engineering}. In \bibinfo{booktitle}{\emph{Proceedings of the 44th International Conference on Software Engineering}} (Pittsburgh, Pennsylvania) \emph{(\bibinfo{series}{ICSE '22})}. \bibinfo{publisher}{Association for Computing Machinery}, \bibinfo{address}{New York, NY, USA}, \bibinfo{pages}{1443–1455}.
\newblock
\showISBNx{9781450392211}
\urldef\tempurl%
\url{https://doi.org/10.1145/3510003.3510049}
\showDOI{\tempurl}


\bibitem[Ahmed and Devanbu(2023)]%
        {ahmed-fewshot}
\bibfield{author}{\bibinfo{person}{Toufique Ahmed} {and} \bibinfo{person}{Premkumar Devanbu}.} \bibinfo{year}{2023}\natexlab{}.
\newblock \showarticletitle{Few-Shot Training LLMs for Project-Specific Code-Summarization}. In \bibinfo{booktitle}{\emph{Proceedings of the 37th IEEE/ACM International Conference on Automated Software Engineering}} (Rochester, MI, USA) \emph{(\bibinfo{series}{ASE '22})}. \bibinfo{publisher}{Association for Computing Machinery}, \bibinfo{address}{New York, NY, USA}, Article \bibinfo{articleno}{177}, \bibinfo{numpages}{5}~pages.
\newblock
\showISBNx{9781450394758}
\urldef\tempurl%
\url{https://doi.org/10.1145/3551349.3559555}
\showDOI{\tempurl}


\bibitem[Ahmed et~al\mbox{.}(2023b)]%
        {ahmed2023towards}
\bibfield{author}{\bibinfo{person}{Toufique Ahmed}, \bibinfo{person}{Dian Yu}, \bibinfo{person}{Chengxuan Huang}, \bibinfo{person}{Cathy Wang}, \bibinfo{person}{Prem Devanbu}, {and} \bibinfo{person}{Kenji Sagae}.} \bibinfo{year}{2023}\natexlab{b}.
\newblock \showarticletitle{Towards Understanding What Code Language Models Learned}.
\newblock \bibinfo{journal}{\emph{arXiv preprint arXiv:2306.11943}} (\bibinfo{year}{2023}).
\newblock


\bibitem[Allamanis et~al\mbox{.}(2016)]%
        {Allamanis2016ACA}
\bibfield{author}{\bibinfo{person}{Miltiadis Allamanis}, \bibinfo{person}{Hao Peng}, {and} \bibinfo{person}{Charles Sutton}.} \bibinfo{year}{2016}\natexlab{}.
\newblock \showarticletitle{A Convolutional Attention Network for Extreme Summarization of Source Code}.
\newblock \bibinfo{journal}{\emph{ArXiv}}  \bibinfo{volume}{abs/1602.03001} (\bibinfo{year}{2016}).
\newblock


\bibitem[Alon et~al\mbox{.}(2019)]%
        {code2seq}
\bibfield{author}{\bibinfo{person}{Uri Alon}, \bibinfo{person}{Omer Levy}, {and} \bibinfo{person}{Eran Yahav}.} \bibinfo{year}{2019}\natexlab{}.
\newblock \showarticletitle{code2seq: Generating Sequences from Structured Representations of Code}. In \bibinfo{booktitle}{\emph{International Conference on Learning Representations}}.
\newblock
\urldef\tempurl%
\url{https://openreview.net/forum?id=H1gKYo09tX}
\showURL{%
\tempurl}


\bibitem[Alon et~al\mbox{.}(2018)]%
        {code2vec}
\bibfield{author}{\bibinfo{person}{Uri Alon}, \bibinfo{person}{Meital Zilberstein}, \bibinfo{person}{Omer Levy}, {and} \bibinfo{person}{Eran Yahav}.} \bibinfo{year}{2018}\natexlab{}.
\newblock \showarticletitle{code2vec: Learning Distributed Representations of Code}.
\newblock \bibinfo{journal}{\emph{CoRR}}  \bibinfo{volume}{abs/1803.09473} (\bibinfo{year}{2018}).
\newblock
\urldef\tempurl%
\url{http://arxiv.org/abs/1803.09473}
\showURL{%
\tempurl}


\bibitem[Arvanitou et~al\mbox{.}(2021)]%
        {ScientificSoftSMS}
\bibfield{author}{\bibinfo{person}{Elvira-Maria Arvanitou}, \bibinfo{person}{Apostolos Ampatzoglou}, \bibinfo{person}{Alexander Chatzigeorgiou}, {and} \bibinfo{person}{Jeffrey~C. Carver}.} \bibinfo{year}{2021}\natexlab{}.
\newblock \showarticletitle{Software engineering practices for scientific software development: A systematic mapping study}.
\newblock \bibinfo{journal}{\emph{Journal of Systems and Software}}  \bibinfo{volume}{172} (\bibinfo{year}{2021}), \bibinfo{pages}{110848}.
\newblock
\showISSN{0164-1212}
\urldef\tempurl%
\url{https://doi.org/10.1016/j.jss.2020.110848}
\showDOI{\tempurl}


\bibitem[Athiwaratkun et~al\mbox{.}(2023)]%
        {Athiwaratkun2023}
\bibfield{author}{\bibinfo{person}{Ben Athiwaratkun}, \bibinfo{person}{Sanjay~Krishna Gouda}, \bibinfo{person}{Zijian Wang}, \bibinfo{person}{Xiaopeng LI}, \bibinfo{person}{Yuchen Tian}, \bibinfo{person}{Ming Tan}, \bibinfo{person}{Wasi Ahmad}, \bibinfo{person}{Shiqi Wang}, \bibinfo{person}{Qing Sun}, \bibinfo{person}{Mingyue Shang}, \bibinfo{person}{Sujan Gonugondla}, \bibinfo{person}{Hantian Ding}, \bibinfo{person}{Varun Kumar}, \bibinfo{person}{Nathan Fulton}, \bibinfo{person}{Arash Farahani}, \bibinfo{person}{Siddhartha Jain}, \bibinfo{person}{Robert Giaquinto}, \bibinfo{person}{Haifeng Qian}, \bibinfo{person}{Murali~Krishna Ramanathan}, \bibinfo{person}{Ramesh Nallapati}, \bibinfo{person}{Baishakhi Ray}, \bibinfo{person}{Parminder Bhatia}, \bibinfo{person}{Sudipta Sengupta}, \bibinfo{person}{Dan Roth}, {and} \bibinfo{person}{Bing Xiang}.} \bibinfo{year}{2023}\natexlab{}.
\newblock \showarticletitle{Multi-lingual evaluation of code generation models}.
\newblock  (\bibinfo{year}{2023}).
\newblock
\urldef\tempurl%
\url{https://www.amazon.science/publications/multi-lingual-evaluation-of-code-generation-models}
\showURL{%
\tempurl}


\bibitem[Babu(2022)]%
        {babu2022qa4r}
\bibfield{author}{\bibinfo{person}{Ganesh Babu}.} \bibinfo{year}{2022}\natexlab{}.
\newblock \showarticletitle{Qa4r: a Question Answering System for R Packages}.
\newblock  (\bibinfo{year}{2022}).
\newblock


\bibitem[Brown et~al\mbox{.}(2020)]%
        {GPT3}
\bibfield{author}{\bibinfo{person}{Tom~B. Brown}, \bibinfo{person}{Benjamin Mann}, \bibinfo{person}{Nick Ryder}, \bibinfo{person}{Melanie Subbiah}, \bibinfo{person}{Jared Kaplan}, \bibinfo{person}{Prafulla Dhariwal}, \bibinfo{person}{Arvind Neelakantan}, \bibinfo{person}{Pranav Shyam}, \bibinfo{person}{Girish Sastry}, \bibinfo{person}{Amanda Askell}, \bibinfo{person}{Sandhini Agarwal}, \bibinfo{person}{Ariel Herbert-Voss}, \bibinfo{person}{Gretchen Krueger}, \bibinfo{person}{Tom Henighan}, \bibinfo{person}{Rewon Child}, \bibinfo{person}{Aditya Ramesh}, \bibinfo{person}{Daniel~M. Ziegler}, \bibinfo{person}{Jeffrey Wu}, \bibinfo{person}{Clemens Winter}, \bibinfo{person}{Christopher Hesse}, \bibinfo{person}{Mark Chen}, \bibinfo{person}{Eric Sigler}, \bibinfo{person}{Mateusz Litwin}, \bibinfo{person}{Scott Gray}, \bibinfo{person}{Benjamin Chess}, \bibinfo{person}{Jack Clark}, \bibinfo{person}{Christopher Berner}, \bibinfo{person}{Sam McCandlish}, \bibinfo{person}{Alec Radford}, \bibinfo{person}{Ilya Sutskever},
  {and} \bibinfo{person}{Dario Amodei}.} \bibinfo{year}{2020}\natexlab{}.
\newblock \showarticletitle{Language models are few-shot learners}. In \bibinfo{booktitle}{\emph{Proceedings of the 34th International Conference on Neural Information Processing Systems}} (Vancouver, BC, Canada) \emph{(\bibinfo{series}{NIPS'20})}. \bibinfo{publisher}{Curran Associates Inc.}, \bibinfo{address}{Red Hook, NY, USA}, Article \bibinfo{articleno}{159}, \bibinfo{numpages}{25}~pages.
\newblock
\showISBNx{9781713829546}


\bibitem[Cassano et~al\mbox{.}(2024)]%
        {cassano2024knowledge}
\bibfield{author}{\bibinfo{person}{Federico Cassano}, \bibinfo{person}{John Gouwar}, \bibinfo{person}{Francesca Lucchetti}, \bibinfo{person}{Claire Schlesinger}, \bibinfo{person}{Carolyn~Jane Anderson}, \bibinfo{person}{{Feldman, Molly Q}}, \bibinfo{person}{Michael Greenberg}, \bibinfo{person}{{Jangda, Abhinav}}, {and} \bibinfo{person}{{Guha, Arjun}}.} \bibinfo{year}{2024}\natexlab{}.
\newblock \showarticletitle{Knowledge {{Transfer}} from {{High-Resource}} to {{Low-Resource Programming Languages}} for {{Code LLMs}}}.
\newblock \bibinfo{journal}{\emph{Proceedings of the ACM on Programming Languages (PACMPL)}} \bibinfo{volume}{8}, \bibinfo{number}{OOPSLA} (\bibinfo{year}{2024}).
\newblock


\bibitem[Cassano et~al\mbox{.}(2022)]%
        {cassano2022multiple}
\bibfield{author}{\bibinfo{person}{Federico Cassano}, \bibinfo{person}{John Gouwar}, \bibinfo{person}{Daniel Nguyen}, \bibinfo{person}{Sydney Nguyen}, \bibinfo{person}{Luna Phipps-Costin}, \bibinfo{person}{Donald Pinckney}, \bibinfo{person}{Ming-Ho Yee}, \bibinfo{person}{Yangtian Zi}, \bibinfo{person}{Carolyn~Jane Anderson}, \bibinfo{person}{Molly~Q Feldman}, \bibinfo{person}{Arjun Guha}, \bibinfo{person}{Michael Greenberg}, {and} \bibinfo{person}{Abhinav Jangda}.} \bibinfo{year}{2022}\natexlab{}.
\newblock \bibinfo{title}{MultiPL-E: A Scalable and Extensible Approach to Benchmarking Neural Code Generation}.
\newblock
\newblock
\showeprint[arxiv]{2208.08227}~[cs.LG]


\bibitem[Cassano et~al\mbox{.}(2023)]%
        {MultiPL-E}
\bibfield{author}{\bibinfo{person}{Federico Cassano}, \bibinfo{person}{John Gouwar}, \bibinfo{person}{Daniel Nguyen}, \bibinfo{person}{Sydney Nguyen}, \bibinfo{person}{Luna Phipps-Costin}, \bibinfo{person}{Donald Pinckney}, \bibinfo{person}{Ming-Ho Yee}, \bibinfo{person}{Yangtian Zi}, \bibinfo{person}{Carolyn~Jane Anderson}, \bibinfo{person}{Molly~Q Feldman}, \bibinfo{person}{Arjun Guha}, \bibinfo{person}{Michael Greenberg}, {and} \bibinfo{person}{Abhinav Jangda}.} \bibinfo{year}{2023}\natexlab{}.
\newblock \showarticletitle{MultiPL-E: A Scalable and Polyglot Approach to Benchmarking Neural Code Generation}.
\newblock \bibinfo{journal}{\emph{IEEE Transactions on Software Engineering}} \bibinfo{volume}{49}, \bibinfo{number}{7} (\bibinfo{year}{2023}), \bibinfo{pages}{3675--3691}.
\newblock
\urldef\tempurl%
\url{https://doi.org/10.1109/TSE.2023.3267446}
\showDOI{\tempurl}


\bibitem[Chandramouli et~al\mbox{.}(2022)]%
        {Chandramouli2022}
\bibfield{author}{\bibinfo{person}{Pranav Chandramouli}, \bibinfo{person}{Zadia Codabux}, {and} \bibinfo{person}{Melina Vidoni}.} \bibinfo{year}{2022}\natexlab{}.
\newblock \showarticletitle{{analyzeR: A SonarQube plugin for analyzing object-oriented R Packages}}.
\newblock \bibinfo{journal}{\emph{SoftwareX}}  \bibinfo{volume}{19} (\bibinfo{year}{2022}), \bibinfo{pages}{101113}.
\newblock
\showISSN{2352-7110}
\urldef\tempurl%
\url{https://doi.org/10.1016/j.softx.2022.101113}
\showDOI{\tempurl}


\bibitem[Chen et~al\mbox{.}(2021)]%
        {humanevaluating}
\bibfield{author}{\bibinfo{person}{Mark Chen}, \bibinfo{person}{Jerry Tworek}, \bibinfo{person}{Heewoo Jun}, \bibinfo{person}{Qiming Yuan}, \bibinfo{person}{Henrique Ponde de~Oliveira Pinto}, \bibinfo{person}{Jared Kaplan}, \bibinfo{person}{Harri Edwards}, \bibinfo{person}{Yuri Burda}, \bibinfo{person}{Nicholas Joseph}, \bibinfo{person}{Greg Brockman}, {et~al\mbox{.}}} \bibinfo{year}{2021}\natexlab{}.
\newblock \showarticletitle{Evaluating large language models trained on code}.
\newblock \bibinfo{journal}{\emph{arXiv preprint arXiv:2107.03374}} (\bibinfo{year}{2021}).
\newblock


\bibitem[Chowdhury(2023)]%
        {chowdhury2023empirical}
\bibfield{author}{\bibinfo{person}{Hemayet~Ahmed Chowdhury}.} \bibinfo{year}{2023}\natexlab{}.
\newblock \emph{\bibinfo{title}{An Empirical Study of API Breaking Changes in Bioconductor}}.
\newblock \bibinfo{thesistype}{Ph.\,D. Dissertation}. \bibinfo{school}{Virginia Tech}.
\newblock


\bibitem[Clark et~al\mbox{.}(2020)]%
        {electra}
\bibfield{author}{\bibinfo{person}{Kevin Clark}, \bibinfo{person}{Minh-Thang Luong}, \bibinfo{person}{Quoc~V. Le}, {and} \bibinfo{person}{Christopher~D. Manning}.} \bibinfo{year}{2020}\natexlab{}.
\newblock \showarticletitle{ELECTRA: Pre-training Text Encoders as Discriminators Rather Than Generators}. In \bibinfo{booktitle}{\emph{International Conference on Learning Representations}}.
\newblock
\urldef\tempurl%
\url{https://openreview.net/forum?id=r1xMH1BtvB}
\showURL{%
\tempurl}


\bibitem[de~F.~Farias et~al\mbox{.}(2016)]%
        {i/e}
\bibfield{author}{\bibinfo{person}{M\'{a}rio~Andr\'{e} de F.~Farias}, \bibinfo{person}{Renato Novais}, \bibinfo{person}{Methanias~Cola\c{c}o J\'{u}nior}, \bibinfo{person}{Lu\'{\i}s~Paulo da Silva~Carvalho}, \bibinfo{person}{Manoel Mendon\c{c}a}, {and} \bibinfo{person}{Rodrigo~Oliveira Sp\'{\i}nola}.} \bibinfo{year}{2016}\natexlab{}.
\newblock \showarticletitle{A systematic mapping study on mining software repositories}. In \bibinfo{booktitle}{\emph{Proceedings of the 31st Annual ACM Symposium on Applied Computing}} (Pisa, Italy) \emph{(\bibinfo{series}{SAC '16})}. \bibinfo{publisher}{Association for Computing Machinery}, \bibinfo{address}{New York, NY, USA}, \bibinfo{pages}{1472–1479}.
\newblock
\showISBNx{9781450337397}
\urldef\tempurl%
\url{https://doi.org/10.1145/2851613.2851786}
\showDOI{\tempurl}


\bibitem[de~Jong et~al\mbox{.}(2022)]%
        {de2022acquisition}
\bibfield{author}{\bibinfo{person}{Roos~M de Jong}, \bibinfo{person}{Manon Alkema}, \bibinfo{person}{Tate Oulton}, \bibinfo{person}{Elin Dumont}, \bibinfo{person}{Karina Teelen}, \bibinfo{person}{Rie Nakajima}, \bibinfo{person}{Rafael~Ramiro de Assis}, \bibinfo{person}{Kathleen W~Dantzler Press}, \bibinfo{person}{Priscilla Ngotho}, \bibinfo{person}{Kevin~KA Tetteh}, {et~al\mbox{.}}} \bibinfo{year}{2022}\natexlab{}.
\newblock \showarticletitle{The acquisition of humoral immune responses targeting Plasmodium falciparum sexual stages in controlled human malaria infections}.
\newblock \bibinfo{journal}{\emph{Frontiers in immunology}}  \bibinfo{volume}{13} (\bibinfo{year}{2022}), \bibinfo{pages}{930956}.
\newblock


\bibitem[de~Santana et~al\mbox{.}(2022)]%
        {de2022bug}
\bibfield{author}{\bibinfo{person}{Taijara~Loiola de Santana}, \bibinfo{person}{Paulo Anselmo da Mota~Silveira Neto}, \bibinfo{person}{Eduardo~Santana de Almeida}, {and} \bibinfo{person}{Iftekhar Ahmed}.} \bibinfo{year}{2022}\natexlab{}.
\newblock \showarticletitle{Bug Analysis in Jupyter Notebook Projects: An Empirical Study}.
\newblock \bibinfo{journal}{\emph{arXiv preprint arXiv:2210.06893}} (\bibinfo{year}{2022}).
\newblock


\bibitem[Decan et~al\mbox{.}(2015)]%
        {EMSE}
\bibfield{author}{\bibinfo{person}{Alexandre Decan}, \bibinfo{person}{Tom Mens}, \bibinfo{person}{Maelick Claes}, {and} \bibinfo{person}{Philippe Grosjean}.} \bibinfo{year}{2015}\natexlab{}.
\newblock \showarticletitle{On the Development and Distribution of R Packages: An Empirical Analysis of the R Ecosystem}. In \bibinfo{booktitle}{\emph{Proceedings of the 2015 European Conference on Software Architecture Workshops}} (Dubrovnik, Cavtat, Croatia) \emph{(\bibinfo{series}{ECSAW '15})}. \bibinfo{publisher}{Association for Computing Machinery}, \bibinfo{address}{New York, NY, USA}, Article \bibinfo{articleno}{41}, \bibinfo{numpages}{6}~pages.
\newblock
\showISBNx{9781450333931}
\urldef\tempurl%
\url{https://doi.org/10.1145/2797433.2797476}
\showDOI{\tempurl}


\bibitem[Decan et~al\mbox{.}(2016)]%
        {packageDependency}
\bibfield{author}{\bibinfo{person}{Alexandre Decan}, \bibinfo{person}{Tom Mens}, \bibinfo{person}{Maëlick Claes}, {and} \bibinfo{person}{Philippe Grosjean}.} \bibinfo{year}{2016}\natexlab{}.
\newblock \showarticletitle{When GitHub Meets CRAN: An Analysis of Inter-Repository Package Dependency Problems}. In \bibinfo{booktitle}{\emph{2016 IEEE 23rd International Conference on Software Analysis, Evolution, and Reengineering (SANER)}}, Vol.~\bibinfo{volume}{1}. \bibinfo{pages}{493--504}.
\newblock
\urldef\tempurl%
\url{https://doi.org/10.1109/SANER.2016.12}
\showDOI{\tempurl}


\bibitem[Devlin et~al\mbox{.}(2019)]%
        {bert}
\bibfield{author}{\bibinfo{person}{Jacob Devlin}, \bibinfo{person}{Ming-Wei Chang}, \bibinfo{person}{Kenton Lee}, {and} \bibinfo{person}{Kristina Toutanova}.} \bibinfo{year}{2019}\natexlab{}.
\newblock \bibinfo{title}{BERT: Pre-training of Deep Bidirectional Transformers for Language Understanding}.
\newblock
\newblock
\showeprint[arxiv]{1810.04805}~[cs.CL]


\bibitem[Dong et~al\mbox{.}(2021)]%
        {splitting}
\bibfield{author}{\bibinfo{person}{Helen Dong}, \bibinfo{person}{Shurui Zhou}, \bibinfo{person}{Jin~L.C. Guo}, {and} \bibinfo{person}{Christian Kästner}.} \bibinfo{year}{2021}\natexlab{}.
\newblock \showarticletitle{Splitting, Renaming, Removing: A Study of Common Cleaning Activities in Jupyter Notebooks}. In \bibinfo{booktitle}{\emph{2021 36th IEEE/ACM International Conference on Automated Software Engineering Workshops (ASEW)}}. \bibinfo{pages}{114--119}.
\newblock
\urldef\tempurl%
\url{https://doi.org/10.1109/ASEW52652.2021.00032}
\showDOI{\tempurl}


\bibitem[Feng et~al\mbox{.}(2020)]%
        {CodeBERT}
\bibfield{author}{\bibinfo{person}{Zhangyin Feng}, \bibinfo{person}{Daya Guo}, \bibinfo{person}{Duyu Tang}, \bibinfo{person}{Nan Duan}, \bibinfo{person}{Xiaocheng Feng}, \bibinfo{person}{Ming Gong~(YIMING)}, \bibinfo{person}{Linjun Shou}, \bibinfo{person}{Bing Qin}, \bibinfo{person}{Ting Liu}, \bibinfo{person}{Daxin Jiang}, {and} \bibinfo{person}{Ming Zhou}.} \bibinfo{year}{2020}\natexlab{}.
\newblock \showarticletitle{CodeBERT: A Pre-Trained Model for Programming and Natural Languages}. In \bibinfo{booktitle}{\emph{Findings of EMNLP 2020}}.
\newblock
\urldef\tempurl%
\url{https://www.microsoft.com/en-us/research/publication/codebert-a-pre-trained-model-for-programming-and-natural-languages/}
\showURL{%
\tempurl}


\bibitem[Fox and Leanage(2016)]%
        {RImpotent}
\bibfield{author}{\bibinfo{person}{John Fox} {and} \bibinfo{person}{Allison Leanage}.} \bibinfo{year}{2016}\natexlab{}.
\newblock \showarticletitle{R and the Journal of Statistical Software}.
\newblock \bibinfo{journal}{\emph{Journal of Statistical Software}}  \bibinfo{volume}{73} (\bibinfo{year}{2016}), \bibinfo{pages}{1--13}.
\newblock


\bibitem[Giorgi et~al\mbox{.}(2022)]%
        {lifeR}
\bibfield{author}{\bibinfo{person}{Federico~M. Giorgi}, \bibinfo{person}{Carmine Ceraolo}, {and} \bibinfo{person}{Daniele Mercatelli}.} \bibinfo{year}{2022}\natexlab{}.
\newblock \showarticletitle{The R Language: An Engine for Bioinformatics and Data Science}.
\newblock \bibinfo{journal}{\emph{Life}} \bibinfo{volume}{12}, \bibinfo{number}{5} (\bibinfo{year}{2022}).
\newblock
\showISSN{2075-1729}
\urldef\tempurl%
\url{https://doi.org/10.3390/life12050648}
\showDOI{\tempurl}


\bibitem[Goble(2014)]%
        {betterSoftBetterResearch}
\bibfield{author}{\bibinfo{person}{Carole Goble}.} \bibinfo{year}{2014}\natexlab{}.
\newblock \showarticletitle{Better Software, Better Research}.
\newblock \bibinfo{journal}{\emph{IEEE Internet Computing}} \bibinfo{volume}{18}, \bibinfo{number}{5} (\bibinfo{year}{2014}), \bibinfo{pages}{4--8}.
\newblock
\urldef\tempurl%
\url{https://doi.org/10.1109/MIC.2014.88}
\showDOI{\tempurl}


\bibitem[Guo et~al\mbox{.}(2022)]%
        {unixcoder}
\bibfield{author}{\bibinfo{person}{Daya Guo}, \bibinfo{person}{Shuai Lu}, \bibinfo{person}{Nan Duan}, \bibinfo{person}{Yanlin Wang}, \bibinfo{person}{Ming Zhou}, {and} \bibinfo{person}{Jian Yin}.} \bibinfo{year}{2022}\natexlab{}.
\newblock \showarticletitle{UniXcoder: Unified Cross-Modal Pre-training for Code Representation}. \bibinfo{pages}{7212--7225}.
\newblock
\urldef\tempurl%
\url{https://doi.org/10.18653/v1/2022.acl-long.499}
\showDOI{\tempurl}


\bibitem[Guo et~al\mbox{.}(2020)]%
        {GraphCodeBERT}
\bibfield{author}{\bibinfo{person}{Daya Guo}, \bibinfo{person}{Shuo Ren}, \bibinfo{person}{Shuai Lu}, \bibinfo{person}{Zhangyin Feng}, \bibinfo{person}{Duyu Tang}, \bibinfo{person}{Shujie Liu}, \bibinfo{person}{Long Zhou}, \bibinfo{person}{Nan Duan}, \bibinfo{person}{Alexey Svyatkovskiy}, \bibinfo{person}{Shengyu Fu}, \bibinfo{person}{Michele Tufano}, \bibinfo{person}{Shao~Kun Deng}, \bibinfo{person}{Colin Clement}, \bibinfo{person}{Dawn Drain}, \bibinfo{person}{Neel Sundaresan}, \bibinfo{person}{Jian Yin}, \bibinfo{person}{Daxin Jiang}, {and} \bibinfo{person}{Ming Zhou}.} \bibinfo{year}{2020}\natexlab{}.
\newblock \bibinfo{title}{GraphCodeBERT: Pre-training Code Representations with Data Flow}.
\newblock
\newblock
\urldef\tempurl%
\url{https://doi.org/10.48550/ARXIV.2009.08366}
\showDOI{\tempurl}


\bibitem[Han et~al\mbox{.}(2021)]%
        {PTM}
\bibfield{author}{\bibinfo{person}{Xu Han}, \bibinfo{person}{Zhengyan Zhang}, \bibinfo{person}{Ning Ding}, \bibinfo{person}{Yuxian Gu}, \bibinfo{person}{Xiao Liu}, \bibinfo{person}{Yuqi Huo}, \bibinfo{person}{Jiezhong Qiu}, \bibinfo{person}{Yuan Yao}, \bibinfo{person}{Ao Zhang}, \bibinfo{person}{Liang Zhang}, \bibinfo{person}{Wentao Han}, \bibinfo{person}{Minlie Huang}, \bibinfo{person}{Qin Jin}, \bibinfo{person}{Yanyan Lan}, \bibinfo{person}{Yang Liu}, \bibinfo{person}{Zhiyuan Liu}, \bibinfo{person}{Zhiwu Lu}, \bibinfo{person}{Xipeng Qiu}, \bibinfo{person}{Ruihua Song}, \bibinfo{person}{Jie Tang}, \bibinfo{person}{Ji-Rong Wen}, \bibinfo{person}{Jinhui Yuan}, \bibinfo{person}{Wayne~Xin Zhao}, {and} \bibinfo{person}{Jun Zhu}.} \bibinfo{year}{2021}\natexlab{}.
\newblock \showarticletitle{Pre-trained models: Past, present and future}.
\newblock \bibinfo{journal}{\emph{AI Open}}  \bibinfo{volume}{2} (\bibinfo{year}{2021}), \bibinfo{pages}{225--250}.
\newblock
\showISSN{2666-6510}
\urldef\tempurl%
\url{https://doi.org/10.1016/j.aiopen.2021.08.002}
\showDOI{\tempurl}


\bibitem[Huang and Yang(2022)]%
        {huang2022empirical}
\bibfield{author}{\bibinfo{person}{Tian-Yuan Huang} {and} \bibinfo{person}{Liying Yang}.} \bibinfo{year}{2022}\natexlab{}.
\newblock \showarticletitle{An empirical exploration of the vibrant R ecosystem}.
\newblock \bibinfo{journal}{\emph{arXiv preprint arXiv:2201.04982}} (\bibinfo{year}{2022}).
\newblock


\bibitem[Husain et~al\mbox{.}(2019)]%
        {codesearchnet}
\bibfield{author}{\bibinfo{person}{Hamel Husain}, \bibinfo{person}{Ho-Hsiang Wu}, \bibinfo{person}{Tiferet Gazit}, \bibinfo{person}{Miltiadis Allamanis}, {and} \bibinfo{person}{Marc Brockschmidt}.} \bibinfo{year}{2019}\natexlab{}.
\newblock \showarticletitle{Codesearchnet challenge: Evaluating the state of semantic code search}.
\newblock \bibinfo{journal}{\emph{arXiv preprint arXiv:1909.09436}} (\bibinfo{year}{2019}).
\newblock


\bibitem[Ihaka and Gentleman(1996)]%
        {Rlanguage}
\bibfield{author}{\bibinfo{person}{Ross Ihaka} {and} \bibinfo{person}{Robert Gentleman}.} \bibinfo{year}{1996}\natexlab{}.
\newblock \showarticletitle{R: A Language for Data Analysis and Graphics}.
\newblock \bibinfo{journal}{\emph{Journal of Computational and Graphical Statistics}} \bibinfo{volume}{5}, \bibinfo{number}{3} (\bibinfo{year}{1996}), \bibinfo{pages}{299--314}.
\newblock
\showISSN{10618600}
\urldef\tempurl%
\url{http://www.jstor.org/stable/1390807}
\showURL{%
\tempurl}


\bibitem[Ivie and Thain(2018)]%
        {reproducibility}
\bibfield{author}{\bibinfo{person}{Peter Ivie} {and} \bibinfo{person}{Douglas Thain}.} \bibinfo{year}{2018}\natexlab{}.
\newblock \showarticletitle{Reproducibility in Scientific Computing}.
\newblock \bibinfo{journal}{\emph{ACM Comput. Surv.}} \bibinfo{volume}{51}, \bibinfo{number}{3}, Article \bibinfo{articleno}{63} (\bibinfo{date}{jul} \bibinfo{year}{2018}), \bibinfo{numpages}{36}~pages.
\newblock
\showISSN{0360-0300}
\urldef\tempurl%
\url{https://doi.org/10.1145/3186266}
\showDOI{\tempurl}


\bibitem[Jiang et~al\mbox{.}(2022)]%
        {JupyterMaintenance}
\bibfield{author}{\bibinfo{person}{Yuan Jiang}, \bibinfo{person}{Christian Kästner}, {and} \bibinfo{person}{Shurui Zhou}.} \bibinfo{year}{2022}\natexlab{}.
\newblock \showarticletitle{Elevating Jupyter Notebook Maintenance Tooling by Identifying and Extracting Notebook Structures}. In \bibinfo{booktitle}{\emph{2022 IEEE International Conference on Software Maintenance and Evolution (ICSME)}}. \bibinfo{pages}{399--403}.
\newblock
\urldef\tempurl%
\url{https://doi.org/10.1109/ICSME55016.2022.00047}
\showDOI{\tempurl}


\bibitem[Jimenez et~al\mbox{.}(2023)]%
        {jimenez2023swe}
\bibfield{author}{\bibinfo{person}{Carlos~E Jimenez}, \bibinfo{person}{John Yang}, \bibinfo{person}{Alexander Wettig}, \bibinfo{person}{Shunyu Yao}, \bibinfo{person}{Kexin Pei}, \bibinfo{person}{Ofir Press}, {and} \bibinfo{person}{Karthik Narasimhan}.} \bibinfo{year}{2023}\natexlab{}.
\newblock \showarticletitle{Swe-bench: Can language models resolve real-world github issues?}
\newblock \bibinfo{journal}{\emph{arXiv preprint arXiv:2310.06770}} (\bibinfo{year}{2023}).
\newblock


\bibitem[Karmakar and Robbes(2021)]%
        {romain}
\bibfield{author}{\bibinfo{person}{Anjan Karmakar} {and} \bibinfo{person}{Romain Robbes}.} \bibinfo{year}{2021}\natexlab{}.
\newblock \showarticletitle{What do pre-trained code models know about code?}. In \bibinfo{booktitle}{\emph{2021 36th IEEE/ACM International Conference on Automated Software Engineering (ASE)}}. \bibinfo{pages}{1332--1336}.
\newblock
\urldef\tempurl%
\url{https://doi.org/10.1109/ASE51524.2021.9678927}
\showDOI{\tempurl}


\bibitem[Korkmaz et~al\mbox{.}(2018)]%
        {impactofR}
\bibfield{author}{\bibinfo{person}{Gizem Korkmaz}, \bibinfo{person}{Claire Kelling}, \bibinfo{person}{Carol Robbins}, {and} \bibinfo{person}{Sallie~A. Keller}.} \bibinfo{year}{2018}\natexlab{}.
\newblock \showarticletitle{Modeling the Impact of R Packages Using Dependency and Contributor Networks}. In \bibinfo{booktitle}{\emph{2018 IEEE/ACM International Conference on Advances in Social Networks Analysis and Mining (ASONAM)}}. \bibinfo{pages}{511--514}.
\newblock
\urldef\tempurl%
\url{https://doi.org/10.1109/ASONAM.2018.8508255}
\showDOI{\tempurl}


\bibitem[LeClair and McMillan(2019)]%
        {mcmillan2019}
\bibfield{author}{\bibinfo{person}{Alexander LeClair} {and} \bibinfo{person}{Collin McMillan}.} \bibinfo{year}{2019}\natexlab{}.
\newblock \showarticletitle{"ecommendations for Datasets for Source Code Summarization"}. In \bibinfo{booktitle}{\emph{Proceedings of the 2019 Conference of the North {A}merican Chapter of the Association for Computational Linguistics: Human Language Technologies, Volume 1 (Long and Short Papers)}}. \bibinfo{publisher}{Association for Computational Linguistics}, \bibinfo{address}{Minneapolis, Minnesota}, \bibinfo{pages}{3931--3937}.
\newblock
\urldef\tempurl%
\url{https://doi.org/10.18653/v1/N19-1394}
\showDOI{\tempurl}


\bibitem[Li et~al\mbox{.}(2023)]%
        {starcoder}
\bibfield{author}{\bibinfo{person}{Raymond Li}, \bibinfo{person}{Loubna~Ben Allal}, \bibinfo{person}{Yangtian Zi}, \bibinfo{person}{Niklas Muennighoff}, \bibinfo{person}{Denis Kocetkov}, \bibinfo{person}{Chenghao Mou}, \bibinfo{person}{Marc Marone}, \bibinfo{person}{Christopher Akiki}, \bibinfo{person}{Jia Li}, \bibinfo{person}{Jenny Chim}, \bibinfo{person}{Qian Liu}, \bibinfo{person}{Evgenii Zheltonozhskii}, \bibinfo{person}{Terry~Yue Zhuo}, \bibinfo{person}{Thomas Wang}, \bibinfo{person}{Olivier Dehaene}, \bibinfo{person}{Mishig Davaadorj}, \bibinfo{person}{Joel Lamy-Poirier}, \bibinfo{person}{João Monteiro}, \bibinfo{person}{Oleh Shliazhko}, \bibinfo{person}{Nicolas Gontier}, \bibinfo{person}{Nicholas Meade}, \bibinfo{person}{Armel Zebaze}, \bibinfo{person}{Ming-Ho Yee}, \bibinfo{person}{Logesh~Kumar Umapathi}, \bibinfo{person}{Jian Zhu}, \bibinfo{person}{Benjamin Lipkin}, \bibinfo{person}{Muhtasham Oblokulov}, \bibinfo{person}{Zhiruo Wang}, \bibinfo{person}{Rudra Murthy}, \bibinfo{person}{Jason
  Stillerman}, \bibinfo{person}{Siva~Sankalp Patel}, \bibinfo{person}{Dmitry Abulkhanov}, \bibinfo{person}{Marco Zocca}, \bibinfo{person}{Manan Dey}, \bibinfo{person}{Zhihan Zhang}, \bibinfo{person}{Nour Fahmy}, \bibinfo{person}{Urvashi Bhattacharyya}, \bibinfo{person}{Wenhao Yu}, \bibinfo{person}{Swayam Singh}, \bibinfo{person}{Sasha Luccioni}, \bibinfo{person}{Paulo Villegas}, \bibinfo{person}{Maxim Kunakov}, \bibinfo{person}{Fedor Zhdanov}, \bibinfo{person}{Manuel Romero}, \bibinfo{person}{Tony Lee}, \bibinfo{person}{Nadav Timor}, \bibinfo{person}{Jennifer Ding}, \bibinfo{person}{Claire Schlesinger}, \bibinfo{person}{Hailey Schoelkopf}, \bibinfo{person}{Jan Ebert}, \bibinfo{person}{Tri Dao}, \bibinfo{person}{Mayank Mishra}, \bibinfo{person}{Alex Gu}, \bibinfo{person}{Jennifer Robinson}, \bibinfo{person}{Carolyn~Jane Anderson}, \bibinfo{person}{Brendan Dolan-Gavitt}, \bibinfo{person}{Danish Contractor}, \bibinfo{person}{Siva Reddy}, \bibinfo{person}{Daniel Fried}, \bibinfo{person}{Dzmitry Bahdanau},
  \bibinfo{person}{Yacine Jernite}, \bibinfo{person}{Carlos~Muñoz Ferrandis}, \bibinfo{person}{Sean Hughes}, \bibinfo{person}{Thomas Wolf}, \bibinfo{person}{Arjun Guha}, \bibinfo{person}{Leandro von Werra}, {and} \bibinfo{person}{Harm de Vries}.} \bibinfo{year}{2023}\natexlab{}.
\newblock \bibinfo{title}{StarCoder: may the source be with you!}
\newblock
\newblock
\showeprint[arxiv]{2305.06161}~[cs.CL]
\urldef\tempurl%
\url{https://arxiv.org/abs/2305.06161}
\showURL{%
\tempurl}


\bibitem[Lin and Och(2004)]%
        {lin-och-2004-orange}
\bibfield{author}{\bibinfo{person}{Chin-Yew Lin} {and} \bibinfo{person}{Franz~Josef Och}.} \bibinfo{year}{2004}\natexlab{}.
\newblock \showarticletitle{{ORANGE}: a Method for Evaluating Automatic Evaluation Metrics for Machine Translation}. In \bibinfo{booktitle}{\emph{{COLING} 2004: Proceedings of the 20th International Conference on Computational Linguistics}}. \bibinfo{publisher}{COLING}, \bibinfo{address}{Geneva, Switzerland}, \bibinfo{pages}{501--507}.
\newblock
\urldef\tempurl%
\url{https://aclanthology.org/C04-1072}
\showURL{%
\tempurl}


\bibitem[Liu et~al\mbox{.}(2021)]%
        {liu2021haconvgnn}
\bibfield{author}{\bibinfo{person}{Xuye Liu}, \bibinfo{person}{Dakuo Wang}, \bibinfo{person}{April Wang}, \bibinfo{person}{Yufang Hou}, {and} \bibinfo{person}{Lingfei Wu}.} \bibinfo{year}{2021}\natexlab{}.
\newblock \showarticletitle{HAConvGNN: Hierarchical attention based convolutional graph neural network for code documentation generation in jupyter notebooks}.
\newblock \bibinfo{journal}{\emph{arXiv preprint arXiv:2104.01002}} (\bibinfo{year}{2021}).
\newblock


\bibitem[Liu et~al\mbox{.}(2019)]%
        {roberta}
\bibfield{author}{\bibinfo{person}{Yinhan Liu}, \bibinfo{person}{Myle Ott}, \bibinfo{person}{Naman Goyal}, \bibinfo{person}{Jingfei Du}, \bibinfo{person}{Mandar Joshi}, \bibinfo{person}{Danqi Chen}, \bibinfo{person}{Omer Levy}, \bibinfo{person}{Mike Lewis}, \bibinfo{person}{Luke Zettlemoyer}, {and} \bibinfo{person}{Veselin Stoyanov}.} \bibinfo{year}{2019}\natexlab{}.
\newblock \bibinfo{title}{RoBERTa: A Robustly Optimized BERT Pretraining Approach}.
\newblock
\newblock
\showeprint[arxiv]{1907.11692}~[cs.CL]


\bibitem[Medeiros~Mirra et~al\mbox{.}(2023)]%
        {MedeirosMirra2023}
\bibfield{author}{\bibinfo{person}{Renata Medeiros~Mirra}, \bibinfo{person}{Jim~O. Vafidis}, \bibinfo{person}{Jeremy~A. Smith}, {and} \bibinfo{person}{Robert~J. Thomas}.} \bibinfo{year}{2023}\natexlab{}.
\newblock \bibinfo{booktitle}{\emph{Teaching Data Analysis to Life Scientists Using ``R'' Statistical Software: Challenges, Opportunities, and Effective Methods}}.
\newblock \bibinfo{publisher}{Springer International Publishing}, \bibinfo{address}{Cham}, \bibinfo{pages}{167--187}.
\newblock
\showISBNx{978-3-031-26010-0}
\urldef\tempurl%
\url{https://doi.org/10.1007/978-3-031-26010-0_12}
\showDOI{\tempurl}


\bibitem[Miceli-Barone et~al\mbox{.}(2023)]%
        {miceli2023larger}
\bibfield{author}{\bibinfo{person}{Antonio~Valerio Miceli-Barone}, \bibinfo{person}{Fazl Barez}, \bibinfo{person}{Ioannis Konstas}, {and} \bibinfo{person}{Shay~B Cohen}.} \bibinfo{year}{2023}\natexlab{}.
\newblock \showarticletitle{The Larger They Are, the Harder They Fail: Language Models do not Recognize Identifier Swaps in Python}.
\newblock \bibinfo{journal}{\emph{ACL}} (\bibinfo{year}{2023}).
\newblock


\bibitem[Morandat et~al\mbox{.}(2012)]%
        {RDesign}
\bibfield{author}{\bibinfo{person}{Flor{\'e}al Morandat}, \bibinfo{person}{Brandon Hill}, \bibinfo{person}{Leo Osvald}, {and} \bibinfo{person}{Jan Vitek}.} \bibinfo{year}{2012}\natexlab{}.
\newblock \showarticletitle{Evaluating the Design of the R Language}. In \bibinfo{booktitle}{\emph{ECOOP 2012 -- Object-Oriented Programming}}, \bibfield{editor}{\bibinfo{person}{James Noble}} (Ed.). \bibinfo{publisher}{Springer Berlin Heidelberg}, \bibinfo{address}{Berlin, Heidelberg}, \bibinfo{pages}{104--131}.
\newblock
\showISBNx{978-3-642-31057-7}


\bibitem[Nabi et~al\mbox{.}(2022)]%
        {ma15175910}
\bibfield{author}{\bibinfo{person}{Rao Adeel~Un Nabi}, \bibinfo{person}{Muhammad~Yasin Naz}, \bibinfo{person}{Shazia Shukrullah}, \bibinfo{person}{Madiha Ghamkhar}, \bibinfo{person}{Najeeb~Ur Rehman}, \bibinfo{person}{Muhammad Irfan}, \bibinfo{person}{Ali~O. Alqarni}, \bibinfo{person}{Stanisław Legutko}, \bibinfo{person}{Izabela Kruszelnicka}, \bibinfo{person}{Dobrochna Ginter-Kramarczyk}, \bibinfo{person}{Marek Ochowiak}, \bibinfo{person}{Sylwia Włodarczak}, \bibinfo{person}{Andżelika Krupińska}, {and} \bibinfo{person}{Magdalena Matuszak}.} \bibinfo{year}{2022}\natexlab{}.
\newblock \showarticletitle{Analysis of Statistically Predicted Rate Constants for Pyrolysis of High-Density Plastic Using R Software}.
\newblock \bibinfo{journal}{\emph{Materials}} \bibinfo{volume}{15}, \bibinfo{number}{17} (\bibinfo{year}{2022}).
\newblock
\showISSN{1996-1944}
\urldef\tempurl%
\url{https://doi.org/10.3390/ma15175910}
\showDOI{\tempurl}


\bibitem[Nashid et~al\mbox{.}(2023)]%
        {nashid2023retrieval}
\bibfield{author}{\bibinfo{person}{Noor Nashid}, \bibinfo{person}{Mifta Sintaha}, {and} \bibinfo{person}{Ali Mesbah}.} \bibinfo{year}{2023}\natexlab{}.
\newblock \showarticletitle{Retrieval-based prompt selection for code-related few-shot learning}. In \bibinfo{booktitle}{\emph{2023 IEEE/ACM 45th International Conference on Software Engineering (ICSE)}}. IEEE, \bibinfo{pages}{2450--2462}.
\newblock


\bibitem[Nijkamp et~al\mbox{.}(2022)]%
        {CodeGen}
\bibfield{author}{\bibinfo{person}{Erik Nijkamp}, \bibinfo{person}{Bo Pang}, \bibinfo{person}{Hiroaki Hayashi}, \bibinfo{person}{Lifu Tu}, \bibinfo{person}{Haiquan Wang}, \bibinfo{person}{Yingbo Zhou}, \bibinfo{person}{Silvio Savarese}, {and} \bibinfo{person}{Caiming Xiong}.} \bibinfo{year}{2022}\natexlab{}.
\newblock \showarticletitle{CodeGen: An Open Large Language Model for Code with Multi-Turn Program Synthesis}. In \bibinfo{booktitle}{\emph{International Conference on Learning Representations}}.
\newblock
\urldef\tempurl%
\url{https://api.semanticscholar.org/CorpusID:252668917}
\showURL{%
\tempurl}


\bibitem[Ningthoujam et~al\mbox{.}(2023)]%
        {ningthoujam2023r}
\bibfield{author}{\bibinfo{person}{Sanjoy~Singh Ningthoujam}, \bibinfo{person}{Rajat Nath}, \bibinfo{person}{Sibashish Kityania}, \bibinfo{person}{Pranab~Behari Mazumder}, \bibinfo{person}{Manabendra Dutta~Choudhury}, \bibinfo{person}{Anupam~Das Talukdar}, \bibinfo{person}{Lutfun Nahar}, {and} \bibinfo{person}{Satyajit~D Sarker}.} \bibinfo{year}{2023}\natexlab{}.
\newblock \showarticletitle{R software for QSAR analysis in phytopharmacological studies}.
\newblock \bibinfo{journal}{\emph{Phytochemical Analysis}} (\bibinfo{year}{2023}).
\newblock


\bibitem[{Niu} et~al\mbox{.}(2023)]%
        {comparison2023}
\bibfield{author}{\bibinfo{person}{Changan {Niu}}, \bibinfo{person}{Chuanyi {Li}}, \bibinfo{person}{Vincent {Ng}}, \bibinfo{person}{Dongxiao {Chen}}, \bibinfo{person}{Jidong {Ge}}, {and} \bibinfo{person}{Bin {Luo}}.} \bibinfo{year}{2023}\natexlab{}.
\newblock \showarticletitle{{An Empirical Comparison of Pre-Trained Models of Source Code}}.
\newblock \bibinfo{journal}{\emph{arXiv e-prints}}, Article \bibinfo{articleno}{arXiv:2302.04026} (\bibinfo{date}{Feb.} \bibinfo{year}{2023}), \bibinfo{numpages}{arXiv:2302.04026}~pages.
\newblock
\urldef\tempurl%
\url{https://doi.org/10.48550/arXiv.2302.04026}
\showDOI{\tempurl}
\showeprint[arxiv]{2302.04026}~[cs.SE]


\bibitem[Nordmann et~al\mbox{.}(2022)]%
        {Rit}
\bibfield{author}{\bibinfo{person}{Emily Nordmann}, \bibinfo{person}{Phil McAleer}, \bibinfo{person}{Wilhelmiina Toivo}, \bibinfo{person}{Helena Paterson}, {and} \bibinfo{person}{Lisa~M. DeBruine}.} \bibinfo{year}{2022}\natexlab{}.
\newblock \showarticletitle{Data Visualization Using R for Researchers Who Do Not Use R}.
\newblock \bibinfo{journal}{\emph{Advances in Methods and Practices in Psychological Science}} \bibinfo{volume}{5}, \bibinfo{number}{2} (\bibinfo{year}{2022}), \bibinfo{pages}{25152459221074654}.
\newblock
\urldef\tempurl%
\url{https://doi.org/10.1177/25152459221074654}
\showDOI{\tempurl}
\showeprint{https://doi.org/10.1177/25152459221074654}


\bibitem[Ooms(2013)]%
        {DependencyVersioningR}
\bibfield{author}{\bibinfo{person}{Jeroen Ooms}.} \bibinfo{year}{2013}\natexlab{}.
\newblock \showarticletitle{Possible Directions for Improving Dependency Versioning in {R}}.
\newblock \bibinfo{journal}{\emph{CoRR}}  \bibinfo{volume}{abs/1303.2140} (\bibinfo{year}{2013}).
\newblock
\showeprint[arXiv]{1303.2140}
\urldef\tempurl%
\url{http://arxiv.org/abs/1303.2140}
\showURL{%
\tempurl}


\bibitem[Ouyang et~al\mbox{.}(2024)]%
        {10.1145/3697010}
\bibfield{author}{\bibinfo{person}{Shuyin Ouyang}, \bibinfo{person}{Jie~M. Zhang}, \bibinfo{person}{Mark Harman}, {and} \bibinfo{person}{Meng Wang}.} \bibinfo{year}{2024}\natexlab{}.
\newblock \showarticletitle{An Empirical Study of the Non-determinism of ChatGPT in Code Generation}.
\newblock \bibinfo{journal}{\emph{ACM Trans. Softw. Eng. Methodol.}} (\bibinfo{date}{Sept.} \bibinfo{year}{2024}).
\newblock
\showISSN{1049-331X}
\urldef\tempurl%
\url{https://doi.org/10.1145/3697010}
\showDOI{\tempurl}
\newblock
\shownote{Just Accepted}.


\bibitem[Petukhov et~al\mbox{.}(2022)]%
        {Methodnameimport}
\bibfield{author}{\bibinfo{person}{Maxim Petukhov}, \bibinfo{person}{Evelina Gudauskayte}, \bibinfo{person}{Arman Kaliyev}, \bibinfo{person}{Mikhail Oskin}, \bibinfo{person}{Dmitry Ivanov}, {and} \bibinfo{person}{Qianxiang Wang}.} \bibinfo{year}{2022}\natexlab{}.
\newblock \showarticletitle{Method Name Prediction for Automatically Generated Unit Tests}. In \bibinfo{booktitle}{\emph{2022 International Conference on Code Quality (ICCQ)}}. \bibinfo{pages}{29--38}.
\newblock
\urldef\tempurl%
\url{https://doi.org/10.1109/ICCQ53703.2022.9763112}
\showDOI{\tempurl}


\bibitem[Plakidas et~al\mbox{.}(2017)]%
        {PLAKIDAS2017119}
\bibfield{author}{\bibinfo{person}{Konstantinos Plakidas}, \bibinfo{person}{Daniel Schall}, {and} \bibinfo{person}{Uwe Zdun}.} \bibinfo{year}{2017}\natexlab{}.
\newblock \showarticletitle{Evolution of the R software ecosystem: Metrics, relationships, and their impact on qualities}.
\newblock \bibinfo{journal}{\emph{Journal of Systems and Software}}  \bibinfo{volume}{132} (\bibinfo{year}{2017}), \bibinfo{pages}{119--146}.
\newblock
\showISSN{0164-1212}
\urldef\tempurl%
\url{https://doi.org/10.1016/j.jss.2017.06.095}
\showDOI{\tempurl}


\bibitem[Popov et~al\mbox{.}(2021)]%
        {CodeCompletionModel}
\bibfield{author}{\bibinfo{person}{Artem Popov}, \bibinfo{person}{Dmitrii Orekhov}, \bibinfo{person}{Denis Litvinov}, \bibinfo{person}{Nikolay Korolev}, {and} \bibinfo{person}{Gleb Morgachev}.} \bibinfo{year}{2021}\natexlab{}.
\newblock \showarticletitle{Time-Efficient Code Completion Model for the {R} Programming Language}. In \bibinfo{booktitle}{\emph{Proceedings of the 1st Workshop on Natural Language Processing for Programming (NLP4Prog 2021)}}. \bibinfo{publisher}{Association for Computational Linguistics}, \bibinfo{address}{Online}, \bibinfo{pages}{34--39}.
\newblock
\urldef\tempurl%
\url{https://doi.org/10.18653/v1/2021.nlp4prog-1.4}
\showDOI{\tempurl}


\bibitem[Radford et~al\mbox{.}(2018)]%
        {GPT2}
\bibfield{author}{\bibinfo{person}{Alec Radford}, \bibinfo{person}{Jeffrey Wu}, \bibinfo{person}{Rewon Child}, \bibinfo{person}{David Luan}, \bibinfo{person}{Dario Amodei}, {and} \bibinfo{person}{Ilya Sutskever}.} \bibinfo{year}{2018}\natexlab{}.
\newblock \showarticletitle{Language Models are Unsupervised Multitask Learners}.
\newblock  (\bibinfo{year}{2018}).
\newblock
\urldef\tempurl%
\url{https://d4mucfpksywv.cloudfront.net/better-language-models/language-models.pdf}
\showURL{%
\tempurl}


\bibitem[Ridnik et~al\mbox{.}(2024)]%
        {ridnik2024code}
\bibfield{author}{\bibinfo{person}{Tal Ridnik}, \bibinfo{person}{Dedy Kredo}, {and} \bibinfo{person}{Itamar Friedman}.} \bibinfo{year}{2024}\natexlab{}.
\newblock \bibinfo{title}{Code Generation with AlphaCodium: From Prompt Engineering to Flow Engineering}.
\newblock
\newblock
\showeprint[arxiv]{2401.08500}~[cs.LG]


\bibitem[Robertson and Zaragoza(2009)]%
        {BM25}
\bibfield{author}{\bibinfo{person}{Stephen Robertson} {and} \bibinfo{person}{Hugo Zaragoza}.} \bibinfo{year}{2009}\natexlab{}.
\newblock \showarticletitle{The Probabilistic Relevance Framework: BM25 and Beyond}.
\newblock \bibinfo{journal}{\emph{Found. Trends Inf. Retr.}} \bibinfo{volume}{3}, \bibinfo{number}{4} (\bibinfo{date}{April} \bibinfo{year}{2009}), \bibinfo{pages}{333–389}.
\newblock
\showISSN{1554-0669}
\urldef\tempurl%
\url{https://doi.org/10.1561/1500000019}
\showDOI{\tempurl}


\bibitem[Robinson et~al\mbox{.}(2022)]%
        {errorIdentification}
\bibfield{author}{\bibinfo{person}{Derek Robinson}, \bibinfo{person}{Neil~A. Ernst}, \bibinfo{person}{Enrique~Larios Vargas}, {and} \bibinfo{person}{Margaret-Anne~D. Storey}.} \bibinfo{year}{2022}\natexlab{}.
\newblock \showarticletitle{Error Identification Strategies for Python Jupyter Notebooks}. In \bibinfo{booktitle}{\emph{Proceedings of the 30th IEEE/ACM International Conference on Program Comprehension}} (Virtual Event) \emph{(\bibinfo{series}{ICPC '22})}. \bibinfo{publisher}{Association for Computing Machinery}, \bibinfo{address}{New York, NY, USA}, \bibinfo{pages}{253–263}.
\newblock
\showISBNx{9781450392983}
\urldef\tempurl%
\url{https://doi.org/10.1145/3524610.3529156}
\showDOI{\tempurl}


\bibitem[Roziere et~al\mbox{.}(2023)]%
        {codellama}
\bibfield{author}{\bibinfo{person}{Baptiste Roziere}, \bibinfo{person}{Jonas Gehring}, \bibinfo{person}{Fabian Gloeckle}, \bibinfo{person}{Sten Sootla}, \bibinfo{person}{Itai Gat}, \bibinfo{person}{Xiaoqing~Ellen Tan}, \bibinfo{person}{Yossi Adi}, \bibinfo{person}{Jingyu Liu}, \bibinfo{person}{Tal Remez}, \bibinfo{person}{J{\'e}r{\'e}my Rapin}, {et~al\mbox{.}}} \bibinfo{year}{2023}\natexlab{}.
\newblock \showarticletitle{Code llama: Open foundation models for code}.
\newblock \bibinfo{journal}{\emph{arXiv preprint arXiv:2308.12950}} (\bibinfo{year}{2023}).
\newblock


\bibitem[Shahbazi and Fard(2023)]%
        {apicontext2com}
\bibfield{author}{\bibinfo{person}{R. Shahbazi} {and} \bibinfo{person}{F. Fard}.} \bibinfo{year}{2023}\natexlab{}.
\newblock \showarticletitle{APIContext2Com: Code Comment Generation by Incorporating Pre-Defined API Documentation}. In \bibinfo{booktitle}{\emph{2023 IEEE/ACM 31st International Conference on Program Comprehension (ICPC)}}. \bibinfo{publisher}{IEEE Computer Society}, \bibinfo{address}{Los Alamitos, CA, USA}, \bibinfo{pages}{13--24}.
\newblock
\urldef\tempurl%
\url{https://doi.org/10.1109/ICPC58990.2023.00012}
\showDOI{\tempurl}


\bibitem[Sharma et~al\mbox{.}(2022a)]%
        {lamner}
\bibfield{author}{\bibinfo{person}{R. Sharma}, \bibinfo{person}{F. Chen}, {and} \bibinfo{person}{F. Fard}.} \bibinfo{year}{2022}\natexlab{a}.
\newblock \showarticletitle{LAMNER: Code Comment Generation Using Character Language Model and Named Entity Recognition}. In \bibinfo{booktitle}{\emph{2022 IEEE/ACM 30th International Conference on Program Comprehension (ICPC)}}. \bibinfo{publisher}{IEEE Computer Society}, \bibinfo{address}{Los Alamitos, CA, USA}, \bibinfo{pages}{48--59}.
\newblock
\urldef\tempurl%
\url{https://doi.org/10.1145/3524610.3527924}
\showDOI{\tempurl}


\bibitem[Sharma et~al\mbox{.}(2022b)]%
        {Sharma2022}
\bibfield{author}{\bibinfo{person}{Rishab Sharma}, \bibinfo{person}{Ramin Shahbazi}, \bibinfo{person}{Fatemeh~H. Fard}, \bibinfo{person}{Zadia Codabux}, {and} \bibinfo{person}{Melina Vidoni}.} \bibinfo{year}{2022}\natexlab{b}.
\newblock \showarticletitle{{Self-Admitted Technical Debt in R: Detection and Causes}}.
\newblock \bibinfo{journal}{\emph{Automated Software Engineering}} \bibinfo{volume}{29}, \bibinfo{number}{2} (\bibinfo{date}{25 Aug} \bibinfo{year}{2022}), \bibinfo{pages}{53}.
\newblock
\showISSN{1573-7535}
\urldef\tempurl%
\url{https://doi.org/10.1007/s10515-022-00358-6}
\showDOI{\tempurl}


\bibitem[Staples(2023)]%
        {evolutionR}
\bibfield{author}{\bibinfo{person}{Timothy~L Staples}.} \bibinfo{year}{2023}\natexlab{}.
\newblock \showarticletitle{Expansion and evolution of the R programming language}.
\newblock \bibinfo{journal}{\emph{Royal Society Open Science}} \bibinfo{volume}{10}, \bibinfo{number}{4} (\bibinfo{year}{2023}), \bibinfo{pages}{221550}.
\newblock


\bibitem[Titov et~al\mbox{.}(2022)]%
        {structureImprovement}
\bibfield{author}{\bibinfo{person}{Sergey Titov}, \bibinfo{person}{Yaroslav Golubev}, {and} \bibinfo{person}{Timofey Bryksin}.} \bibinfo{year}{2022}\natexlab{}.
\newblock \showarticletitle{ReSplit: Improving the Structure of Jupyter Notebooks by Re-Splitting Their Cells}. In \bibinfo{booktitle}{\emph{2022 IEEE International Conference on Software Analysis, Evolution and Reengineering (SANER)}}. \bibinfo{pages}{492--496}.
\newblock
\urldef\tempurl%
\url{https://doi.org/10.1109/SANER53432.2022.00066}
\showDOI{\tempurl}


\bibitem[Touvron et~al\mbox{.}(2023)]%
        {llama2}
\bibfield{author}{\bibinfo{person}{Hugo Touvron}, \bibinfo{person}{Louis Martin}, \bibinfo{person}{Kevin Stone}, \bibinfo{person}{Peter Albert}, \bibinfo{person}{Amjad Almahairi}, \bibinfo{person}{Yasmine Babaei}, \bibinfo{person}{Nikolay Bashlykov}, \bibinfo{person}{Soumya Batra}, \bibinfo{person}{Prajjwal Bhargava}, \bibinfo{person}{Shruti Bhosale}, {et~al\mbox{.}}} \bibinfo{year}{2023}\natexlab{}.
\newblock \showarticletitle{Llama 2: Open foundation and fine-tuned chat models}.
\newblock \bibinfo{journal}{\emph{arXiv preprint arXiv:2307.09288}} (\bibinfo{year}{2023}).
\newblock


\bibitem[Traykov et~al\mbox{.}(2018)]%
        {Rrisk}
\bibfield{author}{\bibinfo{person}{Metodi Traykov}, \bibinfo{person}{Miglena Trencheva}, \bibinfo{person}{Elena Stavrova}, \bibinfo{person}{Radoslav Mavrevski}, {and} \bibinfo{person}{Ivan Trenchev}.} \bibinfo{year}{2018}\natexlab{}.
\newblock \showarticletitle{Risk analysis in the economics through R Language}.
\newblock  (\bibinfo{date}{05} \bibinfo{year}{2018}).
\newblock


\bibitem[Troshin and Chirkova(2022)]%
        {troshin2022probing}
\bibfield{author}{\bibinfo{person}{Sergey Troshin} {and} \bibinfo{person}{Nadezhda Chirkova}.} \bibinfo{year}{2022}\natexlab{}.
\newblock \showarticletitle{Probing pretrained models of source code}.
\newblock \bibinfo{journal}{\emph{arXiv preprint arXiv:2202.08975}} (\bibinfo{year}{2022}).
\newblock


\bibitem[Vaswani et~al\mbox{.}(2017)]%
        {Attention}
\bibfield{author}{\bibinfo{person}{Ashish Vaswani}, \bibinfo{person}{Noam Shazeer}, \bibinfo{person}{Niki Parmar}, \bibinfo{person}{Jakob Uszkoreit}, \bibinfo{person}{Llion Jones}, \bibinfo{person}{Aidan~N. Gomez}, \bibinfo{person}{Lukasz Kaiser}, {and} \bibinfo{person}{Illia Polosukhin}.} \bibinfo{year}{2017}\natexlab{}.
\newblock \bibinfo{title}{Attention Is All You Need}.
\newblock
\newblock
\urldef\tempurl%
\url{https://doi.org/10.48550/ARXIV.1706.03762}
\showDOI{\tempurl}


\bibitem[Vidoni(2022)]%
        {Vidoni2022b}
\bibfield{author}{\bibinfo{person}{Melina Vidoni}.} \bibinfo{year}{2022}\natexlab{}.
\newblock \showarticletitle{{Understanding Roxygen Package Documentation in R}}.
\newblock \bibinfo{journal}{\emph{{Journal of Systems and Software}}}  \bibinfo{volume}{188} (\bibinfo{year}{2022}), \bibinfo{pages}{111265}.
\newblock
\showISSN{0164-1212}
\urldef\tempurl%
\url{https://doi.org/10.1016/j.jss.2022.111265}
\showDOI{\tempurl}


\bibitem[Villanueva and Chen(2019)]%
        {villanueva2019ggplot2}
\bibfield{author}{\bibinfo{person}{Randle Aaron~M Villanueva} {and} \bibinfo{person}{Zhuo~Job Chen}.} \bibinfo{year}{2019}\natexlab{}.
\newblock \bibinfo{title}{ggplot2: elegant graphics for data analysis}.
\newblock
\newblock


\bibitem[Wan et~al\mbox{.}(2022)]%
        {whatdotheycapture}
\bibfield{author}{\bibinfo{person}{Yao Wan}, \bibinfo{person}{Wei Zhao}, \bibinfo{person}{Hongyu Zhang}, \bibinfo{person}{Yulei Sui}, \bibinfo{person}{Guandong Xu}, {and} \bibinfo{person}{Hai Jin}.} \bibinfo{year}{2022}\natexlab{}.
\newblock \showarticletitle{What Do They Capture? A Structural Analysis of Pre-Trained Language Models for Source Code}. In \bibinfo{booktitle}{\emph{Proceedings of the 44th International Conference on Software Engineering}} (Pittsburgh, Pennsylvania) \emph{(\bibinfo{series}{ICSE '22})}. \bibinfo{publisher}{Association for Computing Machinery}, \bibinfo{address}{New York, NY, USA}, \bibinfo{pages}{2377–2388}.
\newblock
\showISBNx{9781450392211}
\urldef\tempurl%
\url{https://doi.org/10.1145/3510003.3510050}
\showDOI{\tempurl}


\bibitem[Wang et~al\mbox{.}(2021a)]%
        {wang2021makes}
\bibfield{author}{\bibinfo{person}{April Wang}, \bibinfo{person}{Dakuo Wang}, \bibinfo{person}{Jaimie Drozdal}, \bibinfo{person}{Xuye Liu}, \bibinfo{person}{Soya Park}, \bibinfo{person}{Steve Oney}, {and} \bibinfo{person}{Christopher Brooks}.} \bibinfo{year}{2021}\natexlab{a}.
\newblock \showarticletitle{What Makes a Well Documented Notebook? A Case Study of Data Scientists' Documentation Practices in Kaggle Notebooks}. ACM| CHI Conference on Human Factors in Computing Systems Extended Abstracts.
\newblock


\bibitem[Wang et~al\mbox{.}(2023)]%
        {wang}
\bibfield{author}{\bibinfo{person}{Deze Wang}, \bibinfo{person}{Boxing Chen}, \bibinfo{person}{Shanshan Li}, \bibinfo{person}{Wei Luo}, \bibinfo{person}{Shaoliang Peng}, \bibinfo{person}{Wei Dong}, {and} \bibinfo{person}{Xiangke Liao}.} \bibinfo{year}{2023}\natexlab{}.
\newblock \showarticletitle{One Adapter for All Programming Languages? Adapter Tuning for Code Search and Summarization}. In \bibinfo{booktitle}{\emph{Proceedings of the 45th International Conference on Software Engineering}} (Melbourne, Victoria, Australia) \emph{(\bibinfo{series}{ICSE '23})}. \bibinfo{publisher}{IEEE Press}, \bibinfo{pages}{5–16}.
\newblock
\showISBNx{9781665457019}
\urldef\tempurl%
\url{https://doi.org/10.1109/ICSE48619.2023.00013}
\showDOI{\tempurl}


\bibitem[Wang et~al\mbox{.}(2021b)]%
        {codet5}
\bibfield{author}{\bibinfo{person}{Yue Wang}, \bibinfo{person}{Weishi Wang}, \bibinfo{person}{Shafiq Joty}, {and} \bibinfo{person}{Steven~C.H. Hoi}.} \bibinfo{year}{2021}\natexlab{b}.
\newblock \showarticletitle{{C}ode{T}5: Identifier-aware Unified Pre-trained Encoder-Decoder Models for Code Understanding and Generation}. In \bibinfo{booktitle}{\emph{Proceedings of the 2021 Conference on Empirical Methods in Natural Language Processing}}. \bibinfo{publisher}{Association for Computational Linguistics}, \bibinfo{address}{Online and Punta Cana, Dominican Republic}, \bibinfo{pages}{8696--8708}.
\newblock
\urldef\tempurl%
\url{https://doi.org/10.18653/v1/2021.emnlp-main.685}
\showDOI{\tempurl}


\bibitem[Wickham(2014)]%
        {advancedR}
\bibfield{author}{\bibinfo{person}{H. Wickham}.} \bibinfo{year}{2014}\natexlab{}.
\newblock \bibinfo{booktitle}{\emph{Advanced R}}.
\newblock \bibinfo{publisher}{Taylor \& Francis}.
\newblock
\showISBNx{9781466586963}
\showLCCN{2012278240}
\urldef\tempurl%
\url{https://books.google.ca/books?id=PFHFNAEACAAJ}
\showURL{%
\tempurl}


\bibitem[Wickham et~al\mbox{.}(2019)]%
        {wickham2019welcome}
\bibfield{author}{\bibinfo{person}{Hadley Wickham}, \bibinfo{person}{Mara Averick}, \bibinfo{person}{Jennifer Bryan}, \bibinfo{person}{Winston Chang}, \bibinfo{person}{Lucy~D'Agostino McGowan}, \bibinfo{person}{Romain Fran{\c{c}}ois}, \bibinfo{person}{Garrett Grolemund}, \bibinfo{person}{Alex Hayes}, \bibinfo{person}{Lionel Henry}, \bibinfo{person}{Jim Hester}, {et~al\mbox{.}}} \bibinfo{year}{2019}\natexlab{}.
\newblock \showarticletitle{Welcome to the Tidyverse}.
\newblock \bibinfo{journal}{\emph{Journal of open source software}} \bibinfo{volume}{4}, \bibinfo{number}{43} (\bibinfo{year}{2019}), \bibinfo{pages}{1686}.
\newblock


\bibitem[Wrenn et~al\mbox{.}(2023)]%
        {wrenn2023dependently}
\bibfield{author}{\bibinfo{person}{John Wrenn}, \bibinfo{person}{Anjali Pal}, \bibinfo{person}{Alexa VanHattum}, {and} \bibinfo{person}{Shriram Krishnamurthi}.} \bibinfo{year}{2023}\natexlab{}.
\newblock \showarticletitle{Dependently Typing R Vectors, Arrays, and Matrices}.
\newblock \bibinfo{journal}{\emph{arXiv preprint arXiv:2304.04265}} (\bibinfo{year}{2023}).
\newblock


\bibitem[Yang et~al\mbox{.}(2021)]%
        {subtleBugs}
\bibfield{author}{\bibinfo{person}{Chenyang Yang}, \bibinfo{person}{Shurui Zhou}, \bibinfo{person}{Jin~L.C. Guo}, {and} \bibinfo{person}{Christian Kästner}.} \bibinfo{year}{2021}\natexlab{}.
\newblock \showarticletitle{Subtle Bugs Everywhere: Generating Documentation for Data Wrangling Code}. In \bibinfo{booktitle}{\emph{2021 36th IEEE/ACM International Conference on Automated Software Engineering (ASE)}}. \bibinfo{pages}{304--316}.
\newblock
\urldef\tempurl%
\url{https://doi.org/10.1109/ASE51524.2021.9678520}
\showDOI{\tempurl}


\bibitem[Yun et~al\mbox{.}(2024)]%
        {YUN2024112149}
\bibfield{author}{\bibinfo{person}{Shangbo Yun}, \bibinfo{person}{Shuhuai Lin}, \bibinfo{person}{Xiaodong Gu}, {and} \bibinfo{person}{Beijun Shen}.} \bibinfo{year}{2024}\natexlab{}.
\newblock \showarticletitle{Project-specific code summarization with in-context learning}.
\newblock \bibinfo{journal}{\emph{Journal of Systems and Software}}  \bibinfo{volume}{216} (\bibinfo{year}{2024}), \bibinfo{pages}{112149}.
\newblock
\showISSN{0164-1212}
\urldef\tempurl%
\url{https://doi.org/10.1016/j.jss.2024.112149}
\showDOI{\tempurl}


\bibitem[Zanella and Liu(2020)]%
        {zanella2020social}
\bibfield{author}{\bibinfo{person}{Gianluca Zanella} {and} \bibinfo{person}{Charles~Z Liu}.} \bibinfo{year}{2020}\natexlab{}.
\newblock \showarticletitle{A social network perspective on the success of open source software: the case of R packages}.
\newblock  (\bibinfo{year}{2020}).
\newblock


\bibitem[Zhang et~al\mbox{.}(2016a)]%
        {10.1145/2884781.2884839}
\bibfield{author}{\bibinfo{person}{Feng Zhang}, \bibinfo{person}{Quan Zheng}, \bibinfo{person}{Ying Zou}, {and} \bibinfo{person}{Ahmed~E. Hassan}.} \bibinfo{year}{2016}\natexlab{a}.
\newblock \showarticletitle{Cross-project defect prediction using a connectivity-based unsupervised classifier}. In \bibinfo{booktitle}{\emph{Proceedings of the 38th International Conference on Software Engineering}} (Austin, Texas) \emph{(\bibinfo{series}{ICSE '16})}. \bibinfo{publisher}{Association for Computing Machinery}, \bibinfo{address}{New York, NY, USA}, \bibinfo{pages}{309–320}.
\newblock
\showISBNx{9781450339001}
\urldef\tempurl%
\url{https://doi.org/10.1145/2884781.2884839}
\showDOI{\tempurl}


\bibitem[Zhang et~al\mbox{.}(2016b)]%
        {cross}
\bibfield{author}{\bibinfo{person}{Feng Zhang}, \bibinfo{person}{Quan Zheng}, \bibinfo{person}{Ying Zou}, {and} \bibinfo{person}{Ahmed~E. Hassan}.} \bibinfo{year}{2016}\natexlab{b}.
\newblock \showarticletitle{Cross-project defect prediction using a connectivity-based unsupervised classifier}. In \bibinfo{booktitle}{\emph{Proceedings of the 38th International Conference on Software Engineering}} (Austin, Texas) \emph{(\bibinfo{series}{ICSE '16})}. \bibinfo{publisher}{Association for Computing Machinery}, \bibinfo{address}{New York, NY, USA}, \bibinfo{pages}{309–320}.
\newblock
\showISBNx{9781450339001}
\urldef\tempurl%
\url{https://doi.org/10.1145/2884781.2884839}
\showDOI{\tempurl}


\bibitem[Zhang et~al\mbox{.}(2022)]%
        {zhang2022coral}
\bibfield{author}{\bibinfo{person}{Ge Zhang}, \bibinfo{person}{Mike~A Merrill}, \bibinfo{person}{Yang Liu}, \bibinfo{person}{Jeffrey Heer}, {and} \bibinfo{person}{Tim Althoff}.} \bibinfo{year}{2022}\natexlab{}.
\newblock \showarticletitle{Coral: Code representation learning with weakly-supervised transformers for analyzing data analysis}.
\newblock \bibinfo{journal}{\emph{EPJ Data Science}} \bibinfo{volume}{11}, \bibinfo{number}{1} (\bibinfo{year}{2022}), \bibinfo{pages}{14}.
\newblock


\bibitem[Zhao et~al\mbox{.}(2021)]%
        {zhao2021propensity}
\bibfield{author}{\bibinfo{person}{Qin-Yu Zhao}, \bibinfo{person}{Jing-Chao Luo}, \bibinfo{person}{Ying Su}, \bibinfo{person}{Yi-Jie Zhang}, \bibinfo{person}{Guo-Wei Tu}, {and} \bibinfo{person}{Zhe Luo}.} \bibinfo{year}{2021}\natexlab{}.
\newblock \showarticletitle{Propensity score matching with R: conventional methods and new features}.
\newblock \bibinfo{journal}{\emph{Annals of translational medicine}} \bibinfo{volume}{9}, \bibinfo{number}{9} (\bibinfo{year}{2021}).
\newblock


\bibitem[Zhou et~al\mbox{.}(2022)]%
        {zhoudocprompting}
\bibfield{author}{\bibinfo{person}{Shuyan Zhou}, \bibinfo{person}{Uri Alon}, \bibinfo{person}{Frank~F Xu}, \bibinfo{person}{Zhengbao Jiang}, {and} \bibinfo{person}{Graham Neubig}.} \bibinfo{year}{2022}\natexlab{}.
\newblock \showarticletitle{DocPrompting: Generating Code by Retrieving the Docs}. In \bibinfo{booktitle}{\emph{The Eleventh International Conference on Learning Representations}}.
\newblock


\bibitem[Zhu and Pan(2019)]%
        {Zhu2019AutomaticCS}
\bibfield{author}{\bibinfo{person}{Yuxiang Zhu} {and} \bibinfo{person}{Minxue Pan}.} \bibinfo{year}{2019}\natexlab{}.
\newblock \showarticletitle{Automatic Code Summarization: A Systematic Literature Review}.
\newblock \bibinfo{journal}{\emph{ArXiv}}  \bibinfo{volume}{abs/1909.04352} (\bibinfo{year}{2019}).
\newblock


\end{thebibliography}

\end{document}